\ifx\pdfsuppresswarningpagegroup\undefined\else\pdfsuppresswarningpagegroup=1\fi\relax
\RequirePackage{silence}
\PassOptionsToPackage{unicode,pdfprintscaling=None,colorlinks,hypertexnames=false}{hyperref}
\documentclass[aps,prx,reprint,nofootinbib,floatfix,unsortedaddress,superscriptaddress,longbibliography,preprintnumbers]{revtex4-2}

\usepackage[normalem]{ulem}
\usepackage{orcidlink,multirow,siunitx,colortbl,dcolumn,ulem,tikz,graphicx,latexsym,amsmath,amssymb,amsfonts,bm,bbm,microtype,xspace,placeins,float,mathtools,xcolor,makecell,tabularx, booktabs, longtable, supertabular,subfigure}

\usepackage[export]{adjustbox}
\usepackage{textgreek}
\usepackage{hyperref}
\usepackage[capitalise]{cleveref}
\usepackage{array}

\usetikzlibrary{decorations.markings, arrows, decorations.pathmorphing}
\graphicspath{{pic/}}
\allowdisplaybreaks

\newlength{\FigureWidth}
\newlength{\RowHeight}
\newcolumntype{L}[1]{>{\raggedright\arraybackslash\rule{0pt}{\RowHeight}}p{#1}}
\newcolumntype{C}[1]{>{\centering\arraybackslash\rule{0pt}{\RowHeight}}p{#1}}
\newcolumntype{R}[1]{>{\raggedleft\arraybackslash\rule{0pt}{\RowHeight}}p{#1}}
\newcolumntype{M}[1]{>{\centering\arraybackslash\rule{0pt}{\RowHeight}}m{#1}}
\newcommand\TopRule{\Xhline{0.08em}}

\newcommand\MidRule{\Xhline{0.03em}}
\newcommand\BotRule{\Xhline{0.08em}}

\makeatletter
\newcommand\showtitleinbib{{\escapechar=`\\ \immediate\write\@auxout{%
\csname citation{REVTEX42Control}\endcsname^^J%
\csname citation{apsrev42Control}\endcsname
}}}
\makeatother

\newcommand{\sing}{{\mathbf{1}}}
\newcommand{\trip}{{\mathbf{3}}}
\newcommand{\atrip}{{\mathbf{\bar{3}}}}

\newcommand{\cH}{{\cal H}}

\newcommand{\ungroup}[1]{#1}
\newcommand{\withbreak}[1]{\expandafter\ungroup#1}

\def\dounbracket[#1]{#1}
\newcommand{\txtw}{{\mathsf{w}}}

\newcommand{\SU}{{\mathrm{SU}}}

\def\ket#1{\left| #1\right\rangle}

\renewcommand\vec\mathbf

\newcommand\redsout{\bgroup\markoverwith{\textcolor{red}{\rule[0.5ex]{2pt}{1.4pt}}}\ULon}

\newcommand\dootimesall[2]{\ifx0#1\else\mathbf{#1}\ifx0#2\else\def\mytmp{\otimes\dootimesall{#2}}\expandafter\expandafter\expandafter\mytmp\fi\fi}

\newcommand{\RN}[1]{%
  \textup{\uppercase\expandafter{\romannumeral#1}}%
}

\usepackage{relsize}

\usepackage[acronyms, nohypertypes={acronym}, nopostdot, style=super, nonumberlist, toc]{glossaries}
\setacronymstyle{long-sc-short}
\glsenableentrycount

\newcommand\ac[1]{\gls{#1}}

\newcommand\acp[1]{\glspl{#1}}

\newacronym{PNA}{pna}{particle-number-algorithm}
\newacronym{SFA}{sfa}{spin-flip-algorithm}

\newacronym{WF}{wf}{Wilson--Fisher}
\newacronym{AF}{af}{asymptotically free}

\newacronym{RG}{rg}{renormalization group}

\newacronym{QIS}{qis}{Quantum Information Science}
\newacronym{PPT}{ppt}{positive-semidefinite partial transpose}
\newacronym{KS}{ks}{Kogut--Susskind}
\newacronym{NPT}{npt}{negative partial transpose}

\newacronym{AS}{as}{Anti-Symmetric}

\newacronym[longplural={conformal field theories}]{CFT}{cft}{conformal field theory}
\newacronym[longplural={lattice field theories}]{LFT}{lft}{lattice field theory}
\newacronym[longplural={effective field theories}]{EFT}{eft}{effective field theory}
\newacronym[longplural={quantum field theories}]{QFT}{qft}{quantum field theory}
\newacronym[longplural={lattice gauge theories}]{LGT}{lgt}{lattice gauge theory}
\newacronym[longplural={monomer-dimer tensor-networks}]{MDTN}{mdtn}{monomer-dimer tensor-network}

\newacronym{YM}{ym}{Yang--Mills}

\newacronym{DMRG}{dmrg}{Density Matrix Renormalization Group}
\newacronym{TFIM}{tfim}{Transverse Field Ising Model}
\newacronym{TSPM}{tspm}{Three State Potts Model}
\newacronym{ICFT}{icft}{Ising-{\acrshort{CFT}}}
\newacronym{E8QFT}{e8qft}{$E_8$-{\acrshort{QFT}}}

\newacronym[]{LOCC}{locc}{Local Operations and Classical Communicaton}
\newacronym[]{OBC}{obc}{open boundary conditions}

\newacronym{MPS}{mps}{matrix product state}

\newacronym{JLP}{jlp}{Jordan--Lee--Preskill}

\newacronym{BBN}{bbn}{big bang nucleosynthesis}

\newacronym{LEC}{lec}{low-energy constant}

\newacronym{QCD}{qcd}{quantum chromodynamics}
\newacronym{MC}{mc}{Monte Carlo}

\newacronym{IR}{ir}{infrared}
\newacronym{UV}{uv}{ultraviolet}

\newacronym{QED}{qed}{quantum electrodynamics}
\newacronym{SNR}{snr}{signal-to-noise ratio}

\newacronym{NLSM}{nlsm}{nonlinear sigma model}

\newacronym{CL}{cl}{Complex Langevin}

\newacronym{CSA}{csa}{Cartan subalgebra}

\newacronym{SSB}{ssb}{spontaneous symmetry breaking}

\newacronym{AFQMC}{afqmc}{auxiliary field quantum Monte Carlo}
\newacronym{iHMC}{ihmc}{imaginary-mass Hybrid Monte Carlo}

\newacronym{MCMC}{mcmc}{Markov Chain Monte Carlo}

\newacronym{QI}{qi}{quantum information}

\newacronym{irrep}{{\rm irrep}}{unitary irreducible representation}

\newacronym{ASQR}{asqr}{antisymmetric qubit regularization}

\begin{document}

\title{Continuum limit of a qubit-regularized SU(3) lattice gauge theory on a plaquette chain}

\author{Rui Xian Siew\,\orcidlink{0000-0002-0745-8853}}%
 \email{ruixian.siew@duke.edu}
\affiliation{ Department of Physics, Box 90305, Duke University, Durham, North Carolina 27708, USA}
\author{Shailesh Chandrasekharan\,\orcidlink{0000-0002-3711-4998}}
\email{sch27@duke.edu}
\affiliation{ Department of Physics, Box 90305, Duke University, Durham, North Carolina 27708, USA}
\author{Tanmoy Bhattacharya\,\orcidlink{0000-0002-1060-652X}}
\email{tanmoy@lanl.gov}
\affiliation{Theoretical Division, Los Alamos National Laboratory, Los Alamos, New Mexico 87545, USA}

\date{\today}

\begin{abstract}
We show that a simple qubit-regularized $\SU(3)$ lattice gauge theory on a plaquette chain admits a continuum limit with massive gauge-invariant excitations, providing a minimal toy model of strong interactions without quarks. By mapping the plaquette-chain Hamiltonian to the three-state quantum clock model in a magnetic field, we demonstrate that the theory can be tuned to a continuum limit governed at short distances by the $\mathbb{Z}_3$ parafermion conformal field theory, which serves as the ultraviolet fixed point. A small relevant magnetic perturbation then drives the system to a massive continuum quantum field theory in the infrared, featuring relativistic particles. In this continuum theory we compute two universal ratios: the mass ratio of the lightest massive particles with opposite charge conjugation, $m^{-}/m^{+} = 1.459(2)$, and $\sqrt{\sigma}/m^{+} = 0.2648(2)$, where $\sigma$ denotes the string tension between a static quark and antiquark.
\end{abstract}

\preprint{LA-UR-26-21250}

\maketitle

Gluons mediate the strong force between quarks, binding them into hadrons such as protons and neutrons. Unlike photons, gluons also interact among themselves and, in the absence of quarks, can form massive bound states known as glueballs. Their dynamics are described by $\SU(3)$ \ac{YM} theory, a non-Abelian gauge theory in the continuum with eight gluons as fundamental degrees of freedom. Despite its conceptual simplicity, determining the glueball spectrum is notoriously difficult. The standard approach is to discretize the theory on a lattice, following Wilson, and compute the energy eigenvalues of the resulting Hamiltonian while taking the thermodynamic and continuum limits~\cite{Athenodorou:2020ani}.

Recently, a new class of \acp{LGT} with finite-dimensional local Hilbert spaces has been proposed using the \ac{MDTN} basis of
conventional \acp{LGT}~\cite{Chandrasekharan:2025smw}. We refer to such constructions as qubit regularizations~\cite{Chandrasekharan:2025Cb}, since they can be encoded in qubits and potentially solved on a quantum computer~\cite{Bauer:2023qgm}. In the philosophy of qubit regularization, a gauge theory is constructed from microscopic lattice quantum degrees of freedom that transform under local gauge transformations. The dynamics is then restricted to the gauge-invariant physical Hilbert space. This construction makes it clear that gauge symmetry is not really a symmetry of the theory, but rather a constraint on the local degrees of freedom in a given basis.  
Indeed, examples abound in highly supersymmetric theories in which the same theory can arise from distinct descriptions using different gauge groups, a property that is probably related to deeper mathematical structures~\cite{Kapustin:2006pk}.
From this perspective, the gauge group itself is not a fundamental characteristic of the physical theory. Rather, gauge theories are defined using a set of quantum degrees of freedom associated with connections between different points in space. These degrees of freedom transform nontrivially under a gauge group, whereas the physical content resides in the gauge-invariant sector. 

It is well known that \acp{LGT} are not uniquely determined by their gauge symmetry alone~\cite{Kogut:1974ag}. Qubit regularization highlights a previously underappreciated source of this non-uniqueness, beyond the lattice geometry and the Hamiltonian couplings: the choice of representation of the gauge group carried by the local lattice fields. Since every second-order critical point defines a continuum quantum field theory, qubit-regularized gauge theories suggest the existence of a richer landscape of possible continuum limits.

Because the space of consistent continuum quantum field theories is expected to be smaller than the space of \acp{LGT}, it is natural to expect that distinct lattice formulations may flow to the same continuum theory. One exciting possibility is that a sufficiently simple qubit-regularized theory could provide an alternative route to studying $(3+1)$-dimensional \ac{YM} theory. Equally plausible, however, is the possibility that qubit-regularized \acp{LGT} do not generically admit nontrivial continuum limits. Even when they do, the resulting continuum theories may belong to universality classes distinct from that of conventional continuum \ac{YM} theory. Our work represents a small step toward understanding this broader landscape.

The first step in determining whether a qubit-regularized \ac{LGT} yields a nontrivial continuum \ac{QFT}, potentially in the universality class of the conventional \ac{YM} theory, is to establish the existence of a relativistic quantum critical point. 
Properties such as asymptotic freedom can arise at such quantum critical points through standard \ac{RG} flows or through more exotic \ac{RG} flows, as discovered recently~\cite{Bhattacharya:2020gpm,Maiti:2023kpn}. 
Assuming such a quantum critical point exists in a pure qubit-regularized \ac{LGT} without matter fields in $3+1$ dimensions, a relevant perturbation would naturally yield massive relativistic particles in the continuum limit. 
Identifying these immediately as the glueballs of conventional \ac{YM} theory would be premature without detailed quantitative comparisons. Nevertheless, such relativistic massive excitations can be defined operationally as ``glueball-like'' particles: they are massive gauge-invariant particle excitations that emerge near a relativistic continuum limit of a microscopic pure non-Abelian \ac{LGT} without matter fields. This terminology emphasizes that the excitations are generated purely by gauge fields of the microscopic lattice theory. We emphasize that glueball-like states are not necessarily identifiable with glueballs of conventional \ac{YM} theory.

Since studying $3+1$ dimensions is difficult, in the present work we focus on constructing an example of a continuum limit in a qubit-regularized $\SU(3)$ \ac{LGT} in a tractable setting. The model we consider is defined on a plaquette chain. We show that it possesses an \ac{UV} fixed point and use it to define a massive relativistic continuum limit with glueball-like particles as defined above. We compute the ratio of the lowest charge-conjugation-odd and charge-conjugation-even masses and, in the presence of static matter fields, determine a dimensionless ratio involving the string tension. 

A similar massive continuum limit was recently identified in a qubit-regularized $\SU(2)$ \ac{LGT}~\cite{2cy4-cp7f}. The present work shows that analogous ideas extend to an $\SU(3)$-motivated construction with the corresponding changes in global symmetry. These results motivate the conjecture that continuum limits of qubit-regularized gauge theories in higher dimensions may emerge as quantum critical points separating confined and deconfined phases.  

Although our model is defined on a ladder geometry, its continuum limit should not be identified with pure \ac{YM} theory in one spatial dimension, which has no local propagating degrees of freedom. By contrast, the plaquette-chain geometry retains a finite transverse structure consisting of two longitudinal chains connected by rungs.
The rung links carry information about the transverse gauge degrees of freedom, which can also be interpreted as adjoint scalars. Gauge-invariant operators involving both the rung and longitudinal links can therefore create local excitations that propagate along the longitudinal direction. Such excitations are present in both the conventional Kogut--Susskind \ac{LGT} and its qubit-regularized counterpart; they are not artifacts of the qubit regularization. It is possible that both theories can be tuned to a continuum limit described by a \ac{YM} theory coupled to adjoint scalar fields in one spatial dimension.

There is an interesting aspect of these adjoint matter degrees of freedom as additional chains are coupled and the transverse direction becomes more fully resolved. The microscopic transverse gauge degrees of freedom of the lattice theory, which are interpreted as adjoint scalars in the finite-width geometry, become the true transverse gauge degrees of freedom of the higher-dimensional theory. For example, in electromagnetism, these scalar-like degrees of freedom can be identified with the fundamental modes propagating along a thin transmission line, or along waveguides with appropriate boundary conditions, whose quanta are often referred to as one-dimensional photons~\cite{wgphoton}. In this sense, the adjoint matter fields can be viewed as remnants of the gauge bosons of the higher-dimensional theory. Since these lowest-energy modes remain unaffected by the smallness of the extra dimension, the connection of these adjoint scalar modes with the gauge bosons of the higher-dimensional theory can persist even when all the other modes are pushed to energies near the \ac{UV} cutoff scale. Thus, in our calculations on the plaquette chain, the rung degrees of freedom retain remnants of the transverse gluons. Such remnants are also familiar in Kaluza–Klein settings. 
For example, five-dimensional \ac{YM} theories compactified on \(S^1\) describe a four-dimensional gauge theory coupled to a scalar field in the adjoint representation of the gauge group~\cite{DelDebbio:2012mr}.
Bound states of such remnants in non-Abelian theories are then expected to capture rudimentary glueball-like physics of the higher-dimensional theory. This interpretation has been discussed previously in a different context~\cite{Antonuccio:1995rb}.

\begin{figure}[t]
\centering
\includegraphics[width=0.48\textwidth]{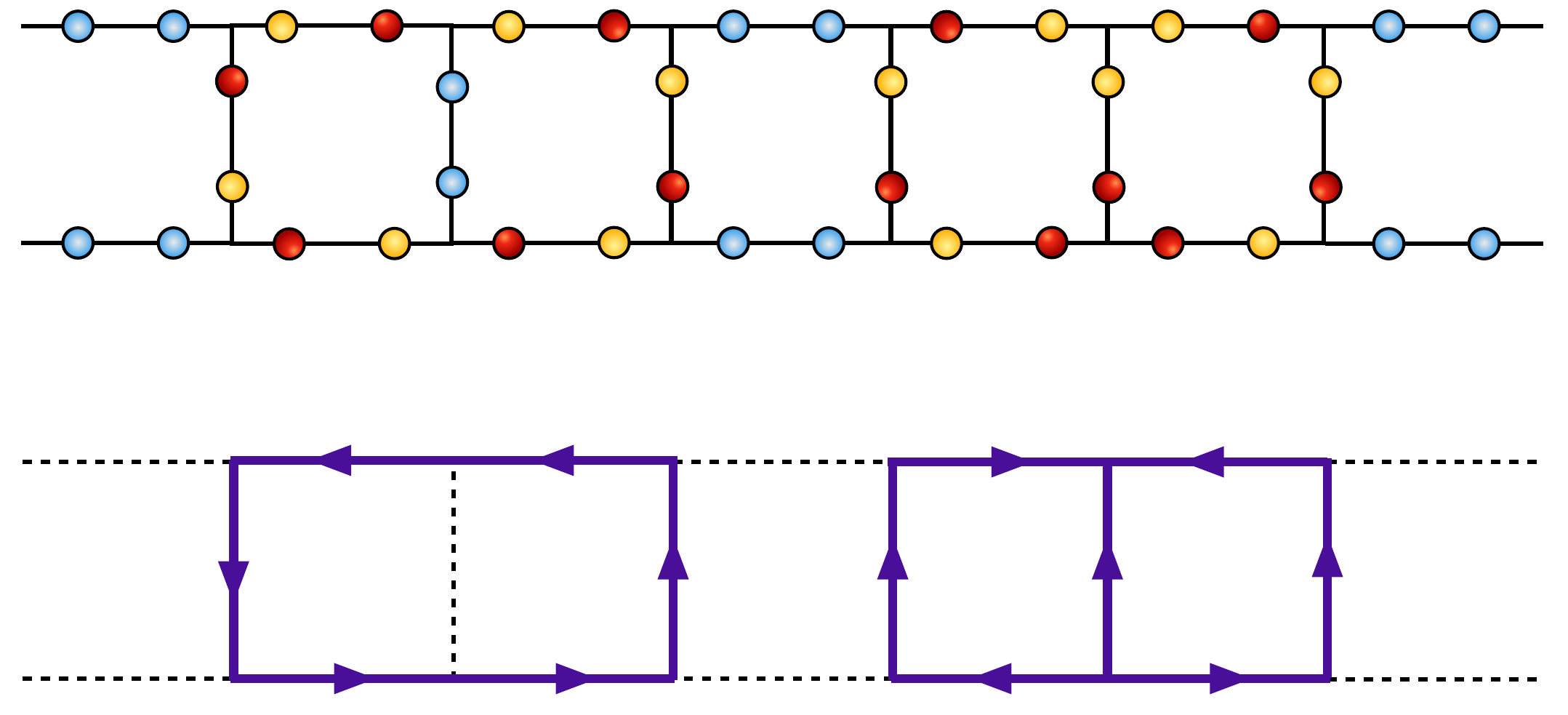}
\caption{Physical basis state of the qubit-regularized $\mathrm{SU}(3)$ \ac{LGT} on a plaquette chain in the \ac{MDTN} basis. Links are represented by dimer tensor states, which may equivalently be viewed as color flux states $\ket{\lambda}$, where $\lambda=\sing$ (blue), $\trip$ (red), and $\atrip$ (yellow) are irreducible representations of $\mathrm{SU}(3)$. The two indices of each tensor are associated with the lattice sites connected by the link and transform in the irreps $\lambda$ and its conjugate $\bar{\lambda}$. Gauge invariance enforces that the irreps meeting at each site combine to a singlet, implying for a plaquette chain the only possibilities are (i) a singlet, a triplet, and an anti-triplet index, (ii) three triplet indices, or (iii) three anti-triplet indices meeting at a site (top). An equivalent representation in terms of color fluxes is shown in the bottom panel, where a dashed line denotes the state $\ket{\sing}$, arrows pointing to the right or upward denote $\ket{\trip}$, and oppositely oriented arrows denote $\ket{\atrip}$.}
\label{fig:su3pchain}
\end{figure}

The physical Hilbert space of the $\mathrm{SU}(3)$ plaquette chain is constructed in the \ac{MDTN} basis~\cite{Chandrasekharan:2025smw}, where gauge degrees of freedom reside on links connecting neighboring sites and are labeled by dimer tensor (equivalently, color-flux) states $\ket{\lambda}$ with $\lambda=\sing,\trip,\atrip$, the irreducible representations of $\mathrm{SU}(3)$. 
A representative physical basis state is shown in \cref{fig:su3pchain}.

The ladder geometry of the plaquette chain allows us to distinguish links along the chain and those along the rungs, which we denote by $\ell_c$ and $\ell_r$, respectively. Elementary plaquettes $P$ are defined as the squares formed by four such links. Owing to this geometry, the physical Hilbert space of the theory naturally splits into three topologically distinct sectors. The first sector, $\cH_{\RN{1}}$, consists of basis states in which the top and bottom chain links of every plaquette $P$ are either $\ket{\sing,\sing}$, $\ket{\trip,\atrip}$, or $\ket{\atrip,\trip}$. The second sector, $\cH_{\RN{2}}$, contains plaquettes in the states $\ket{\atrip,\atrip}$, $\ket{\sing,\trip}$, or $\ket{\trip,\sing}$, while the third sector, $\cH_{\RN{3}}$, contains plaquettes in the states $\ket{\trip,\trip}$, $\ket{\sing,\atrip}$, or $\ket{\atrip,\sing}$. Under charge conjugation $\hat{\mathcal{C}}$, $\trip$ and $\atrip$ are exchanged. Consequently, $\cH_{\RN{1}}$ is invariant under charge conjugation, whereas the sectors $\cH_{\RN{2}}$ and $\cH_{\RN{3}}$ are exchanged into one another. 
Thus, charge conjugation is not a symmetry within either of the latter two sectors separately.

\begin{figure}[t]
\centering
\includegraphics[width=0.48\textwidth]{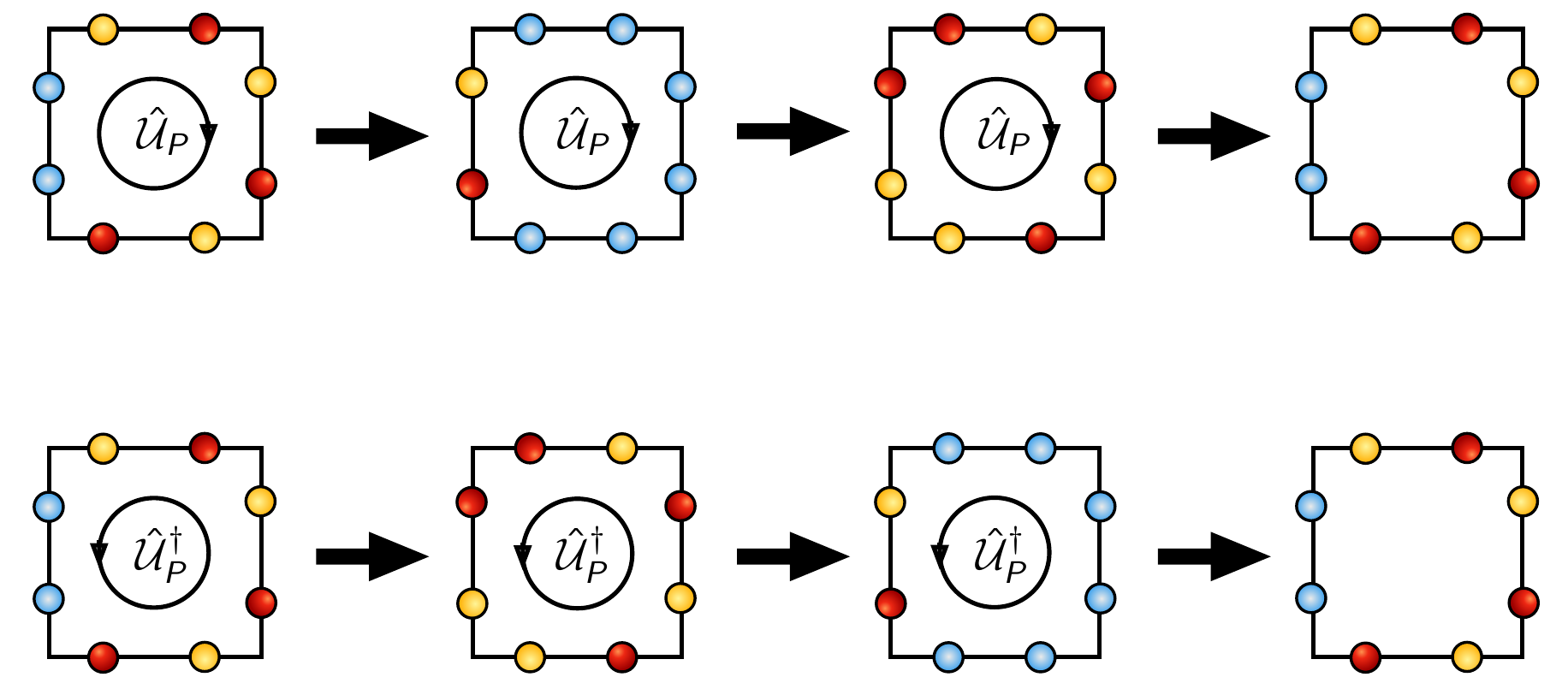}
\caption{ Action of the plaquette operators $\hat{\mathcal{U}}_P$ and $\hat{\mathcal{U}}_P^\dagger$ on a representative \ac{MDTN} state. Traversing the plaquette clockwise or counterclockwise updates the links according to the cyclic rule $\sing \rightarrow \trip \rightarrow \atrip \rightarrow \sing$, applied to the irrep at the first site of each link; the irrep at the opposite end is then fixed.
}
\label{fig:plaqop}
\end{figure}

Plaquette chains have been used as a simplified class of quasi-one-dimensional lattice geometries for studying \acp{LGT}, particularly in work motivated by quantum computation~\cite{Nyhegn:2020cxu,Pradhan:2022lzo,Yao:2023pht,PhysRevD.110.014505,Popov:2025byi}. 
Specializing to $\SU(3)$, the above simple Hilbert space has also been explored previously \cite{PhysRevD.103.094501,PhysRevD.111.074516,Chen:2026hnh}. Most earlier work has focused on models based on the conventional Kogut--Susskind Hamiltonian \cite{Kogut:1974ag}, which typically suffer from sign problems. This makes it difficult to study continuum limits using Monte Carlo methods in higher dimensions. To overcome this difficulty, in Ref.~\cite{Chandrasekharan:2025smw} we proposed a class of simple gauge-invariant Hamiltonians for $\SU(N)$ \acp{LGT} that are free of sign problems. Here we adapt that construction to the $\SU(3)$ plaquette chain, introducing additional couplings to demonstrate the existence of a nontrivial continuum limit. Our plaquette-chain Hamiltonian is given by
\begin{align}
H_{pc} \;=\; \kappa_c \sum_{\ell_c} \hat{\mathcal{E}}_{\ell_c}
      + \kappa_r \sum_{\ell_r} \hat{\mathcal{E}}_{\ell_r}
      - g \sum_{P} \left( \hat{\mathcal{U}}_P + \hat{\mathcal{U}}_P^\dagger \right)\, ,
\label{eq:Hchain}
\end{align}
where $\hat{\mathcal{E}}_\ell$ is diagonal in the link basis $\ket{\lambda}$ associated with link $\ell$, with eigenvalue $(1-\delta_{\lambda,\mathbf{1}})$. 
It assigns equal energies to the $\trip$ and $\atrip$ irreps, which are typically higher than the energy assigned to the $\sing$ irrep on each link, implying $\kappa_c,\kappa_r \ge 0$. 
In contrast, the plaquette operator $\hat{\mathcal{U}}_P$ and its Hermitian conjugate are off-diagonal in the \ac{MDTN} basis and modify the irreps on the four links of a plaquette while preserving gauge invariance. 
Their action is illustrated in \cref{fig:plaqop}. 
By restricting to $g \ge 0$, we can avoid sign problems and study the model with Monte Carlo methods. Finally, we note that $[H_{pc}, \hat{\mathcal{C}}] = 0$, where $\hat{\mathcal{C}}$ is the charge-conjugation operator introduced earlier.
Since the Hamiltonian is local and gauge invariant, it does not mix the three Hilbert spaces $\cH_{\RN{1}}$, $\cH_{\RN{2}}$, and $\cH_{\RN{3}}$. We identify $\cH_{\RN{1}}$ as the vacuum sector. Introducing heavy static quarks and antiquarks at two sites generates a string of plaquettes connecting them that necessarily lies in one of the other two sectors. Thus, while the Hamiltonian retains the same form as in \cref{eq:Hchain}, the presence or absence of static charges alters the accessible Hilbert space and consequently affects the energy spectrum.

Since the plaquettes admit only three quantum states within each Hilbert space sector, the $\SU(3)$ plaquette chain is equivalent, in each sector, to a one-dimensional three-state quantum clock model~\cite{Chen:2026hnh}. The latter is defined in terms of the operators
\begin{align}
\hat{X}=\begin{pmatrix}
0 & 1 & 0\\
0 & 0 & 1\\
1 & 0 & 0
\end{pmatrix}\,, \qquad
\hat{Z}=\begin{pmatrix}
1 & 0 & 0\\
0 & e^{i2\pi/3} & 0\\
0 & 0 & e^{-i2\pi/3}
\end{pmatrix}\,,
\end{align}
which act on the three states associated with each plaquette $P$. Restricting the Hamiltonian in \cref{eq:Hchain} to the sector $\cH_{\RN{1}}$ yields
\begin{align}
H^{\RN{1}}_{\rm qcm}
= - \sum_P \Big\{\, J\, \hat{Z}_P \hat{Z}^\dagger_{P+1}
+ g\, \hat{X}_P
+ \frac{h}{3}\, \hat{Z}_P
+ \text{h.c.} \,\Big\},
\label{eq:qcm}
\end{align}
with $J=\kappa_r/3$ and $h=2\,\kappa_c$, up to an overall constant energy shift $E_0$. In the remaining sectors $\cH_{\RN{2}}$ and $\cH_{\RN{3}}$, the Hamiltonians $H^{\RN{2}}_{\rm qcm}$ and $H^{\RN{3}}_{\rm qcm}$ take the same form, except that $h$ is replaced by $h=-\kappa_c$.

The quantum clock model defined in \cref{eq:qcm} is critical at $J=g=1$ and $h=0$, where its quantum critical behavior belongs to the universality class of the 2D three-state classical Potts model~\cite{PhysRevD.19.3698}. 
At this lattice critical point, the \ac{IR} physics is described by a relativistic $\mathbb{Z}_3$ parafermion \ac{CFT} with central charge $c=4/5$~\cite{Fateev:1985mm}. 
Via the mapping from \cref{eq:Hchain} to \cref{eq:qcm}, the same quantum critical point is realized in the qubit-regularized $\SU(3)$ plaquette chain at $\kappa_r=3$, $g=1$, and $\kappa_c=0$, where we interpret the parafermionic degrees of freedom as describing the emergent continuum gauge degrees of freedom\footnote{A strictly one-dimensional gauge theory has no local propagating degrees of freedom. Our model has rungs connecting two chains, and allows transverse gauge degrees of freedom that are expected to be described as adjoint one-dimensional scalars. The parafermions in the \ac{UV} presumably describe the physics of these bosons.}. 
Relevant perturbations of the parafermion \ac{CFT}, in the form of a thermal coupling $\tau>0$ with $\tau=g-1$ and a magnetic coupling $h>0$, generate massive \acp{QFT}~\cite{Lepori:2009ip}. 
Once the critical theory is perturbed to define a massive continuum limit, the same \ac{CFT} serves as the \ac{UV} fixed point of the resulting continuum \ac{QFT}.
In this work, we focus on the case $\tau=0$ and $h>0$ (equivalently, $\kappa_c>0$), which arises naturally in the vacuum sector of the plaquette chain. 

Since the correlation length associated with the magnetic perturbation diverges as $\xi \sim h^{-15/28}$ for $h\to 0$~\cite{cardy1996scaling,Mong_2014,Lepori:2009ip}, universal quantities in a finite system of $L$ plaquettes arranged on a circle are functions of the dimensionless variable $\mu = h^{15/28} L$. One particularly interesting universal quantity is the mass ratio $m^{-}/m^{+}$, where $m^{-}$ and $m^{+}$ denote the masses of the lightest stable $\hat{\mathcal{C}}$-odd and $\hat{\mathcal{C}}$-even particles, respectively. 
This mass ratio can be extracted from the eigenvalues of \cref{eq:qcm} as a function of $\mu$ from the \ac{UV} regime where $\mu\rightarrow 0$ to the \ac{IR} regime where $\mu\to\infty$. We will be interested in the \ac{IR} regime where the lattice model describes the massive \ac{QFT}.

\begin{figure}[t]
\centering
\includegraphics[width=0.45\textwidth]{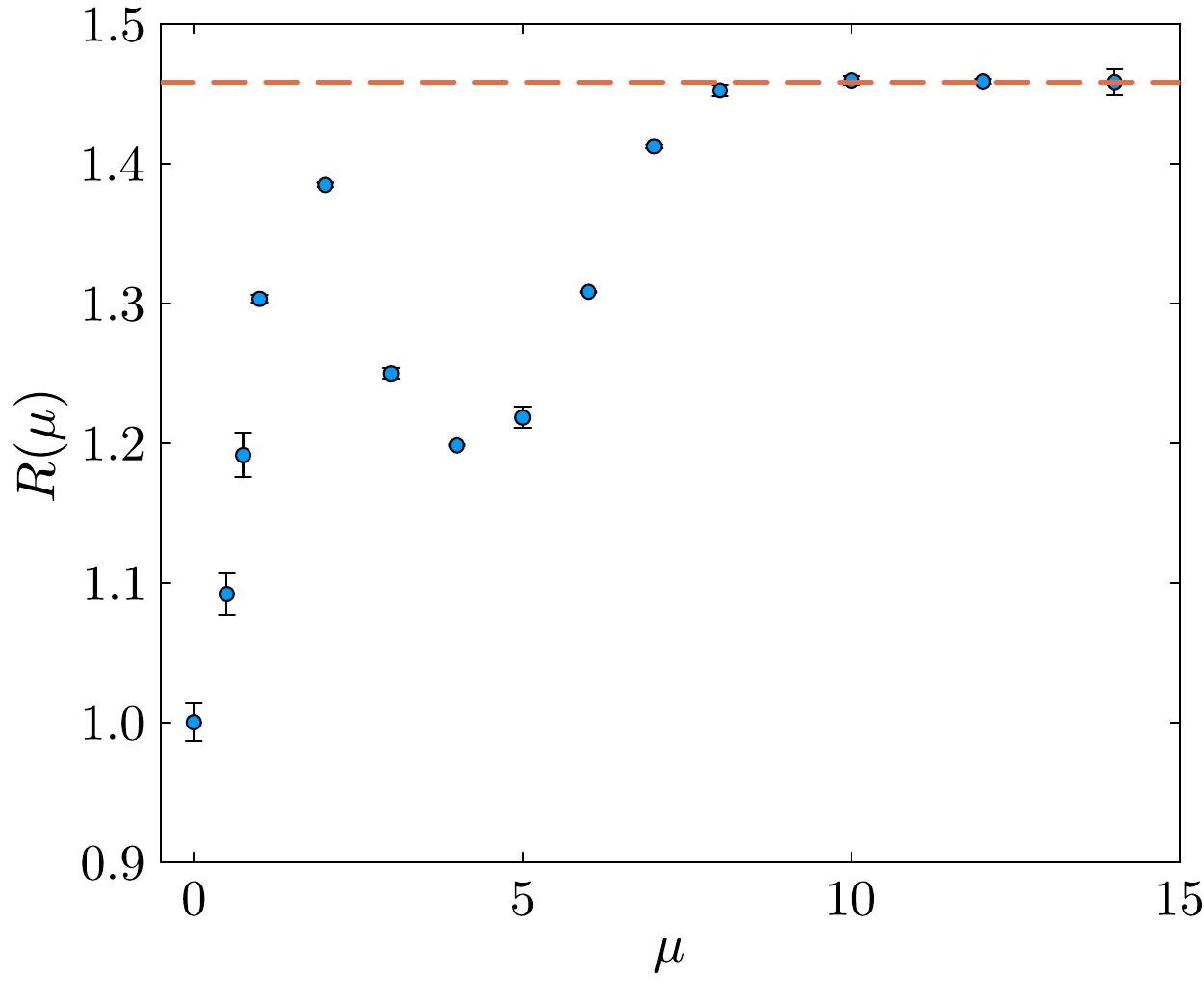}
\caption{Plot of the mass ratio $R(\mu)$ in \cref{eq:mass_scaling_ansatz} as a function of the scaling variable $\mu = h^{15/28}L$. 
The data points are obtained by extrapolating the \ac{DMRG} data at each fixed value of $\mu$ to the joint limit $L \to \infty$ and $h \to 0$. The \ac{UV} and \ac{IR} regimes correspond to $\mu \to 0$ and $\mu \to \infty$, respectively. In the \ac{UV} limit, the degeneracy of the lightest spin primaries related by charge conjugation at the $\mathbb{Z}_3$ parafermion \ac{CFT} leads to the result that $R(0) = 1$ \cite{CARDY1986186,Lepori:2009ip}. In the \ac{IR} limit, no closed-form expression for $R(\infty)$ is known since the theory is non-integrable. 
}
\label{fig:mratio}
\end{figure}

In this work, we use density matrix renormalization group (\ac{DMRG}) to determine $m^{-}/m^{+}$ as a function of $\mu$. We compute the low-energy spectrum of \cref{eq:qcm} at $J=g=1$ for various values of $\mu$. For each $\mu$, we consider five lattice sizes $L=32,48,64,80,$ and $96$, which enables a controlled extrapolation to the thermodynamic limit. In each case, $h$ is tuned to realize the desired value of $\mu$. We denote the ground-state energy at fixed $(L,\mu)$ by $\mathtt{E}^+_0(L,\mu)$, and the excited energies by $\mathtt{E}^C_i(L,\mu)$, where $C=\pm$ labels the charge-conjugation eigenvalue and $i=1,2,\dots$ orders the states by increasing energy within each sector. Our \ac{DMRG} computation was performed using the ITensor software library in Julia~\cite{ITensor,ITensor-r0.3}. Further details of the calculation can be found in the Supplementary Material~\cite{supp}.

\begin{figure}[t]
\centering
\includegraphics[width=0.45\textwidth]{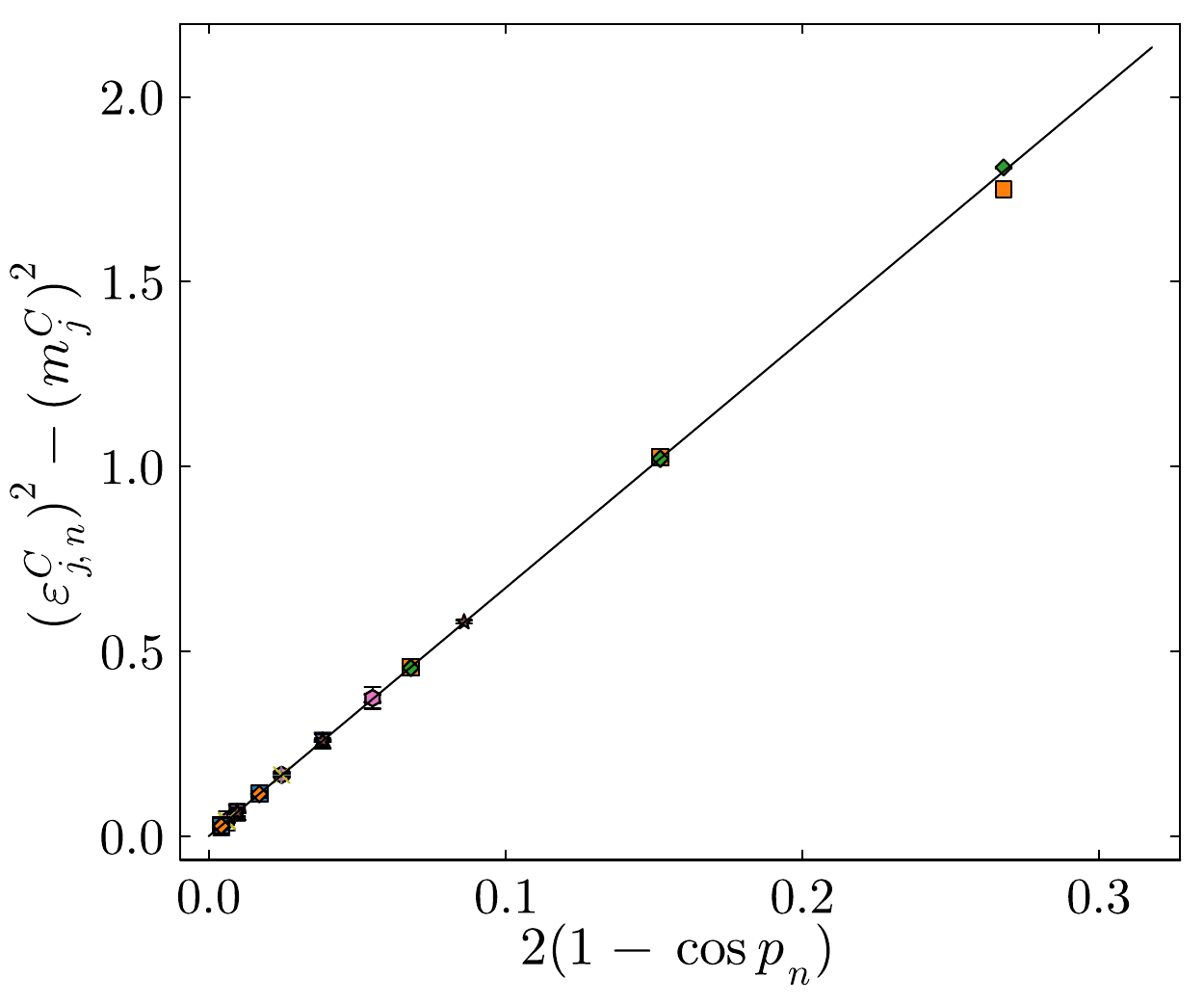}
\caption{Plot of the relativistic dispersion relation defined in \cref{eq:dispersion}. The straight line is a linear fit to all 87 data points shown, including lattice sizes $48\le L\le 96$ and values of $\mu$ in the range $10.0\le \mu\le 14.0$. The data include particles from both $C=\pm$ sectors with quantum numbers $0\le j\le 2$ and $0\le n\le 3$. The slope of the fitted line is $\zeta^2$, yielding $\zeta = 2.5907(2)$ with $\chi^2/\mathrm{d.o.f.}=1.73$. The data underlying this plot, including all points used in the fit, are tabulated in \cite{supp}.
\label{fig:dispersion}}
\end{figure}

We extract the lowest particle masses in the charge-conjugation-even and -odd sectors as
\begin{align}
m^C(L,\mu) = \mathtt{E}^{C}_1(L,\mu) - \mathtt{E}^{+}_0(L,\mu)\,.
\end{align}
To obtain the dimensionless ratio $m^{-}/m^{+}$ in the large-$L$ limit, we assume that for sufficiently large $L$,
\begin{align}
\frac{m^{-}(L,\mu)}{m^{+}(L,\mu)}
= R(\mu) - \frac{A(\mu)}{L}\,,
\label{eq:mass_scaling_ansatz}
\end{align}
where $R(\mu)$ and $A(\mu)$ are $\mu$-dependent coefficients. This ansatz provides a good fit at all values of $\mu$~\cite{supp}. The extracted $R(\mu)$ is shown in \cref{fig:mratio}. We observe non-monotonic behavior at intermediate $\mu$, while for $\mu \ge 8$ the ratio begins to saturate toward its \ac{IR} value. Although we do not have a simple explanation for this non-monotonicity, similar behavior has been observed in conventional lattice calculations of glueball masses, where it is attributed to the presence of a critical point in the fundamental–adjoint coupling plane at which the scalar glueball mass vanishes~\cite{Heller:1995bz}.
The extracted coefficients in the large $\mu$ regime are listed in \cref{tab:m_ratio}, yielding the estimate
\begin{align}
m^{-}/m^{+} \approx 1.459(2)\,.
\end{align}
This ratio was previously computed numerically using the truncated conformal space approach in Ref.~\cite{Lepori:2009ip}, since the massive \ac{QFT} emerging in the \ac{IR} is non-integrable~\cite{Fateev91}. 
Because the earlier results were presented graphically, extracting a precise numerical value is difficult; however, a rough estimate indicates consistency with our result.

\begin{table}[h]
\centering
\renewcommand{\arraystretch}{1.4}
\setlength{\tabcolsep}{4pt}
\begin{tabular}{|c|c|c|c|}
\TopRule
$\mu$ & $A(\mu)$ & $R(\mu)$  & $\sqrt{\sigma}/m^+$\\
\MidRule
8.0 & 1.1(1) & 1.452(4) & 0.2676(4)\\
10.0 & 1.3(1) & 1.459(3) & 0.2649(2)\\
12.0 & 1.57(8) & 1.459(2) & 0.2648(2)\\
14.0 & 1.8(3) & 1.458(9) & 0.264(1)\\
\BotRule 
\end{tabular}
\caption{\label{tab:m_ratio}
Fixed-$\mu$ scaling-limit estimates of $A(\mu)$ and $R(\mu)$ appearing in
\cref{eq:mass_scaling_ansatz}. The corresponding dimensionless
string-tension ratio, $\sqrt{\sigma}/m^+$, is also shown. The
extrapolations use $L=32,48,64,80,$ and $96$.
}
\end{table}

\begin{figure}[t]
\centering
\includegraphics[width=0.45\textwidth]{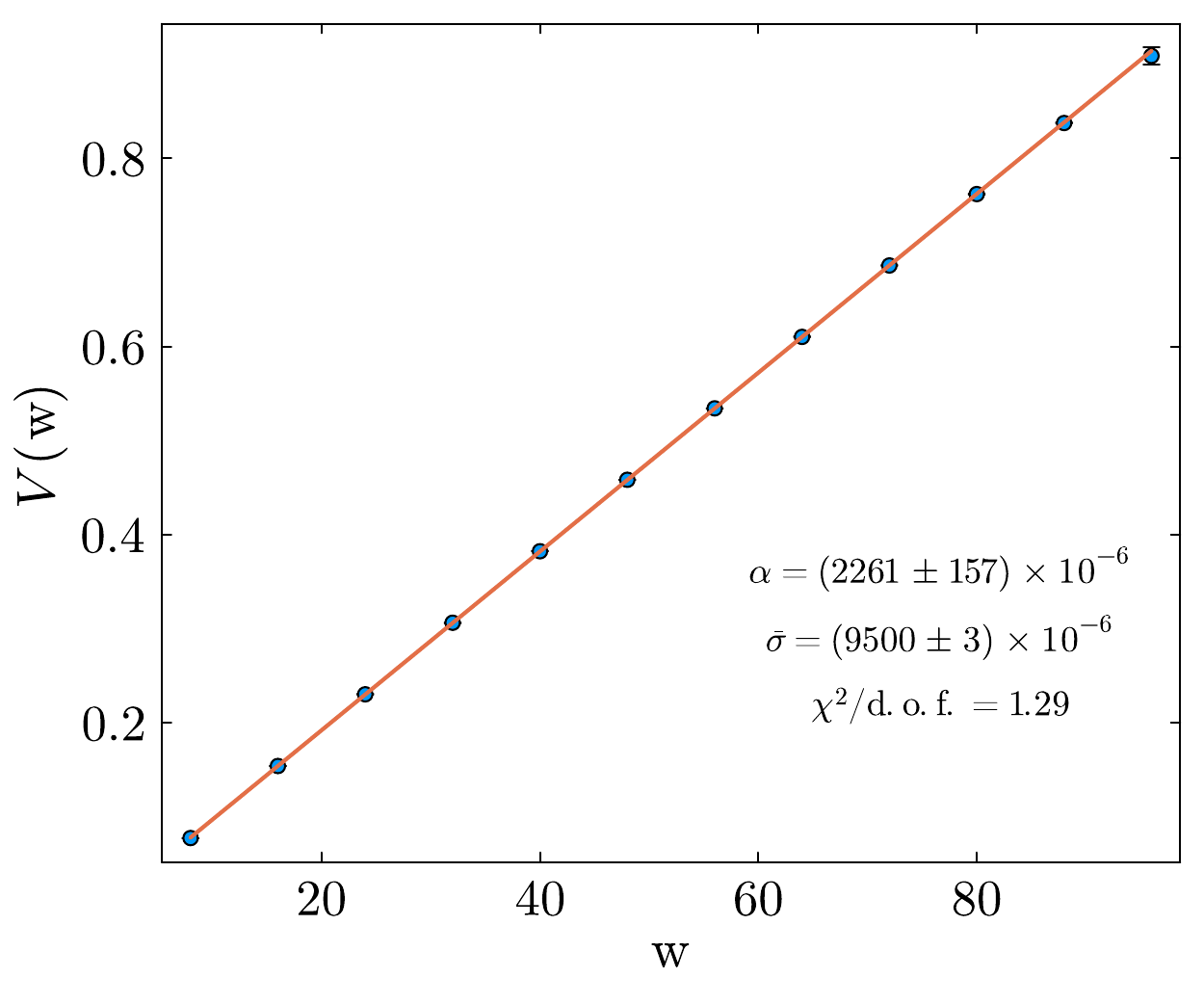}
\caption{Static quark potential $V(\txtw)$ versus separation $\txtw$ for $L=96$ and $\mu=14.0$. The solid line is a linear fit whose slope gives $\bar{\sigma} = 0.009500(3)$.}
\label{fig:stringtension}
\end{figure}

To confirm that the massive continuum \ac{QFT} emerging in the \ac{IR} is relativistic, we study the dispersion relation of the low-energy spectrum of \cref{eq:qcm}. Since the system is invariant under discrete translations, each energy eigenvalue $\mathtt{E}^C_i(L,\mu)$ carries a definite lattice momentum 
$p_n = 2\pi n/L$. Single-particle excitations of distinct particle species below the two-particle threshold can be identified as non-degenerate states at zero momentum. Using this insight, we reorganize the low-energy spectrum $\mathtt{E}^C_i(L,\mu)$ by relabeling the index $i$ as a pair $(j,n)$. 
We assign the vacuum energy $\mathtt{E}^+_0(L,\mu)$ the label $(j=0,n=0)$. 
The higher-energy states (in both sectors) correspond to the $j$-th massive particle with lattice momentum $p_n$. We denote the corresponding energies by $\mathtt{E}^C_{j,n}(L,\mu)$~\cite{supp}.

We then define the single-particle excitation energies by subtracting the vacuum energy,
\begin{align}
    \epsilon^C_{j,n}(L,\mu) = \mathtt{E}^C_{j,n}(L,\mu) - \mathtt{E}^+_0(L,\mu)\,.
\end{align}
The mass of the $j$-th particle is then defined as $m^C_j = \epsilon^C_{j,0}(L,\mu)$. The dispersion relation 
\begin{align}
    \big(\epsilon^C_{j,n}\big)^2 - (m^C_j)^2
    = 2\big(1-\cos p_n\big)\zeta^2
    + \mathcal{O}(n^4 L^{-4})\,, 
\label{eq:dispersion}
\end{align}
where $\zeta$ is a constant (independent of $n$, $L$, and $j$) that plays the role of $\hbar c$ converting inverse lattice spacing into energy units, is a hallmark of a relativistic particle. As shown in \cref{fig:dispersion}, our data are well described by this dispersion relation, yielding $\zeta = 2.5907(2)$.

For sufficiently large $L$ and $\mu$, we can also compute the string tension between a static quark--antiquark pair by placing the quark at site $x=0$ and the antiquark at $x=\txtw$, both on the lower chain. Although the Hamiltonian remains that of \cref{eq:Hchain}, the Hilbert space is modified: the plaquettes between $x=0$ and $x=\txtw$ are forced into sector $\RN{2}$, while the remaining plaquettes remain in the vacuum sector $\RN{1}$. In the effective quantum clock model of \cref{eq:qcm}, this corresponds to a modification of the coupling $h$: on the plaquettes between $x=0$ and $x=\txtw$, $h$ is reduced by a factor of two and changes sign relative to its value on the remaining plaquettes.

The heavy quark--antiquark potential is defined as
\begin{align}
V(\txtw) = {\mathtt{E}}_0^{+}(\txtw) - {\mathtt{E}}_0^{+}\,,
\end{align}
where ${\mathtt{E}}_0^{+}(\txtw)$ denotes the ground-state energy in the presence of a static quark--antiquark pair separated by $\txtw$, and ${\mathtt{E}}_0^{+}$ is the vacuum ground-state energy. 
In the confined phase we expect a linear potential,
\begin{align}
V(\txtw) = \alpha + \bar{\sigma}\,\txtw\,,
\end{align}
where $\bar{\sigma}$ is the string tension (energy per unit length). 
\cref{fig:stringtension} illustrates how we extract $\bar{\sigma}$ from our data when $L=96$ and $\mu=14$.

Defining $\sigma = \bar{\sigma}\,\zeta$, which has dimensions of energy squared, and using the value of $\zeta$ extracted from the dispersion analysis, we compute the dimensionless ratio $\sqrt{\sigma}/m^{+}$ for each $\mu$ and $L$. Assuming the finite-size corrections are of the form $L^{-1}$, we extrapolate to the thermodynamic limit~\cite{supp}. 
The resulting estimates of $\sqrt{\sigma}/m^{+}$ are listed in the fourth column of \cref{tab:m_ratio}, from which we obtain
\[
\sqrt{\sigma}/m^{+} = 0.2648(2)\,.
\]

Our work demonstrates that simple qubit-regularized $\SU(3)$ \acp{LGT} can exhibit nontrivial continuum limits with massive relativistic excitations. A similar massive continuum limit was recently identified in a qubit-regularized $\SU(2)$ theory~\cite{2cy4-cp7f}. To the best of our knowledge, the continuum limits of the conventional Kogut--Susskind \ac{LGT} on the plaquette-chain geometry have not yet been investigated. An important open question is therefore whether the continuum limit identified in this work can also be realized within the conventional Kogut--Susskind formulation. In the Euclidean space-time formulation, the Wilson action contains independent couplings multiplying the temporal-chain, temporal-rung, and purely spatial plaquette terms. Exploring the continuum limits obtained by tuning these plaquette couplings would determine whether the conventional and qubit-regularized \acp{LGT} can flow to the same continuum theory or instead belong to distinct universality classes.

The continuum limits that we have discovered here and in Ref.~\cite{2cy4-cp7f} can be viewed as arising from confinement--deconfinement transitions. For small nonzero $h$, the ground state of \cref{eq:qcm} lies in the ordered phase with $g/J<1$, where $h$ selects one of three degenerate vacua, corresponding to confinement. For $g/J>1$, the ground state is disordered and largely insensitive to $h$, corresponding instead to deconfinement. It is plausible that analogous quantum critical points associated with confinement--deconfinement transitions exist in higher dimensions. Identifying such points in simple qubit-regularized $\SU(3)$ gauge theories and determining whether they give rise to massive continuum \acp{QFT} would be an important next step. In three spatial dimensions, such theories would presumably correspond either to conventional \ac{YM} theory or to a more exotic variant enabled by qubit regularization.

\acknowledgments

S.C. conceived of the idea and R.X.S. carried out the investigations of this theory. Both of them created the outline of the manuscript. R.X.S. did all the \ac{DMRG} calculations. The two together produced the first draft of the paper. T.B. checked the correctness of the physics discussed and provided further interpretation of the model in terms of a pseudo-one-dimensional gauge theory. All three authors refined the manuscript and agree with its final version.

We acknowledge the use of ChatGPT (OpenAI; GPT‑5.6 Sol) to assist with language editing, literature organization, and the clarification of portions of the scientific presentation. The authors critically reviewed and independently verified all AI-assisted text, references, calculations, and scientific claims and take full responsibility for the manuscript.
S.C. and R.X.S. are supported in part by the U.S. Department of Energy, Office of Science, Nuclear Physics program under Award No.\ DE-FG02-05ER41368. T.B. was supported by the U.S. Department of Energy, Office of Science, Office of High Energy Physics under contract number KA2401012 (LANLE83G) at Los Alamos National Laboratory operated by Triad National Security, LLC, for the National Nuclear Security Administration of the U.S. Department of Energy (Contract No.\ 89233218CNA000001).

\bibliographystyle{apsrev4-2}

\clearpage
\newpage 

\onecolumngrid
\renewcommand{\thefigure}{S\arabic{figure}}
\renewcommand{\thetable}{S\arabic{table}}
\renewcommand{\thepage}{S\arabic{page}}
\setcounter{page}{1}
\setcounter{table}{0}
\setcounter{figure}{0}
\setcounter{equation}{0}

\section*{Supplementary Material}

\begin{center}
{\bf Continuum limit of a qubit-regularized SU(3) lattice gauge theory on a plaquette chain}

\vskip0.1in

{\sl Rui Xian Siew, Shailesh Chandrasekharan, and Tanmoy Bhattacharya}

\end{center}

\subsection{DMRG details}

The ITensor implementation of the \ac{DMRG} algorithm requires several numerical parameters in addition to the lattice Hamiltonian. One of the most important is the maximum \ac{MPS} bond dimension, $\chi_{\max}$, which sets the upper bound on the bond dimension while allowing it to be adjusted dynamically during the optimization. In all calculations, we used $\chi_{\max}=400$. Another key parameter is the truncation-error cutoff, which controls the accuracy of the calculations. We set this cutoff to $10^{-12}$. The optimization was performed for up to $25$ \ac{DMRG} sweeps. These parameter values were found to perform well in our previous study of the $\SU(2)$ plaquette chain.

For the lattice model, we considered periodic chains with $L=24,32,48,64,80$, and $96$ sites, and separately targeted low-lying states in the $C=+$ and $C=-$ charge-conjugation sectors. To determine the mass ratios, we computed the two lowest-energy states in the even sector and the lowest-energy state in the odd sector. For the relativistic-dispersion analysis, we additionally targeted several higher excited states in both sectors. When targeting excited states, we set the \ac{DMRG} overlap-penalty parameter to $\mathsf{weight}=25$, which suppresses overlap with previously converged states and improves convergence to distinct eigenstates.

It is well known that \ac{DMRG} converges more slowly near criticality, particularly for excited states and systems with periodic boundary conditions. Accordingly, as the system approaches criticality—either by increasing $L$ at fixed $\mu$ (IR criticality) or by decreasing $\mu$ at fixed but large $L$ (UV criticality)—the convergence of higher excited states becomes progressively more difficult. We therefore carefully checked the convergence of the targeted states and included in the dispersion analysis only those data for which we were confident of convergence. In practice, this restricted the dispersion plot to $\mu \geq 10.0$.

To assess the convergence of the \ac{DMRG} calculations, we estimated the uncertainty in each computed energy using the energy standard deviation,
\begin{equation}
\sigma_E=
\sqrt{
\langle\psi|\hat{H}^2|\psi\rangle
-\langle\psi|\hat{H}|\psi\rangle^2
},
\end{equation}
where $\ket{\psi}$ is the corresponding approximate eigenstate obtained from the \ac{DMRG} calculation. For an exact eigenstate, $\sigma_E=0$, so the energy standard deviation provides a direct measure of the residual error in the variational wavefunction.

\begin{figure}[h!]
    \centering
    \subfigure[]{\includegraphics[width=0.44\textwidth]{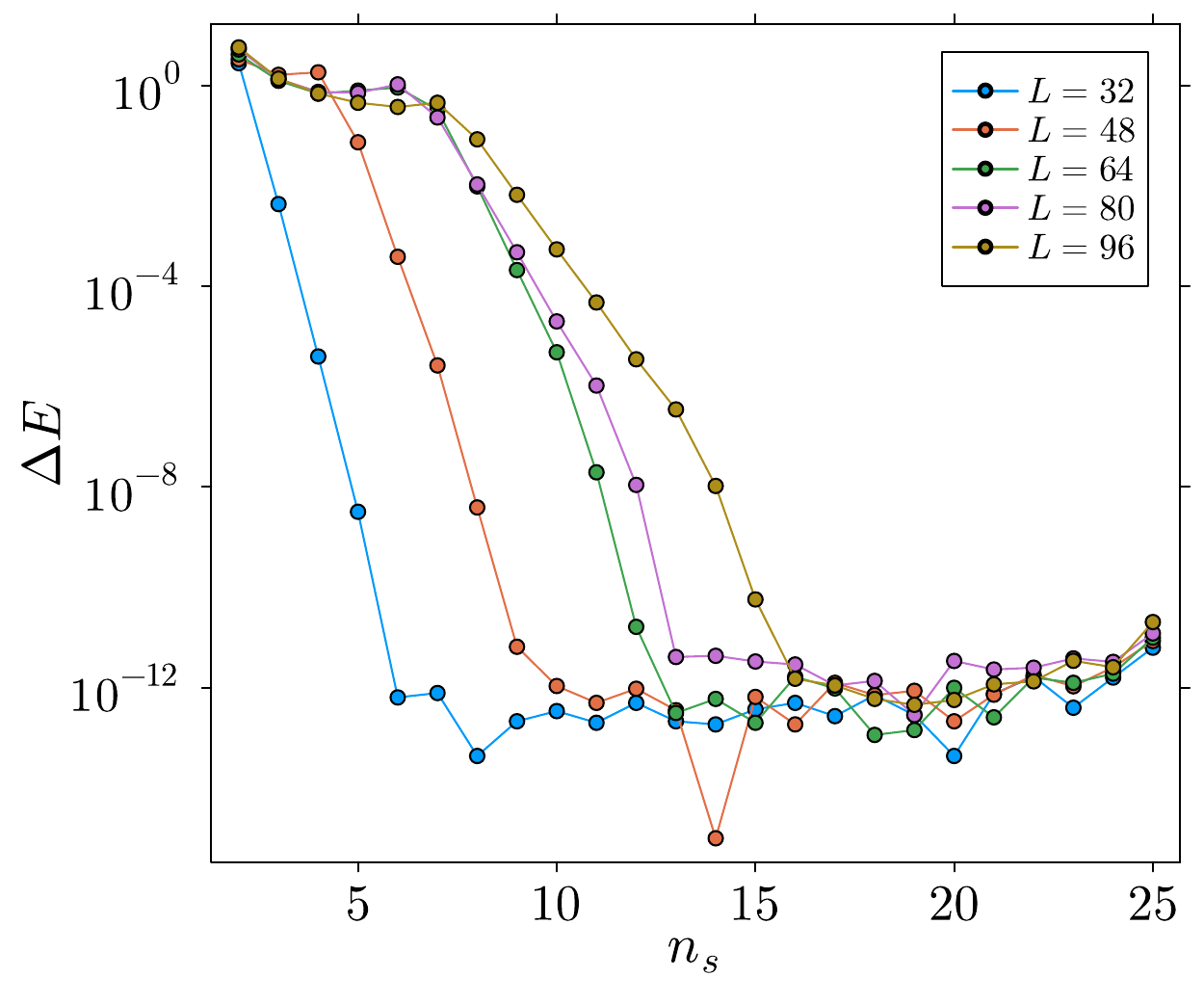}}
    \subfigure[]{\includegraphics[width=0.44\textwidth]{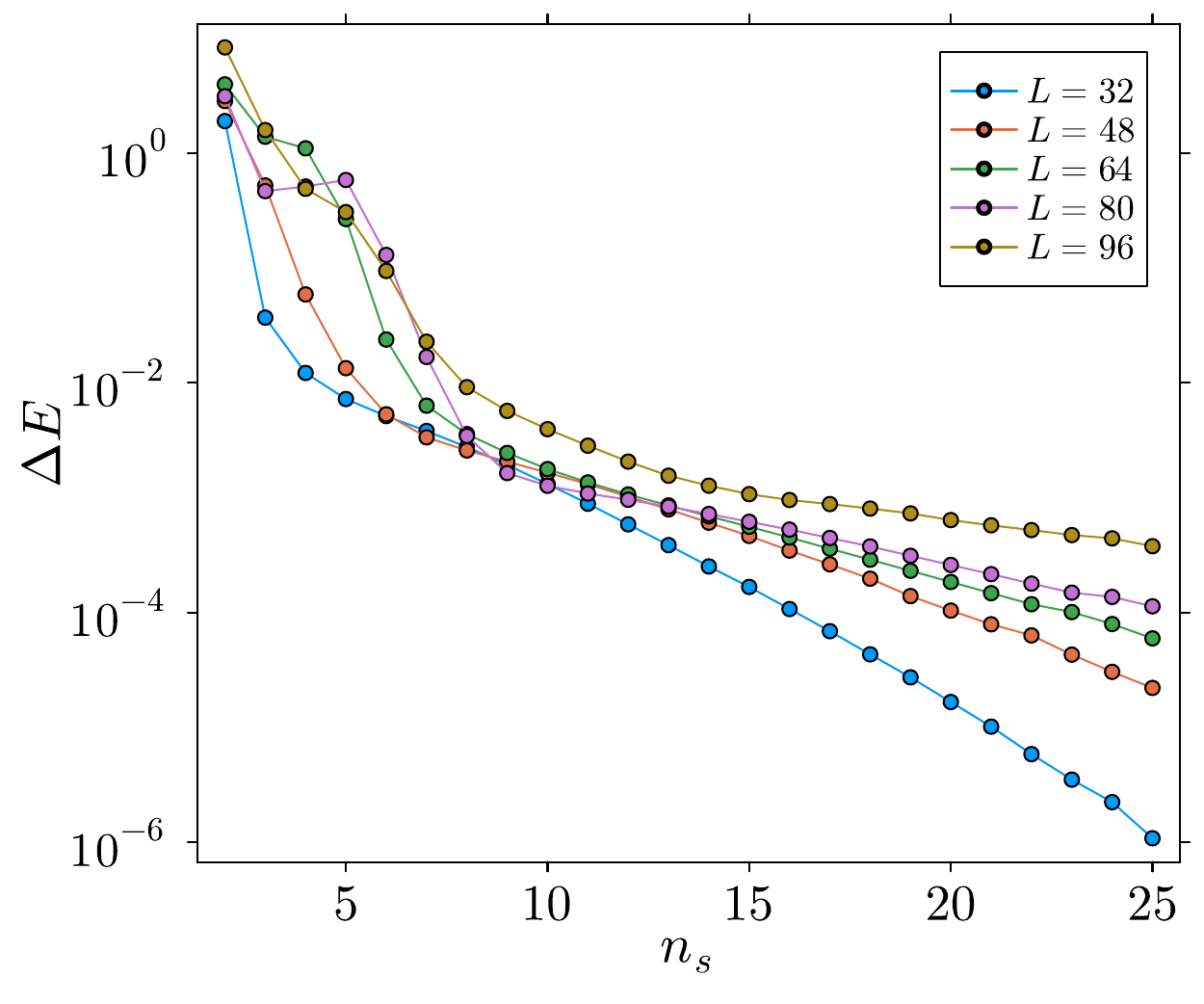}}
    \subfigure[]{\includegraphics[width=0.44\textwidth]{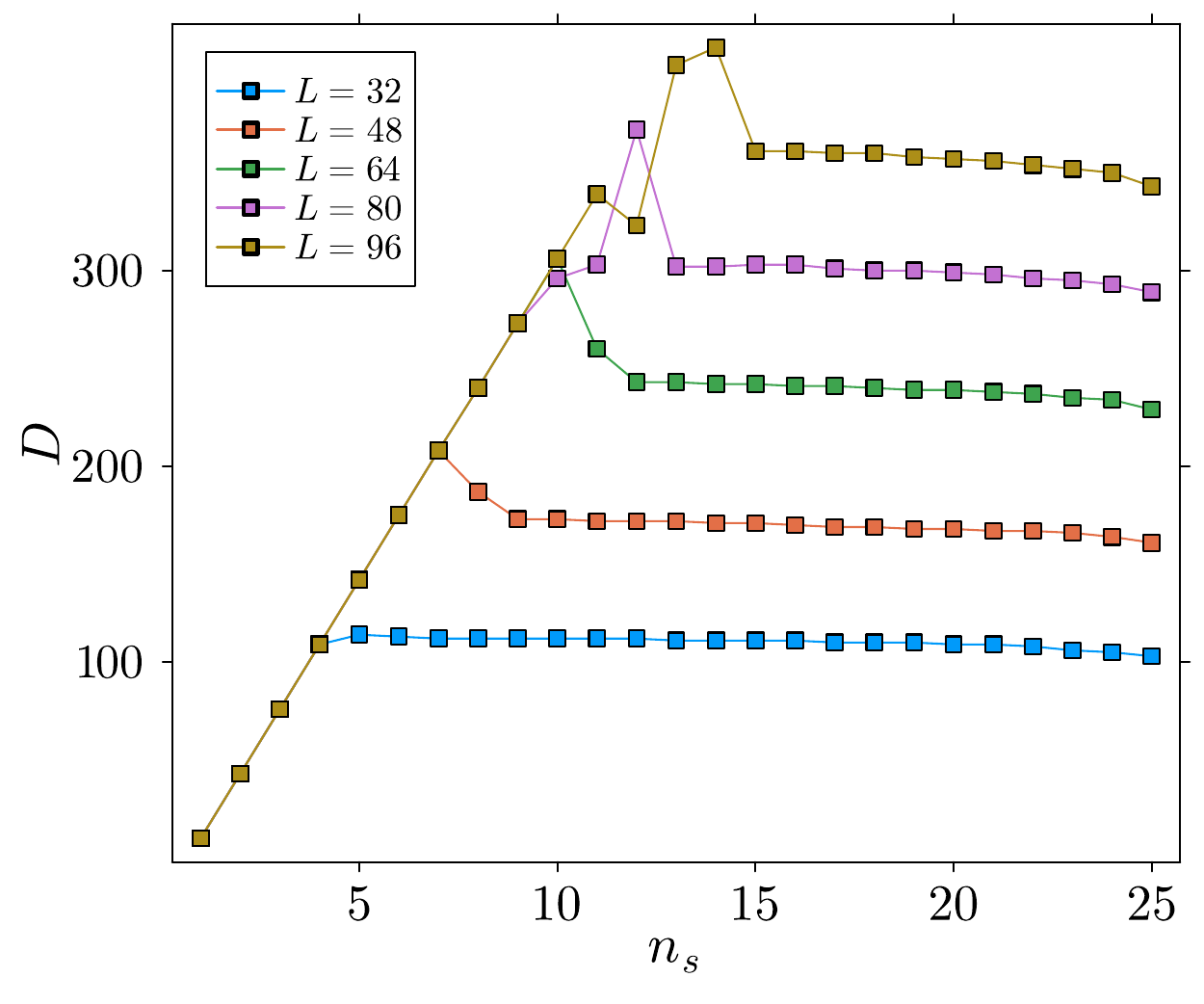}}
    \subfigure[]{\includegraphics[width=0.44\textwidth]{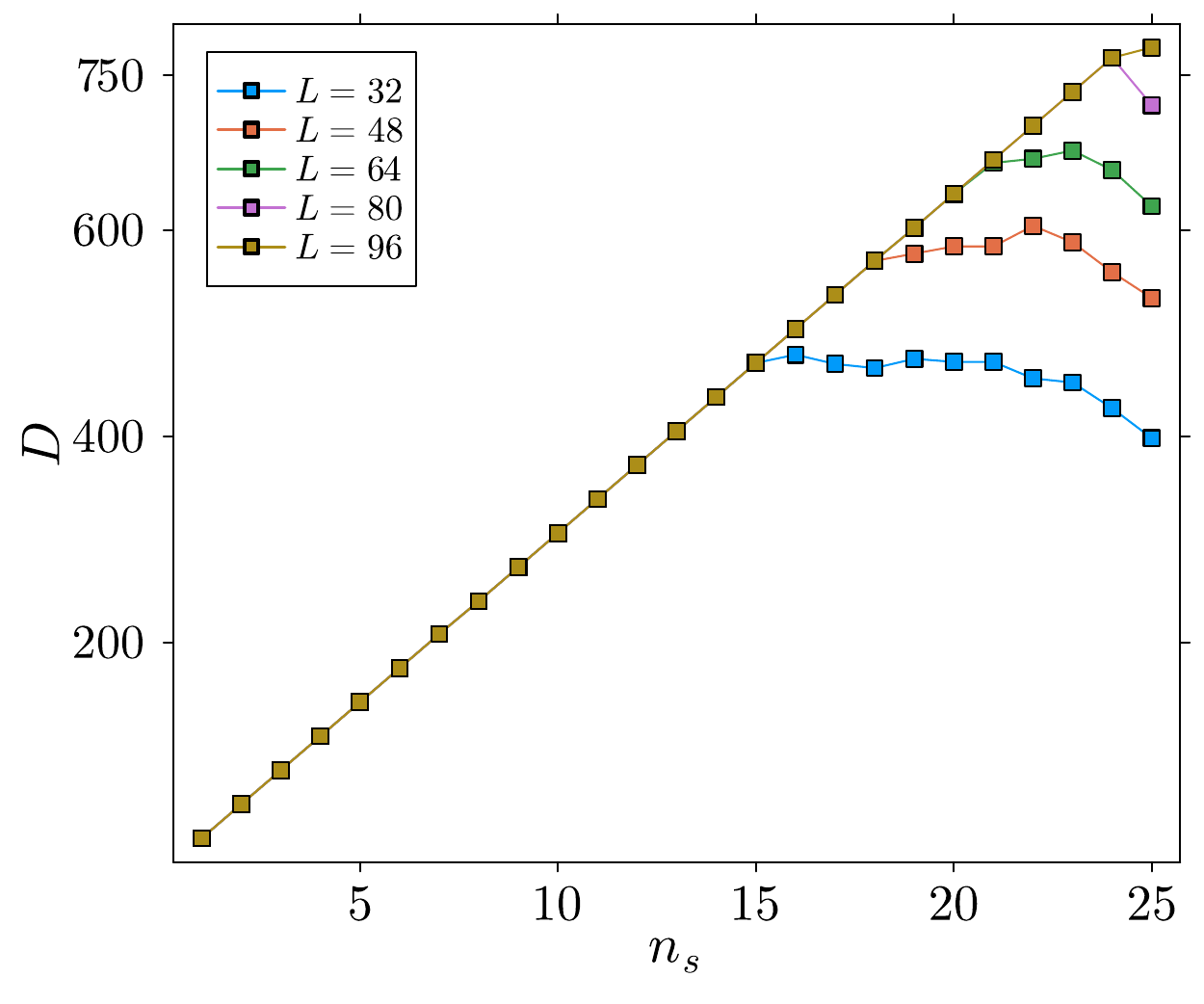}}
    \subfigure[]{\includegraphics[width=0.44\textwidth]{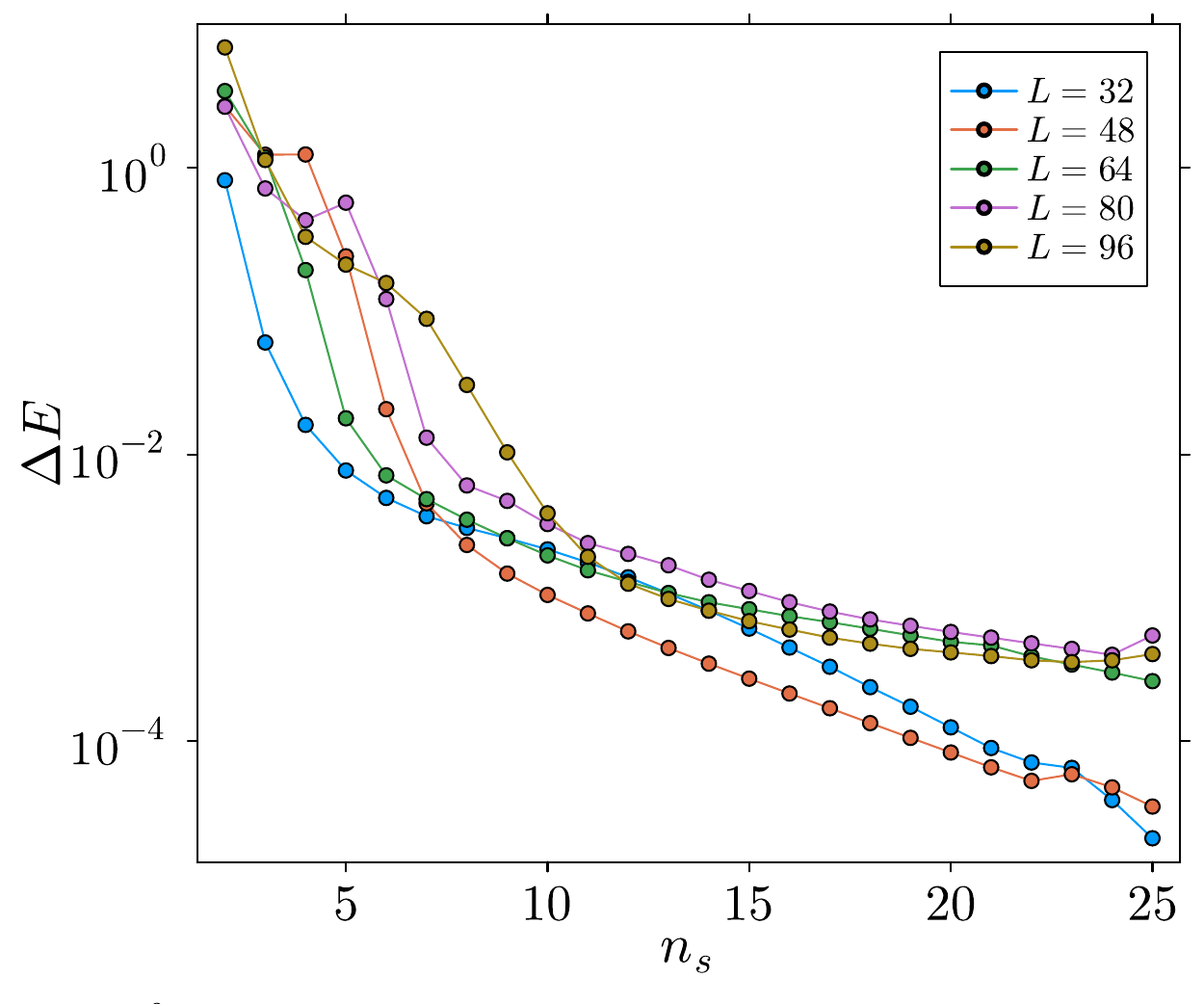}}
    \subfigure[]{\includegraphics[width=0.44\textwidth]{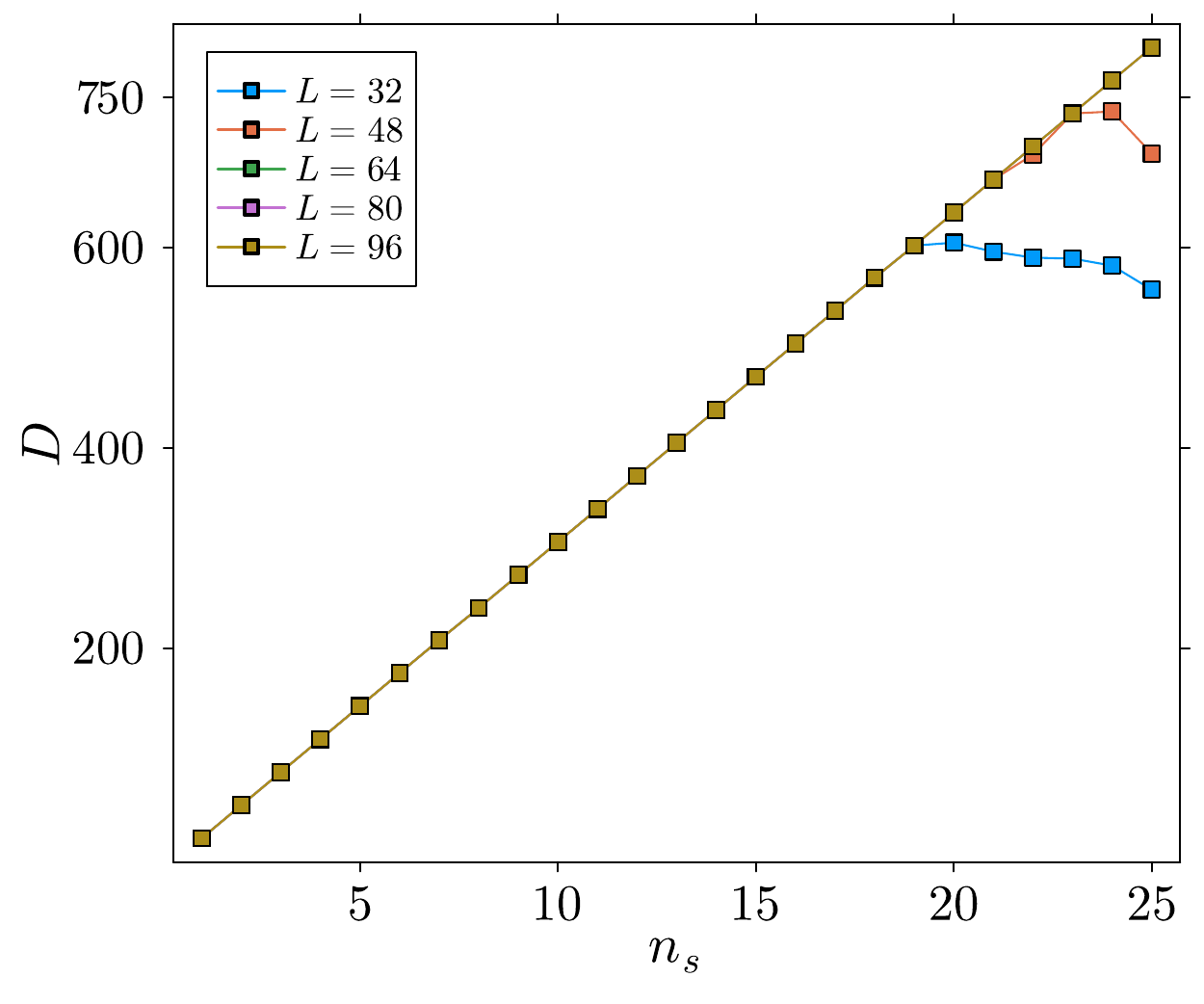}}
    \caption{(Upper row) Fluctuation between the estimated eigenenergy at the $n_s$-th and the $(n_s-1)$-th sweeps for the $C=+$ sector (a) ground state and (b) first excited state. (Middle row) Maximum bond dimension at the $n_s$-th sweep for the (c) ground state and (d) first excited state in the $C=+$ sector.
    (Lower row) (e) Fluctuation between the estimated eigenenergy at the $n_s$-th and the $(n_s-1)$-th sweeps in the $C=-$ sector; (f) maximum bond dimension at the $n_s$-th sweep for the $C=-$ ground state.
    These \ac{DMRG} data all have a truncation error cutoff of $10^{-13}$, a maximum bond dimension cutoff $\chi_{\mathrm{max}}=800$, and $\mu=14.0$.}
    \label{fig:dmrg_error}
\end{figure}

As an example of our convergence analysis, we consider the case $\mu=14.0$, for which we repeated the calculations using a larger maximum bond dimension, $\chi_{\max}=800$, and a smaller truncation-error cutoff of $10^{-13}$. These more stringent parameters changed the computed energies only beyond approximately five significant digits. The corresponding convergence results are shown in \cref{fig:dmrg_error}, where we plot the energy change between successive sweeps,
\begin{equation}
\Delta E(n_s)=E(n_s)-E(n_s-1),
\end{equation}
with $E(n_s)$ denoting the energy after $n_s$ sweeps. We also plot the maximum bond dimension $D$ reached during each sweep. For the ground state in the even sector, $D$ typically saturated below the imposed cutoff after approximately $15$ sweeps. In contrast, for the excited state in the even sector and the lowest-energy state in the odd sector, $D$ did not always saturate below the cutoff for the largest lattice sizes, even when the cutoff was increased to $800$. Nevertheless, the energies obtained with $\chi_{\max}=400$ and $\chi_{\max}=800$ agreed within the estimated uncertainties, indicating that the incomplete saturation of the bond dimension affects the results only at a level beyond the precision required in this work. We observed similar behavior in our previous study of the $\SU(2)$ plaquette chain.

\clearpage
\subsection{Low energy spectrum in the vacuum sector}\label{subsec:lowspectab}

\begin{table*}[h]
\centering
\renewcommand{\arraystretch}{1.3}
\setlength{\tabcolsep}{4pt}

\begin{tabular}{c|c|c|c|c}
\TopRule
\multicolumn{5}{c}{$L=24\,,\ \mu=0.0\,,\ C=+$} \\
\BotRule
$i$ & $\mathtt{E}_{i}^{+}$ & $\delta \mathtt{E}_{i}^{+}$ & $j$ & $n$ \\
\MidRule
0 & -5.850916e+01 & 7.87e-05 & 0 & 0 \\
1 & -5.841808e+01 & 7.48e-05 & 1 & 0 \\
2 & -5.773844e+01 & 8.34e-05 & 1 & 1 \\
3 & -5.773844e+01 & 8.46e-05 & 1 & 1 \\
\BotRule
\end{tabular}
\hfill
\begin{tabular}{c|c|c|c|c}
\TopRule
\multicolumn{5}{c}{$L=24\,,\ \mu=0.0\,,\ C=-$} \\
\BotRule
$i$ & $\mathtt{E}_{i}^{-}$ & $\delta \mathtt{E}_{i}^{-}$ & $j$ & $n$ \\
\MidRule
1 & -5.841808e+01 & 7.53e-05 & 1 & 0 \\
2 & -5.773844e+01 & 7.49e-05 & 1 & 1 \\
3 & -5.773844e+01 & 8.43e-05 & 1 & 1 \\
4 & -5.760537e+01 & 1.05e-04 & 2 & 0 \\
\BotRule
\end{tabular}
\hfill
\begin{tabular}{c|c|c|c|c}
\TopRule
\multicolumn{5}{c}{$L=24\,,\ \mu=0.5\,,\ C=+$} \\
\BotRule
$i$ & $\mathtt{E}_{i}^{+}$ & $\delta \mathtt{E}_{i}^{+}$ & $j$ & $n$ \\
\MidRule
0 & -5.850950e+01 & 7.63e-05 & 0 & 0 \\
1 & -5.842190e+01 & 7.48e-05 & 1 & 0 \\
2 & -5.774067e+01 & 8.45e-05 & 1 & 1 \\
3 & -5.774067e+01 & 7.56e-05 & 1 & 1 \\
\BotRule
\end{tabular}
\\[1em]
\begin{tabular}{c|c|c|c|c}
\TopRule
\multicolumn{5}{c}{$L=24\,,\ \mu=0.5\,,\ C=-$} \\
\BotRule
$i$ & $\mathtt{E}_{i}^{-}$ & $\delta \mathtt{E}_{i}^{-}$ & $j$ & $n$ \\
\MidRule
1 & -5.841395e+01 & 7.57e-05 & 1 & 0 \\
2 & -5.773625e+01 & 8.16e-05 & 1 & 1 \\
3 & -5.773625e+01 & 7.72e-05 & 1 & 1 \\
4 & -5.760569e+01 & 1.05e-04 & 2 & 0 \\
\BotRule
\end{tabular}
\hfill
\begin{tabular}{c|c|c|c|c}
\TopRule
\multicolumn{5}{c}{$L=24\,,\ \mu=0.75\,,\ C=+$} \\
\BotRule
$i$ & $\mathtt{E}_{i}^{+}$ & $\delta \mathtt{E}_{i}^{+}$ & $j$ & $n$ \\
\MidRule
0 & -5.851074e+01 & 7.61e-05 & 0 & 0 \\
1 & -5.842541e+01 & 7.73e-05 & 1 & 0 \\
2 & -5.774322e+01 & 8.41e-05 & 1 & 1 \\
3 & -5.774322e+01 & 7.99e-05 & 1 & 1 \\
\BotRule
\end{tabular}
\hfill
\begin{tabular}{c|c|c|c|c}
\TopRule
\multicolumn{5}{c}{$L=24\,,\ \mu=0.75\,,\ C=-$} \\
\BotRule
$i$ & $\mathtt{E}_{i}^{-}$ & $\delta \mathtt{E}_{i}^{-}$ & $j$ & $n$ \\
\MidRule
1 & -5.840928e+01 & 7.50e-05 & 1 & 0 \\
2 & -5.773380e+01 & 7.51e-05 & 1 & 1 \\
3 & -5.773380e+01 & 8.47e-05 & 1 & 1 \\
4 & -5.760607e+01 & 1.05e-04 & 2 & 0 \\
\BotRule
\end{tabular}
\end{table*}

\begin{table*}[h]
\centering
\renewcommand{\arraystretch}{1.4}
\setlength{\tabcolsep}{4pt}

\begin{tabular}{c|c|c|c|c}
\TopRule
\multicolumn{5}{c}{$L=24\,,\ \mu=1.0\,,\ C=+$} \\
\BotRule
$i$ & $\mathtt{E}_{i}^{+}$ & $\delta \mathtt{E}_{i}^{+}$ & $j$ & $n$ \\
\MidRule
0 & -5.851398e+01 & 7.48e-05 & 0 & 0 \\
1 & -5.842862e+01 & 7.65e-05 & 1 & 0 \\
2 & -5.774670e+01 & 8.33e-05 & 1 & 1 \\
3 & -5.774670e+01 & 8.17e-05 & 1 & 1 \\
\BotRule
\end{tabular}
\hfill
\begin{tabular}{c|c|c|c|c}
\TopRule
\multicolumn{5}{c}{$L=24\,,\ \mu=1.0\,,\ C=-$} \\
\BotRule
$i$ & $\mathtt{E}_{i}^{-}$ & $\delta \mathtt{E}_{i}^{-}$ & $j$ & $n$ \\
\MidRule
1 & -5.840303e+01 & 7.63e-05 & 1 & 0 \\
2 & -5.773058e+01 & 8.33e-05 & 1 & 1 \\
3 & -5.773058e+01 & 7.69e-05 & 1 & 1 \\
4 & -5.707537e+01 & 2.46e-03 & 1 & 2 \\
\BotRule
\end{tabular}
\hfill
\begin{tabular}{c|c|c|c|c}
\TopRule
\multicolumn{5}{c}{$L=24\,,\ \mu=2.0\,,\ C=+$} \\
\BotRule
$i$ & $\mathtt{E}_{i}^{+}$ & $\delta \mathtt{E}_{i}^{+}$ & $j$ & $n$ \\
\MidRule
0 & -5.856628e+01 & 7.62e-05 & 0 & 0 \\
1 & -5.841944e+01 & 7.66e-05 & 1 & 0 \\
2 & -5.777049e+01 & 7.50e-05 & 1 & 1 \\
3 & -5.777049e+01 & 8.23e-05 & 1 & 1 \\
\BotRule
\end{tabular}
\\[1em]
\begin{tabular}{c|c|c|c|c}
\TopRule
\multicolumn{5}{c}{$L=24\,,\ \mu=2.0\,,\ C=-$} \\
\BotRule
$i$ & $\mathtt{E}_{i}^{-}$ & $\delta \mathtt{E}_{i}^{-}$ & $j$ & $n$ \\
\MidRule
1 & -5.836352e+01 & 7.73e-05 & 1 & 0 \\
2 & -5.771176e+01 & 7.97e-05 & 1 & 1 \\
3 & -5.771176e+01 & 8.53e-05 & 1 & 1 \\
4 & -5.709239e+01 & 2.72e-03 & 1 & 2 \\
\BotRule
\end{tabular}
\hfill
\begin{tabular}{c|c|c|c|c}
\TopRule
\multicolumn{5}{c}{$L=24\,,\ \mu=3.0\,,\ C=+$} \\
\BotRule
$i$ & $\mathtt{E}_{i}^{+}$ & $\delta \mathtt{E}_{i}^{+}$ & $j$ & $n$ \\
\MidRule
0 & -5.868241e+01 & 7.31e-05 & 0 & 0 \\
1 & -5.837771e+01 & 7.69e-05 & 1 & 0 \\
2 & -5.781290e+01 & 7.56e-05 & 1 & 1 \\
3 & -5.781290e+01 & 7.82e-05 & 1 & 1 \\
\BotRule
\end{tabular}
\hfill
\begin{tabular}{c|c|c|c|c}
\TopRule
\multicolumn{5}{c}{$L=24\,,\ \mu=3.0\,,\ C=-$} \\
\BotRule
$i$ & $\mathtt{E}_{i}^{-}$ & $\delta \mathtt{E}_{i}^{-}$ & $j$ & $n$ \\
\MidRule
1 & -5.830300e+01 & 7.51e-05 & 1 & 0 \\
2 & -5.768912e+01 & 3.04e-04 & 1 & 1 \\
3 & -5.768912e+01 & 2.51e-04 & 1 & 1 \\
4 & -5.712302e+01 & 2.03e-04 & 1 & 2 \\
\BotRule
\end{tabular}
\caption{Low energy spectrum of the quantum clock model at various values of $\mu = h^{15/28} L$ in the two charge-conjugation sectors at $L=24$.}
\end{table*}

\begin{table*}[h]
\centering
\renewcommand{\arraystretch}{1.3}
\setlength{\tabcolsep}{4pt}

\begin{tabular}{c|c|c|c|c}
\TopRule
\multicolumn{5}{c}{$L=24\,,\ \mu=4.0\,,\ C=+$} \\
\BotRule
$i$ & $\mathtt{E}_{i}^{+}$ & $\delta \mathtt{E}_{i}^{+}$ & $j$ & $n$ \\
\MidRule
0 & -5.885019e+01 & 7.31e-05 & 0 & 0 \\
1 & -5.832573e+01 & 7.55e-05 & 1 & 0 \\
2 & -5.787899e+01 & 7.63e-05 & 1 & 1 \\
3 & -5.787899e+01 & 7.54e-05 & 1 & 1 \\
\BotRule
\end{tabular}
\hfill
\begin{tabular}{c|c|c|c|c}
\TopRule
\multicolumn{5}{c}{$L=24\,,\ \mu=4.0\,,\ C=-$} \\
\BotRule
$i$ & $\mathtt{E}_{i}^{-}$ & $\delta \mathtt{E}_{i}^{-}$ & $j$ & $n$ \\
\MidRule
1 & -5.822457e+01 & 7.74e-05 & 1 & 0 \\
2 & -5.767407e+01 & 4.82e-03 & 1 & 1 \\
3 & -5.767414e+01 & 1.57e-03 & 1 & 1 \\
4 & -5.717168e+01 & 6.85e-04 & 1 & 2 \\
\BotRule
\end{tabular}
\hfill
\begin{tabular}{c|c|c|c|c}
\TopRule
\multicolumn{5}{c}{$L=24\,,\ \mu=7.0\,,\ C=+$} \\
\BotRule
$i$ & $\mathtt{E}_{i}^{+}$ & $\delta \mathtt{E}_{i}^{+}$ & $j$ & $n$ \\
\MidRule
0 & -5.964230e+01 & 7.15e-05 & 0 & 0 \\
1 & -5.845812e+01 & 7.67e-05 & 1 & 0 \\
2 & -5.827234e+01 & 7.55e-05 & 1 & 1 \\
3 & -5.827234e+01 & 7.53e-05 & 1 & 1 \\
\BotRule
\end{tabular}
\\[1em]
\begin{tabular}{c|c|c|c|c}
\TopRule
\multicolumn{5}{c}{$L=24\,,\ \mu=7.0\,,\ C=-$} \\
\BotRule
$i$ & $\mathtt{E}_{i}^{-}$ & $\delta \mathtt{E}_{i}^{-}$ & $j$ & $n$ \\
\MidRule
1 & -5.800538e+01 & 7.49e-05 & 1 & 0 \\
2 & -5.782229e+01 & 7.94e-05 & 1 & 1 \\
3 & -5.782229e+01 & 7.44e-05 & 1 & 1 \\
4 & -5.724916e+01 & 1.88e-04 & 1 & 2 \\
\BotRule
\end{tabular}
\hfill
\begin{tabular}{c|c|c|c|c}
\TopRule
\multicolumn{5}{c}{$L=24\,,\ \mu=8.0\,,\ C=+$} \\
\BotRule
$i$ & $\mathtt{E}_{i}^{+}$ & $\delta \mathtt{E}_{i}^{+}$ & $j$ & $n$ \\
\MidRule
0 & -6.000141e+01 & 7.42e-05 & 0 & 0 \\
1 & -5.864163e+01 & 7.41e-05 & 1 & 0 \\
2 & -5.848177e+01 & 7.56e-05 & 1 & 1 \\
3 & -5.848177e+01 & 7.52e-05 & 1 & 1 \\
\BotRule
\end{tabular}
\hfill
\begin{tabular}{c|c|c|c|c}
\TopRule
\multicolumn{5}{c}{$L=24\,,\ \mu=8.0\,,\ C=-$} \\
\BotRule
$i$ & $\mathtt{E}_{i}^{-}$ & $\delta \mathtt{E}_{i}^{-}$ & $j$ & $n$ \\
\MidRule
1 & -5.808616e+01 & 7.67e-05 & 1 & 0 \\
2 & -5.796069e+01 & 7.60e-05 & 1 & 1 \\
3 & -5.796069e+01 & 7.55e-05 & 1 & 1 \\
4 & -5.765286e+01 & 3.86e-04 & 1 & 2 \\
\BotRule
\end{tabular}
\end{table*}

\begin{table*}[h]
\centering
\renewcommand{\arraystretch}{1.3}
\setlength{\tabcolsep}{4pt}

\begin{tabular}{c|c|c|c|c}
\TopRule
\multicolumn{5}{c}{$L=24\,,\ \mu=10.0\,,\ C=+$} \\
\BotRule
$i$ & $\mathtt{E}_{i}^{+}$ & $\delta \mathtt{E}_{i}^{+}$ & $j$ & $n$ \\
\MidRule
0 & -6.086137e+01 & 3.14e-04 & 0 & 0 \\
1 & -5.915841e+01 & 7.50e-05 & 1 & 0 \\
2 & -5.903077e+01 & 7.58e-05 & 1 & 1 \\
3 & -5.903077e+01 & 7.63e-05 & 1 & 1 \\
\BotRule
\end{tabular}
\hfill
\begin{tabular}{c|c|c|c|c}
\TopRule
\multicolumn{5}{c}{$L=24\,,\ \mu=10.0\,,\ C=-$} \\
\BotRule
$i$ & $\mathtt{E}_{i}^{-}$ & $\delta \mathtt{E}_{i}^{-}$ & $j$ & $n$ \\
\MidRule
1 & -5.846969e+01 & 7.65e-05 & 1 & 0 \\
2 & -5.837711e+01 & 7.63e-05 & 1 & 1 \\
3 & -5.837711e+01 & 7.47e-05 & 1 & 1 \\
4 & -5.812171e+01 & 7.58e-05 & 1 & 2 \\
\BotRule
\end{tabular}
\hfill
\begin{tabular}{c|c|c|c|c}
\TopRule
\multicolumn{5}{c}{$L=24\,,\ \mu=12.0\,,\ C=+$} \\
\BotRule
$i$ & $\mathtt{E}_{i}^{+}$ & $\delta \mathtt{E}_{i}^{+}$ & $j$ & $n$ \\
\MidRule
0 & -6.190946e+01 & 4.63e-04 & 0 & 0 \\
1 & -5.986487e+01 & 7.61e-05 & 1 & 0 \\
2 & -5.975859e+01 & 7.58e-05 & 1 & 1 \\
3 & -5.975859e+01 & 7.63e-05 & 1 & 1 \\
4 & -5.947145e+01 & 7.56e-05 & 1 & 2 \\
5 & -5.947145e+01 & 7.72e-05 & 1 & 2 \\
6 & -5.906790e+01 & 7.43e-05 & 1 & 3 \\
7 & -5.906790e+01 & 7.47e-05 & 1 & 3 \\
\BotRule
\end{tabular}
\\[1em]
\begin{tabular}{c|c|c|c|c}
\TopRule
\multicolumn{5}{c}{$L=24\,,\ \mu=12.0\,,\ C=-$} \\
\BotRule
$i$ & $\mathtt{E}_{i}^{-}$ & $\delta \mathtt{E}_{i}^{-}$ & $j$ & $n$ \\
\MidRule
1 & -5.905819e+01 & 2.00e-04 & 1 & 0 \\
2 & -5.898151e+01 & 1.09e-04 & 1 & 1 \\
3 & -5.898151e+01 & 8.82e-05 & 1 & 1 \\
4 & -5.876468e+01 & 8.54e-05 & 1 & 2 \\
5 & -5.876468e+01 & 7.74e-05 & 1 & 2 \\
6 & -5.843921e+01 & 7.43e-05 & 1 & 3 \\
7 & -5.843921e+01 & 7.79e-05 & 1 & 3 \\
8 & -5.803885e+01 & 1.83e-03 & 1 & 4 \\
\BotRule
\end{tabular}
\hfill
\begin{tabular}{c|c|c|c|c}
\TopRule
\multicolumn{5}{c}{$L=24\,,\ \mu=14.0\,,\ C=+$} \\
\BotRule
$i$ & $\mathtt{E}_{i}^{+}$ & $\delta \mathtt{E}_{i}^{+}$ & $j$ & $n$ \\
\MidRule
0 & -6.314472e+01 & 5.75e-04 & 0 & 0 \\
1 & -6.075872e+01 & 1.26e-04 & 1 & 0 \\
2 & -6.066815e+01 & 7.64e-05 & 1 & 1 \\
3 & -6.066815e+01 & 1.04e-04 & 1 & 1 \\
4 & -6.041801e+01 & 7.95e-05 & 1 & 2 \\
5 & -6.041801e+01 & 8.31e-05 & 1 & 2 \\
6 & -6.005622e+01 & 7.73e-05 & 1 & 3 \\
7 & -6.005622e+01 & 7.74e-05 & 1 & 3 \\
\BotRule
\end{tabular}
\hfill
\begin{tabular}{c|c|c|c|c}
\TopRule
\multicolumn{5}{c}{$L=24\,,\ \mu=14.0\,,\ C=-$} \\
\BotRule
$i$ & $\mathtt{E}_{i}^{-}$ & $\delta \mathtt{E}_{i}^{-}$ & $j$ & $n$ \\
\MidRule
1 & -5.984086e+01 & 8.47e-04 & 1 & 0 \\
2 & -5.977547e+01 & 2.80e-04 & 1 & 1 \\
3 & -5.977547e+01 & 3.85e-04 & 1 & 1 \\
4 & -5.958829e+01 & 1.40e-04 & 1 & 2 \\
5 & -5.958829e+01 & 1.92e-04 & 1 & 2 \\
6 & -5.930195e+01 & 1.02e-04 & 1 & 3 \\
7 & -5.930195e+01 & 9.17e-05 & 1 & 3 \\
8 & -5.894418e+01 & 7.69e-05 & 1 & 4 \\
\BotRule
\end{tabular}
\caption{Low energy spectrum of the quantum clock model at various values of $\mu = h^{15/28} L$ in the two charge-conjugation sectors at $L=24$.}
\end{table*}

\begin{table*}[h]
\centering
\renewcommand{\arraystretch}{1.3}
\setlength{\tabcolsep}{4pt}

\begin{tabular}{c|c|c|c|c}
\TopRule
\multicolumn{5}{c}{$L=32\,,\ \mu=0.0\,,\ C=+$} \\
\BotRule
$i$ & $\mathtt{E}_{i}^{+}$ & $\delta \mathtt{E}_{i}^{+}$ & $j$ & $n$ \\
\MidRule
0 & -7.798574e+01 & 1.23e-04 & 0 & 0 \\
1 & -7.791749e+01 & 1.11e-04 & 1 & 0 \\
2 & -7.740734e+01 & 2.80e-04 & 1 & 1 \\
3 & -7.740734e+01 & 2.24e-04 & 1 & 1 \\
\BotRule
\end{tabular}
\hfill
\begin{tabular}{c|c|c|c|c}
\TopRule
\multicolumn{5}{c}{$L=32\,,\ \mu=0.0\,,\ C=-$} \\
\BotRule
$i$ & $\mathtt{E}_{i}^{-}$ & $\delta \mathtt{E}_{i}^{-}$ & $j$ & $n$ \\
\MidRule
1 & -7.791749e+01 & 9.06e-05 & 1 & 0 \\
2 & -7.740734e+01 & 2.55e-04 & 1 & 1 \\
3 & -7.740734e+01 & 2.38e-04 & 1 & 1 \\
4 & -7.730700e+01 & 3.90e-04 & 2 & 0 \\
\BotRule
\end{tabular}
\hfill
\begin{tabular}{c|c|c|c|c}
\TopRule
\multicolumn{5}{c}{$L=32\,,\ \mu=0.5\,,\ C=+$} \\
\BotRule
$i$ & $\mathtt{E}_{i}^{+}$ & $\delta \mathtt{E}_{i}^{+}$ & $j$ & $n$ \\
\MidRule
0 & -7.798599e+01 & 1.22e-04 & 0 & 0 \\
1 & -7.792037e+01 & 1.12e-04 & 1 & 0 \\
2 & -7.740901e+01 & 2.78e-04 & 1 & 1 \\
3 & -7.740901e+01 & 2.13e-04 & 1 & 1 \\
\BotRule
\end{tabular}
\\[1em]
\begin{tabular}{c|c|c|c|c}
\TopRule
\multicolumn{5}{c}{$L=32\,,\ \mu=0.5\,,\ C=-$} \\
\BotRule
$i$ & $\mathtt{E}_{i}^{-}$ & $\delta \mathtt{E}_{i}^{-}$ & $j$ & $n$ \\
\MidRule
1 & -7.791439e+01 & 9.10e-05 & 1 & 0 \\
2 & -7.740569e+01 & 2.36e-04 & 1 & 1 \\
3 & -7.740569e+01 & 2.59e-04 & 1 & 1 \\
4 & -7.708089e+01 & 5.60e-04 & 1 & 2 \\
\BotRule
\end{tabular}
\hfill
\begin{tabular}{c|c|c|c|c}
\TopRule
\multicolumn{5}{c}{$L=32\,,\ \mu=0.75\,,\ C=+$} \\
\BotRule
$i$ & $\mathtt{E}_{i}^{+}$ & $\delta \mathtt{E}_{i}^{+}$ & $j$ & $n$ \\
\MidRule
0 & -7.798693e+01 & 1.20e-04 & 0 & 0 \\
1 & -7.792300e+01 & 1.16e-04 & 1 & 0 \\
2 & -7.741093e+01 & 2.69e-04 & 1 & 1 \\
3 & -7.741093e+01 & 2.50e-04 & 1 & 1 \\
\BotRule
\end{tabular}
\hfill
\begin{tabular}{c|c|c|c|c}
\TopRule
\multicolumn{5}{c}{$L=32\,,\ \mu=0.75\,,\ C=-$} \\
\BotRule
$i$ & $\mathtt{E}_{i}^{-}$ & $\delta \mathtt{E}_{i}^{-}$ & $j$ & $n$ \\
\MidRule
1 & -7.791088e+01 & 9.10e-05 & 1 & 0 \\
2 & -7.740385e+01 & 2.05e-04 & 1 & 1 \\
3 & -7.740385e+01 & 2.79e-04 & 1 & 1 \\
4 & -7.708089e+01 & 5.80e-04 & 1 & 2 \\
\BotRule
\end{tabular}
\end{table*}

\begin{table*}[h]
\centering
\renewcommand{\arraystretch}{1.3}
\setlength{\tabcolsep}{4pt}

\begin{tabular}{c|c|c|c|c}
\TopRule
\multicolumn{5}{c}{$L=32\,,\ \mu=1.0\,,\ C=+$} \\
\BotRule
$i$ & $\mathtt{E}_{i}^{+}$ & $\delta \mathtt{E}_{i}^{+}$ & $j$ & $n$ \\
\MidRule
0 & -7.798935e+01 & 1.17e-04 & 0 & 0 \\
1 & -7.792541e+01 & 1.18e-04 & 1 & 0 \\
2 & -7.741354e+01 & 2.75e-04 & 1 & 1 \\
3 & -7.741354e+01 & 2.14e-04 & 1 & 1 \\
\BotRule
\end{tabular}
\hfill
\begin{tabular}{c|c|c|c|c}
\TopRule
\multicolumn{5}{c}{$L=32\,,\ \mu=1.0\,,\ C=-$} \\
\BotRule
$i$ & $\mathtt{E}_{i}^{-}$ & $\delta \mathtt{E}_{i}^{-}$ & $j$ & $n$ \\
\MidRule
1 & -7.790619e+01 & 9.15e-05 & 1 & 0 \\
2 & -7.740143e+01 & 2.59e-04 & 1 & 1 \\
3 & -7.740143e+01 & 2.47e-04 & 1 & 1 \\
4 & -7.690542e+01 & 1.26e-02 & 1 & 2 \\
\BotRule
\end{tabular}
\hfill
\begin{tabular}{c|c|c|c|c}
\TopRule
\multicolumn{5}{c}{$L=32\,,\ \mu=2.0\,,\ C=+$} \\
\BotRule
$i$ & $\mathtt{E}_{i}^{+}$ & $\delta \mathtt{E}_{i}^{+}$ & $j$ & $n$ \\
\MidRule
0 & -7.802863e+01 & 9.01e-05 & 0 & 0 \\
1 & -7.791855e+01 & 1.18e-04 & 1 & 0 \\
2 & -7.743143e+01 & 2.06e-04 & 1 & 1 \\
3 & -7.743143e+01 & 2.58e-04 & 1 & 1 \\
\BotRule
\end{tabular}
\\[1em]
\begin{tabular}{c|c|c|c|c}
\TopRule
\multicolumn{5}{c}{$L=32\,,\ \mu=2.0\,,\ C=-$} \\
\BotRule
$i$ & $\mathtt{E}_{i}^{-}$ & $\delta \mathtt{E}_{i}^{-}$ & $j$ & $n$ \\
\MidRule
1 & -7.787651e+01 & 9.05e-05 & 1 & 0 \\
2 & -7.738724e+01 & 2.92e-04 & 1 & 1 \\
3 & -7.738724e+01 & 2.32e-04 & 1 & 1 \\
4 & -7.691892e+01 & 2.05e-02 & 1 & 2 \\
\BotRule
\end{tabular}
\hfill
\begin{tabular}{c|c|c|c|c}
\TopRule
\multicolumn{5}{c}{$L=32\,,\ \mu=3.0\,,\ C=+$} \\
\BotRule
$i$ & $\mathtt{E}_{i}^{+}$ & $\delta \mathtt{E}_{i}^{+}$ & $j$ & $n$ \\
\MidRule
0 & -7.811585e+01 & 8.66e-05 & 0 & 0 \\
1 & -7.788732e+01 & 1.22e-04 & 1 & 0 \\
2 & -7.746336e+01 & 2.21e-04 & 1 & 1 \\
3 & -7.746336e+01 & 2.09e-04 & 1 & 1 \\
\BotRule
\end{tabular}
\hfill
\begin{tabular}{c|c|c|c|c}
\TopRule
\multicolumn{5}{c}{$L=32\,,\ \mu=3.0\,,\ C=-$} \\
\BotRule
$i$ & $\mathtt{E}_{i}^{-}$ & $\delta \mathtt{E}_{i}^{-}$ & $j$ & $n$ \\
\MidRule
1 & -7.783103e+01 & 9.31e-05 & 1 & 0 \\
2 & -7.736999e+01 & 3.06e-04 & 1 & 1 \\
3 & -7.736998e+01 & 2.70e-03 & 1 & 1 \\
4 & -7.694256e+01 & 8.07e-04 & 1 & 2 \\
\BotRule
\end{tabular}

\end{table*}

\begin{table*}[h]
\centering
\renewcommand{\arraystretch}{1.3}
\setlength{\tabcolsep}{4pt}

\begin{tabular}{c|c|c|c|c}
\TopRule
\multicolumn{5}{c}{$L=32\,,\ \mu=4.0\,,\ C=+$} \\
\BotRule
$i$ & $\mathtt{E}_{i}^{+}$ & $\delta \mathtt{E}_{i}^{+}$ & $j$ & $n$ \\
\MidRule
0 & -7.824189e+01 & 8.45e-05 & 0 & 0 \\
1 & -7.784850e+01 & 1.37e-04 & 1 & 0 \\
2 & -7.751316e+01 & 2.02e-04 & 1 & 1 \\
3 & -7.751316e+01 & 1.52e-04 & 1 & 1 \\
\BotRule
\end{tabular}
\hfill
\begin{tabular}{c|c|c|c|c}
\TopRule
\multicolumn{5}{c}{$L=32\,,\ \mu=4.0\,,\ C=-$} \\
\BotRule
$i$ & $\mathtt{E}_{i}^{-}$ & $\delta \mathtt{E}_{i}^{-}$ & $j$ & $n$ \\
\MidRule
1 & -7.777206e+01 & 9.95e-05 & 1 & 0 \\
2 & -7.735808e+01 & 4.53e-04 & 1 & 1 \\
3 & -7.735807e+01 & 2.07e-03 & 1 & 1 \\
4 & -7.697850e+01 & 7.03e-04 & 1 & 2 \\
\BotRule
\end{tabular}
\hfill
\begin{tabular}{c|c|c|c|c}
\TopRule
\multicolumn{5}{c}{$L=32\,,\ \mu=5.0\,,\ C=+$} \\
\BotRule
$i$ & $\mathtt{E}_{i}^{+}$ & $\delta \mathtt{E}_{i}^{+}$ & $j$ & $n$ \\
\MidRule
0 & -7.840443e+01 & 8.44e-05 & 0 & 0 \\
1 & -7.782533e+01 & 1.54e-04 & 1 & 0 \\
2 & -7.758505e+01 & 9.46e-04 & 1 & 1 \\
3 & -7.758505e+01 & 2.00e-04 & 1 & 1 \\
\BotRule
\end{tabular}
\\[1em]
\begin{tabular}{c|c|c|c|c}
\TopRule
\multicolumn{5}{c}{$L=32\,,\ \mu=5.0\,,\ C=-$} \\
\BotRule
$i$ & $\mathtt{E}_{i}^{-}$ & $\delta \mathtt{E}_{i}^{-}$ & $j$ & $n$ \\
\MidRule
1 & -7.770339e+01 & 1.16e-04 & 1 & 0 \\
2 & -7.736346e+01 & 6.83e-04 & 1 & 1 \\
3 & -7.736220e+01 & 2.01e-03 & 1 & 1 \\
4 & -7.703176e+01 & 6.07e-04 & 1 & 2 \\
\BotRule
\end{tabular}
\hfill
\begin{tabular}{c|c|c|c|c}
\TopRule
\multicolumn{5}{c}{$L=32\,,\ \mu=6.0\,,\ C=+$} \\
\BotRule
$i$ & $\mathtt{E}_{i}^{+}$ & $\delta \mathtt{E}_{i}^{+}$ & $j$ & $n$ \\
\MidRule
0 & -7.860298e+01 & 8.60e-05 & 0 & 0 \\
1 & -7.785674e+01 & 1.39e-04 & 1 & 0 \\
2 & -7.768288e+01 & 6.85e-04 & 1 & 1 \\
3 & -7.768288e+01 & 1.24e-04 & 1 & 1 \\
\BotRule
\end{tabular}
\hfill
\begin{tabular}{c|c|c|c|c}
\TopRule
\multicolumn{5}{c}{$L=32\,,\ \mu=6.0\,,\ C=-$} \\
\BotRule
$i$ & $\mathtt{E}_{i}^{-}$ & $\delta \mathtt{E}_{i}^{-}$ & $j$ & $n$ \\
\MidRule
1 & -7.763505e+01 & 1.53e-04 & 1 & 0 \\
2 & -7.739713e+01 & 1.74e-03 & 1 & 1 \\
3 & -7.739688e+01 & 1.20e-03 & 1 & 1 \\
4 & -7.704389e+01 & 2.80e-03 & 1 & 2 \\
\BotRule
\end{tabular}
\caption{Low energy spectrum of the quantum clock model at various values of $\mu = h^{15/28} L$ in the two charge-conjugation sectors at $L=32$.}
\end{table*}

\begin{table*}[h]
\centering
\renewcommand{\arraystretch}{1.3}
\setlength{\tabcolsep}{4pt}

\begin{tabular}{c|c|c|c|c}
\TopRule
\multicolumn{5}{c}{$L=32\,,\ \mu=7.0\,,\ C=+$} \\
\BotRule
$i$ & $\mathtt{E}_{i}^{+}$ & $\delta \mathtt{E}_{i}^{+}$ & $j$ & $n$ \\
\MidRule
0 & -7.883735e+01 & 8.54e-05 & 0 & 0 \\
1 & -7.794991e+01 & 1.11e-04 & 1 & 0 \\
2 & -7.780968e+01 & 1.13e-04 & 1 & 1 \\
3 & -7.780968e+01 & 8.98e-05 & 1 & 1 \\
\BotRule
\end{tabular}
\hfill
\begin{tabular}{c|c|c|c|c}
\TopRule
\multicolumn{5}{c}{$L=32\,,\ \mu=7.0\,,\ C=-$} \\
\BotRule
$i$ & $\mathtt{E}_{i}^{-}$ & $\delta \mathtt{E}_{i}^{-}$ & $j$ & $n$ \\
\MidRule
1 & -7.760402e+01 & 1.99e-04 & 1 & 0 \\
2 & -7.746474e+01 & 2.62e-04 & 1 & 1 \\
3 & -7.746474e+01 & 2.04e-04 & 1 & 1 \\
4 & -7.703667e+01 & 8.02e-04 & 1 & 2 \\
\BotRule
\end{tabular}
\hfill
\begin{tabular}{c|c|c|c|c}
\TopRule
\multicolumn{5}{c}{$L=32\,,\ \mu=8.0\,,\ C=+$} \\
\BotRule
$i$ & $\mathtt{E}_{i}^{+}$ & $\delta \mathtt{E}_{i}^{+}$ & $j$ & $n$ \\
\MidRule
0 & -7.910746e+01 & 1.03e-04 & 0 & 0 \\
1 & -7.808847e+01 & 9.34e-05 & 1 & 0 \\
2 & -7.796758e+01 & 8.98e-05 & 1 & 1 \\
3 & -7.796758e+01 & 9.33e-05 & 1 & 1 \\
\BotRule
\end{tabular}
\\[1em]
\begin{tabular}{c|c|c|c|c}
\TopRule
\multicolumn{5}{c}{$L=32\,,\ \mu=8.0\,,\ C=-$} \\
\BotRule
$i$ & $\mathtt{E}_{i}^{-}$ & $\delta \mathtt{E}_{i}^{-}$ & $j$ & $n$ \\
\MidRule
1 & -7.766163e+01 & 1.76e-04 & 1 & 0 \\
2 & -7.756689e+01 & 2.03e-04 & 1 & 1 \\
3 & -7.756689e+01 & 1.80e-04 & 1 & 1 \\
4 & -7.733444e+01 & 3.12e-04 & 1 & 2 \\
\BotRule
\end{tabular}
\hfill
\begin{tabular}{c|c|c|c|c}
\TopRule
\multicolumn{5}{c}{$L=32\,,\ \mu=10.0\,,\ C=+$} \\
\BotRule
$i$ & $\mathtt{E}_{i}^{+}$ & $\delta \mathtt{E}_{i}^{+}$ & $j$ & $n$ \\
\MidRule
0 & -7.975458e+01 & 1.11e-04 & 0 & 0 \\
1 & -7.847839e+01 & 8.84e-05 & 1 & 0 \\
2 & -7.838149e+01 & 8.98e-05 & 1 & 1 \\
3 & -7.838149e+01 & 9.01e-05 & 1 & 1 \\
\BotRule
\end{tabular}
\hfill
\begin{tabular}{c|c|c|c|c}
\TopRule
\multicolumn{5}{c}{$L=32\,,\ \mu=10.0\,,\ C=-$} \\
\BotRule
$i$ & $\mathtt{E}_{i}^{-}$ & $\delta \mathtt{E}_{i}^{-}$ & $j$ & $n$ \\
\MidRule
1 & -7.794557e+01 & 1.71e-04 & 1 & 0 \\
2 & -7.787576e+01 & 1.14e-04 & 1 & 1 \\
3 & -7.787576e+01 & 1.34e-04 & 1 & 1 \\
4 & -7.768242e+01 & 1.56e-04 & 1 & 2 \\
\BotRule
\end{tabular}

\end{table*}

\begin{table*}[h]
\centering
\renewcommand{\arraystretch}{1.3}
\setlength{\tabcolsep}{4pt}

\begin{tabular}{c|c|c|c|c}
\TopRule
\multicolumn{5}{c}{$L=32\,,\ \mu=12.0\,,\ C=+$} \\
\BotRule
$i$ & $\mathtt{E}_{i}^{+}$ & $\delta \mathtt{E}_{i}^{+}$ & $j$ & $n$ \\
\MidRule
0 & -8.054374e+01 & 6.41e-04 & 0 & 0 \\
1 & -7.901139e+01 & 9.12e-05 & 1 & 0 \\
2 & -7.893035e+01 & 9.26e-05 & 1 & 1 \\
3 & -7.893035e+01 & 9.21e-05 & 1 & 1 \\
4 & -7.871034e+01 & 9.04e-05 & 1 & 2 \\
5 & -7.871034e+01 & 8.99e-05 & 1 & 2 \\
6 & -7.839795e+01 & 9.20e-05 & 1 & 3 \\
7 & -7.839795e+01 & 9.17e-05 & 1 & 3 \\
\BotRule
\end{tabular}
\hfill
\begin{tabular}{c|c|c|c|c}
\TopRule
\multicolumn{5}{c}{$L=32\,,\ \mu=12.0\,,\ C=-$} \\
\BotRule
$i$ & $\mathtt{E}_{i}^{-}$ & $\delta \mathtt{E}_{i}^{-}$ & $j$ & $n$ \\
\MidRule
1 & -7.838342e+01 & 8.79e-05 & 1 & 0 \\
2 & -7.832538e+01 & 9.23e-05 & 1 & 1 \\
3 & -7.832538e+01 & 8.97e-05 & 1 & 1 \\
4 & -7.816059e+01 & 9.71e-05 & 1 & 2 \\
5 & -7.816059e+01 & 9.69e-05 & 1 & 2 \\
6 & -7.791121e+01 & 1.06e-04 & 1 & 3 \\
7 & -7.791121e+01 & 1.04e-04 & 1 & 3 \\
8 & -7.761681e+01 & 2.65e-03 & 1 & 4 \\
\BotRule
\end{tabular}
\hfill
\begin{tabular}{c|c|c|c|c}
\TopRule
\multicolumn{5}{c}{$L=32\,,\ \mu=14.0\,,\ C=+$} \\
\BotRule
$i$ & $\mathtt{E}_{i}^{+}$ & $\delta \mathtt{E}_{i}^{+}$ & $j$ & $n$ \\
\MidRule
0 & -8.147438e+01 & 1.30e-03 & 0 & 0 \\
1 & -7.968590e+01 & 8.14e-04 & 1 & 0 \\
2 & -7.961647e+01 & 5.57e-04 & 1 & 1 \\
3 & -7.961647e+01 & 1.21e-04 & 1 & 1 \\
4 & -7.942376e+01 & 1.74e-04 & 1 & 2 \\
5 & -7.942376e+01 & 1.60e-04 & 1 & 2 \\
6 & -7.914210e+01 & 1.08e-04 & 1 & 3 \\
7 & -7.914210e+01 & 1.04e-04 & 1 & 3 \\
\BotRule
\end{tabular}
\\[1em]
\begin{tabular}{c|c|c|c|c}
\TopRule
\multicolumn{5}{c}{$L=32\,,\ \mu=14.0\,,\ C=-$} \\
\BotRule
$i$ & $\mathtt{E}_{i}^{-}$ & $\delta \mathtt{E}_{i}^{-}$ & $j$ & $n$ \\
\MidRule
1 & -7.896675e+01 & 3.71e-04 & 1 & 0 \\
2 & -7.891701e+01 & 2.13e-04 & 1 & 1 \\
3 & -7.891701e+01 & 1.69e-04 & 1 & 1 \\
4 & -7.877402e+01 & 1.42e-04 & 1 & 2 \\
5 & -7.877402e+01 & 1.43e-04 & 1 & 2 \\
6 & -7.855342e+01 & 9.87e-05 & 1 & 3 \\
7 & -7.855342e+01 & 9.23e-05 & 1 & 3 \\
8 & -7.827426e+01 & 1.93e-04 & 1 & 4 \\
\BotRule
\end{tabular}
\hfill
\begin{tabular}{c|c|c|c|c}
\TopRule
\multicolumn{5}{c}{$L=48\,,\ \mu=0.0\,,\ C=+$} \\
\BotRule
$i$ & $\mathtt{E}_{i}^{+}$ & $\delta \mathtt{E}_{i}^{+}$ & $j$ & $n$ \\
\MidRule
0 & -1.169503e+02 & 5.94e-04 & 0 & 0 \\
1 & -1.169048e+02 & 4.96e-04 & 1 & 0 \\
2 & -1.165645e+02 & 1.32e-03 & 1 & 1 \\
3 & -1.166746e+02 & 1.48e-03 & 2 & 0 \\
\BotRule
\end{tabular}
\hfill
\begin{tabular}{c|c|c|c|c}
\TopRule
\multicolumn{5}{c}{$L=48\,,\ \mu=0.0\,,\ C=-$} \\
\BotRule
$i$ & $\mathtt{E}_{i}^{-}$ & $\delta \mathtt{E}_{i}^{-}$ & $j$ & $n$ \\
\MidRule
1 & -1.169048e+02 & 2.84e-04 & 1 & 0 \\
2 & -1.165645e+02 & 1.07e-03 & 1 & 1 \\
3 & -1.165645e+02 & 1.26e-03 & 1 & 1 \\
4 & -1.164973e+02 & 1.86e-03 & 2 & 0 \\
\BotRule
\end{tabular}
\caption{Low energy spectrum of the quantum clock model at various values of $\mu = h^{15/28} L$ in the two charge-conjugation sectors at $L=32$.}
\end{table*}

\begin{table*}[h]
\centering
\renewcommand{\arraystretch}{1.3}
\setlength{\tabcolsep}{4pt}

\begin{tabular}{c|c|c|c|c}
\TopRule
\multicolumn{5}{c}{$L=48\,,\ \mu=0.5\,,\ C=+$} \\
\BotRule
$i$ & $\mathtt{E}_{i}^{+}$ & $\delta \mathtt{E}_{i}^{+}$ & $j$ & $n$ \\
\MidRule
0 & -1.169504e+02 & 5.93e-04 & 0 & 0 \\
1 & -1.169067e+02 & 5.15e-04 & 1 & 0 \\
2 & -1.165657e+02 & 1.30e-03 & 1 & 1 \\
3 & -1.165657e+02 & 1.06e-03 & 1 & 1 \\
\BotRule
\end{tabular}
\hfill
\begin{tabular}{c|c|c|c|c}
\TopRule
\multicolumn{5}{c}{$L=48\,,\ \mu=0.5\,,\ C=-$} \\
\BotRule
$i$ & $\mathtt{E}_{i}^{-}$ & $\delta \mathtt{E}_{i}^{-}$ & $j$ & $n$ \\
\MidRule
1 & -1.169027e+02 & 2.86e-04 & 1 & 0 \\
2 & -1.165634e+02 & 1.02e-03 & 1 & 1 \\
3 & -1.165634e+02 & 1.30e-03 & 1 & 1 \\
4 & -1.162260e+02 & 7.72e-03 & 1 & 2 \\
\BotRule
\end{tabular}
\hfill
\begin{tabular}{c|c|c|c|c}
\TopRule
\multicolumn{5}{c}{$L=48\,,\ \mu=0.75\,,\ C=+$} \\
\BotRule
$i$ & $\mathtt{E}_{i}^{+}$ & $\delta \mathtt{E}_{i}^{+}$ & $j$ & $n$ \\
\MidRule
0 & -1.169510e+02 & 5.78e-04 & 0 & 0 \\
1 & -1.169085e+02 & 5.58e-04 & 1 & 0 \\
2 & -1.165669e+02 & 1.23e-03 & 1 & 1 \\
3 & -1.165669e+02 & 1.19e-03 & 1 & 1 \\
\BotRule
\end{tabular}
\\[1em]
\begin{tabular}{c|c|c|c|c}
\TopRule
\multicolumn{5}{c}{$L=48\,,\ \mu=0.75\,,\ C=-$} \\
\BotRule
$i$ & $\mathtt{E}_{i}^{-}$ & $\delta \mathtt{E}_{i}^{-}$ & $j$ & $n$ \\
\MidRule
1 & -1.169004e+02 & 2.84e-04 & 1 & 0 \\
2 & -1.165622e+02 & 1.30e-03 & 1 & 1 \\
3 & -1.165622e+02 & 1.04e-03 & 1 & 1 \\
4 & -1.164975e+02 & 1.85e-03 & 2 & 0 \\
\BotRule
\end{tabular}
\hfill
\begin{tabular}{c|c|c|c|c}
\TopRule
\multicolumn{5}{c}{$L=48\,,\ \mu=1.0\,,\ C=+$} \\
\BotRule
$i$ & $\mathtt{E}_{i}^{+}$ & $\delta \mathtt{E}_{i}^{+}$ & $j$ & $n$ \\
\MidRule
0 & -1.169527e+02 & 5.62e-04 & 0 & 0 \\
1 & -1.169101e+02 & 5.65e-04 & 1 & 0 \\
2 & -1.165687e+02 & 1.10e-03 & 1 & 1 \\
3 & -1.165687e+02 & 1.27e-03 & 1 & 1 \\
\BotRule
\end{tabular}
\hfill
\begin{tabular}{c|c|c|c|c}
\TopRule
\multicolumn{5}{c}{$L=48\,,\ \mu=1.0\,,\ C=-$} \\
\BotRule
$i$ & $\mathtt{E}_{i}^{-}$ & $\delta \mathtt{E}_{i}^{-}$ & $j$ & $n$ \\
\MidRule
1 & -1.168973e+02 & 2.91e-04 & 1 & 0 \\
2 & -1.165606e+02 & 1.30e-03 & 1 & 1 \\
3 & -1.165606e+02 & 1.08e-03 & 1 & 1 \\
4 & -1.162259e+02 & 1.85e-02 & 1 & 2 \\
\BotRule
\end{tabular}

\end{table*}

\begin{table*}[h]
\centering
\renewcommand{\arraystretch}{1.3}
\setlength{\tabcolsep}{4pt}

\begin{tabular}{c|c|c|c|c}
\TopRule
\multicolumn{5}{c}{$L=48\,,\ \mu=2.0\,,\ C=+$} \\
\BotRule
$i$ & $\mathtt{E}_{i}^{+}$ & $\delta \mathtt{E}_{i}^{+}$ & $j$ & $n$ \\
\MidRule
0 & -1.169789e+02 & 3.36e-04 & 0 & 0 \\
1 & -1.169055e+02 & 5.57e-04 & 1 & 0 \\
2 & -1.165806e+02 & 1.21e-03 & 1 & 1 \\
3 & -1.165806e+02 & 9.61e-04 & 1 & 1 \\
\BotRule
\end{tabular}
\hfill
\begin{tabular}{c|c|c|c|c}
\TopRule
\multicolumn{5}{c}{$L=48\,,\ \mu=2.0\,,\ C=-$} \\
\BotRule
$i$ & $\mathtt{E}_{i}^{-}$ & $\delta \mathtt{E}_{i}^{-}$ & $j$ & $n$ \\
\MidRule
1 & -1.168774e+02 & 3.12e-04 & 1 & 0 \\
2 & -1.165511e+02 & 1.22e-03 & 1 & 1 \\
3 & -1.165511e+02 & 1.32e-03 & 1 & 1 \\
4 & -1.164995e+02 & 1.85e-03 & 2 & 0 \\
\BotRule
\end{tabular}
\hfill
\begin{tabular}{c|c|c|c|c}
\TopRule
\multicolumn{5}{c}{$L=48\,,\ \mu=3.0\,,\ C=+$} \\
\BotRule
$i$ & $\mathtt{E}_{i}^{+}$ & $\delta \mathtt{E}_{i}^{+}$ & $j$ & $n$ \\
\MidRule
0 & -1.170371e+02 & 1.86e-04 & 0 & 0 \\
1 & -1.168848e+02 & 5.80e-04 & 1 & 0 \\
2 & -1.166020e+02 & 9.27e-04 & 1 & 1 \\
3 & -1.166020e+02 & 1.06e-03 & 1 & 1 \\
\BotRule
\end{tabular}
\\[1em]
\begin{tabular}{c|c|c|c|c}
\TopRule
\multicolumn{5}{c}{$L=48\,,\ \mu=3.0\,,\ C=-$} \\
\BotRule
$i$ & $\mathtt{E}_{i}^{-}$ & $\delta \mathtt{E}_{i}^{-}$ & $j$ & $n$ \\
\MidRule
1 & -1.168470e+02 & 3.51e-04 & 1 & 0 \\
2 & -1.165394e+02 & 1.51e-03 & 1 & 1 \\
3 & -1.165394e+02 & 2.49e-03 & 1 & 1 \\
4 & -1.162528e+02 & 4.28e-03 & 1 & 2 \\
\BotRule
\end{tabular}
\hfill
\begin{tabular}{c|c|c|c|c}
\TopRule
\multicolumn{5}{c}{$L=48\,,\ \mu=4.0\,,\ C=+$} \\
\BotRule
$i$ & $\mathtt{E}_{i}^{+}$ & $\delta \mathtt{E}_{i}^{+}$ & $j$ & $n$ \\
\MidRule
0 & -1.171213e+02 & 1.21e-04 & 0 & 0 \\
1 & -1.168590e+02 & 6.77e-04 & 1 & 0 \\
2 & -1.166354e+02 & 8.23e-04 & 1 & 1 \\
3 & -1.166354e+02 & 9.01e-04 & 1 & 1 \\
\BotRule
\end{tabular}
\hfill
\begin{tabular}{c|c|c|c|c}
\TopRule
\multicolumn{5}{c}{$L=48\,,\ \mu=4.0\,,\ C=-$} \\
\BotRule
$i$ & $\mathtt{E}_{i}^{-}$ & $\delta \mathtt{E}_{i}^{-}$ & $j$ & $n$ \\
\MidRule
1 & -1.168076e+02 & 3.99e-04 & 1 & 0 \\
2 & -1.165309e+02 & 4.85e-03 & 1 & 1 \\
3 & -1.165310e+02 & 1.24e-03 & 1 & 1 \\
4 & -1.162765e+02 & 3.10e-03 & 1 & 2 \\
\BotRule
\end{tabular}

\end{table*}

\begin{table*}[h]
\centering
\renewcommand{\arraystretch}{1.3}
\setlength{\tabcolsep}{4pt}

\begin{tabular}{c|c|c|c|c}
\TopRule
\multicolumn{5}{c}{$L=48\,,\ \mu=5.0\,,\ C=+$} \\
\BotRule
$i$ & $\mathtt{E}_{i}^{+}$ & $\delta \mathtt{E}_{i}^{+}$ & $j$ & $n$ \\
\MidRule
0 & -1.172299e+02 & 1.07e-04 & 0 & 0 \\
1 & -1.168439e+02 & 7.54e-04 & 1 & 0 \\
2 & -1.166836e+02 & 8.06e-04 & 1 & 1 \\
3 & -1.166836e+02 & 5.99e-04 & 1 & 1 \\
\BotRule
\end{tabular}
\hfill
\begin{tabular}{c|c|c|c|c}
\TopRule
\multicolumn{5}{c}{$L=48\,,\ \mu=5.0\,,\ C=-$} \\
\BotRule
$i$ & $\mathtt{E}_{i}^{-}$ & $\delta \mathtt{E}_{i}^{-}$ & $j$ & $n$ \\
\MidRule
1 & -1.167617e+02 & 5.43e-04 & 1 & 0 \\
2 & -1.165336e+02 & 3.27e-03 & 1 & 1 \\
3 & -1.165338e+02 & 1.61e-03 & 1 & 1 \\
4 & -1.165321e+02 & 2.59e-03 & 2 & 0 \\
\BotRule
\end{tabular}
\hfill
\begin{tabular}{c|c|c|c|c}
\TopRule
\multicolumn{5}{c}{$L=48\,,\ \mu=6.0\,,\ C=+$} \\
\BotRule
$i$ & $\mathtt{E}_{i}^{+}$ & $\delta \mathtt{E}_{i}^{+}$ & $j$ & $n$ \\
\MidRule
0 & -1.173626e+02 & 1.06e-04 & 0 & 0 \\
1 & -1.168655e+02 & 6.75e-04 & 1 & 0 \\
2 & -1.167492e+02 & 6.32e-04 & 1 & 1 \\
3 & -1.167492e+02 & 5.03e-04 & 1 & 1 \\
\BotRule
\end{tabular}
\\[1em]
\begin{tabular}{c|c|c|c|c}
\TopRule
\multicolumn{5}{c}{$L=48\,,\ \mu=6.0\,,\ C=-$} \\
\BotRule
$i$ & $\mathtt{E}_{i}^{-}$ & $\delta \mathtt{E}_{i}^{-}$ & $j$ & $n$ \\
\MidRule
1 & -1.167156e+02 & 7.27e-04 & 1 & 0 \\
2 & -1.165550e+02 & 2.35e-03 & 1 & 1 \\
3 & -1.165549e+02 & 1.82e-03 & 1 & 1 \\
4 & -1.165551e+02 & 1.81e-03 & 2 & 0 \\
\BotRule
\end{tabular}
\hfill
\begin{tabular}{c|c|c|c|c}
\TopRule
\multicolumn{5}{c}{$L=48\,,\ \mu=7.0\,,\ C=+$} \\
\BotRule
$i$ & $\mathtt{E}_{i}^{+}$ & $\delta \mathtt{E}_{i}^{+}$ & $j$ & $n$ \\
\MidRule
0 & -1.175193e+02 & 1.11e-04 & 0 & 0 \\
1 & -1.169282e+02 & 5.06e-04 & 1 & 0 \\
2 & -1.168343e+02 & 5.08e-04 & 1 & 1 \\
3 & -1.168343e+02 & 3.62e-04 & 1 & 1 \\
\BotRule
\end{tabular}
\hfill
\begin{tabular}{c|c|c|c|c}
\TopRule
\multicolumn{5}{c}{$L=48\,,\ \mu=7.0\,,\ C=-$} \\
\BotRule
$i$ & $\mathtt{E}_{i}^{-}$ & $\delta \mathtt{E}_{i}^{-}$ & $j$ & $n$ \\
\MidRule
1 & -1.166932e+02 & 9.90e-04 & 1 & 0 \\
2 & -1.165991e+02 & 1.02e-03 & 1 & 1 \\
3 & -1.165991e+02 & 1.29e-03 & 1 & 1 \\
4 & -1.163153e+02 & 3.44e-03 & 1 & 2 \\
\BotRule
\end{tabular}
\caption{Low energy spectrum of the quantum clock model at various values of $\mu = h^{15/28} L$ in the two charge-conjugation sectors at $L=48$.}
\end{table*}

\begin{table*}[h]
\centering
\renewcommand{\arraystretch}{1.3}
\setlength{\tabcolsep}{4pt}

\begin{tabular}{c|c|c|c|c}
\TopRule
\multicolumn{5}{c}{$L=48\,,\ \mu=8.0\,,\ C=+$} \\
\BotRule
$i$ & $\mathtt{E}_{i}^{+}$ & $\delta \mathtt{E}_{i}^{+}$ & $j$ & $n$ \\
\MidRule
0 & -1.176999e+02 & 1.17e-04 & 0 & 0 \\
1 & -1.170213e+02 & 3.67e-04 & 1 & 0 \\
2 & -1.169402e+02 & 3.85e-04 & 1 & 1 \\
3 & -1.169402e+02 & 2.68e-04 & 1 & 1 \\
\BotRule
\end{tabular}
\hfill
\begin{tabular}{c|c|c|c|c}
\TopRule
\multicolumn{5}{c}{$L=48\,,\ \mu=8.0\,,\ C=-$} \\
\BotRule
$i$ & $\mathtt{E}_{i}^{-}$ & $\delta \mathtt{E}_{i}^{-}$ & $j$ & $n$ \\
\MidRule
1 & -1.167296e+02 & 8.97e-04 & 1 & 0 \\
2 & -1.166661e+02 & 9.83e-04 & 1 & 1 \\
3 & -1.166661e+02 & 9.52e-04 & 1 & 1 \\
4 & -1.165107e+02 & 1.44e-03 & 1 & 2 \\
\BotRule
\end{tabular}
\hfill
\begin{tabular}{c|c|c|c|c}
\TopRule
\multicolumn{5}{c}{$L=48\,,\ \mu=10.0\,,\ C=+$} \\
\BotRule
$i$ & $\mathtt{E}_{i}^{+}$ & $\delta \mathtt{E}_{i}^{+}$ & $j$ & $n$ \\
\MidRule
0 & -1.181329e+02 & 1.46e-04 & 0 & 0 \\
1 & -1.172831e+02 & 2.20e-04 & 1 & 0 \\
2 & -1.172179e+02 & 2.13e-04 & 1 & 1 \\
3 & -1.172179e+02 & 1.67e-04 & 1 & 1 \\
\BotRule
\end{tabular}
\\[1em]
\begin{tabular}{c|c|c|c|c}
\TopRule
\multicolumn{5}{c}{$L=48\,,\ \mu=10.0\,,\ C=-$} \\
\BotRule
$i$ & $\mathtt{E}_{i}^{-}$ & $\delta \mathtt{E}_{i}^{-}$ & $j$ & $n$ \\
\MidRule
1 & -1.169165e+02 & 9.40e-04 & 1 & 0 \\
2 & -1.168699e+02 & 5.36e-04 & 1 & 1 \\
3 & -1.168699e+02 & 7.61e-04 & 1 & 1 \\
4 & -1.167404e+02 & 8.35e-04 & 1 & 2 \\
\BotRule
\end{tabular}
\hfill
\begin{tabular}{c|c|c|c|c}
\TopRule
\multicolumn{5}{c}{$L=48\,,\ \mu=12.0\,,\ C=+$} \\
\BotRule
$i$ & $\mathtt{E}_{i}^{+}$ & $\delta \mathtt{E}_{i}^{+}$ & $j$ & $n$ \\
\MidRule
0 & -1.186612e+02 & 1.08e-04 & 0 & 0 \\
1 & -1.176408e+02 & 1.35e-04 & 1 & 0 \\
2 & -1.175861e+02 & 1.27e-04 & 1 & 1 \\
3 & -1.175861e+02 & 1.22e-04 & 1 & 1 \\
4 & -1.174371e+02 & 1.51e-04 & 1 & 2 \\
5 & -1.174371e+02 & 1.63e-04 & 1 & 2 \\
6 & -1.172242e+02 & 1.66e-04 & 1 & 3 \\
7 & -1.172242e+02 & 1.71e-04 & 1 & 3 \\
\BotRule
\end{tabular}
\hfill
\begin{tabular}{c|c|c|c|c}
\TopRule
\multicolumn{5}{c}{$L=48\,,\ \mu=12.0\,,\ C=-$} \\
\BotRule
$i$ & $\mathtt{E}_{i}^{-}$ & $\delta \mathtt{E}_{i}^{-}$ & $j$ & $n$ \\
\MidRule
1 & -1.172062e+02 & 5.92e-04 & 1 & 0 \\
2 & -1.171674e+02 & 6.17e-04 & 1 & 1 \\
3 & -1.171674e+02 & 3.69e-04 & 1 & 1 \\
4 & -1.170568e+02 & 4.66e-04 & 1 & 2 \\
5 & -1.170568e+02 & 5.64e-04 & 1 & 2 \\
6 & -1.168885e+02 & 6.00e-04 & 1 & 3 \\
7 & -1.168885e+02 & 6.18e-04 & 1 & 3 \\
8 & -1.166947e+02 & 2.41e-03 & 1 & 4 \\
\BotRule
\end{tabular}

\end{table*}

\begin{table*}[h]
\centering
\renewcommand{\arraystretch}{1.3}
\setlength{\tabcolsep}{4pt}

\begin{tabular}{c|c|c|c|c}
\TopRule
\multicolumn{5}{c}{$L=48\,,\ \mu=14.0\,,\ C=+$} \\
\BotRule
$i$ & $\mathtt{E}_{i}^{+}$ & $\delta \mathtt{E}_{i}^{+}$ & $j$ & $n$ \\
\MidRule
0 & -1.192846e+02 & 3.26e-03 & 0 & 0 \\
1 & -1.180936e+02 & 2.74e-04 & 1 & 0 \\
2 & -1.180466e+02 & 1.64e-04 & 1 & 1 \\
3 & -1.180466e+02 & 2.66e-04 & 1 & 1 \\
4 & -1.179156e+02 & 1.74e-04 & 1 & 2 \\
5 & -1.179156e+02 & 1.34e-04 & 1 & 2 \\
6 & -1.177228e+02 & 1.34e-04 & 1 & 3 \\
7 & -1.177228e+02 & 1.36e-04 & 1 & 3 \\
\BotRule
\end{tabular}
\hfill
\begin{tabular}{c|c|c|c|c}
\TopRule
\multicolumn{5}{c}{$L=48\,,\ \mu=14.0\,,\ C=-$} \\
\BotRule
$i$ & $\mathtt{E}_{i}^{-}$ & $\delta \mathtt{E}_{i}^{-}$ & $j$ & $n$ \\
\MidRule
1 & -1.175929e+02 & 3.33e-03 & 1 & 0 \\
2 & -1.175595e+02 & 2.33e-03 & 1 & 1 \\
3 & -1.175595e+02 & 5.65e-04 & 1 & 1 \\
4 & -1.174633e+02 & 2.62e-04 & 1 & 2 \\
5 & -1.174633e+02 & 9.86e-04 & 1 & 2 \\
6 & -1.173139e+02 & 3.41e-04 & 1 & 3 \\
7 & -1.173139e+02 & 3.85e-04 & 1 & 3 \\
8 & -1.171233e+02 & 5.23e-04 & 1 & 4 \\
\BotRule
\end{tabular}
\hfill
\begin{tabular}{c|c|c|c|c}
\TopRule
\multicolumn{5}{c}{$L=64\,,\ \mu=0.0\,,\ C=+$} \\
\BotRule
$i$ & $\mathtt{E}_{i}^{+}$ & $\delta \mathtt{E}_{i}^{+}$ & $j$ & $n$ \\
\MidRule
0 & -1.559204e+02 & 1.57e-03 & 0 & 0 \\
1 & -1.558864e+02 & 1.35e-03 & 1 & 0 \\
2 & -1.556312e+02 & 2.57e-03 & 1 & 1 \\
3 & -1.557142e+02 & 3.60e-03 & 2 & 0 \\
\BotRule
\end{tabular}
\\[1em]
\begin{tabular}{c|c|c|c|c}
\TopRule
\multicolumn{5}{c}{$L=64\,,\ \mu=0.0\,,\ C=-$} \\
\BotRule
$i$ & $\mathtt{E}_{i}^{-}$ & $\delta \mathtt{E}_{i}^{-}$ & $j$ & $n$ \\
\MidRule
1 & -1.558864e+02 & 7.38e-04 & 1 & 0 \\
2 & -1.556312e+02 & 2.65e-03 & 1 & 1 \\
3 & -1.556312e+02 & 3.02e-03 & 1 & 1 \\
4 & -1.555806e+02 & 4.44e-03 & 2 & 0 \\
\BotRule
\end{tabular}
\hfill
\begin{tabular}{c|c|c|c|c}
\TopRule
\multicolumn{5}{c}{$L=64\,,\ \mu=0.5\,,\ C=+$} \\
\BotRule
$i$ & $\mathtt{E}_{i}^{+}$ & $\delta \mathtt{E}_{i}^{+}$ & $j$ & $n$ \\
\MidRule
0 & -1.559206e+02 & 1.56e-03 & 0 & 0 \\
1 & -1.558878e+02 & 1.37e-03 & 1 & 0 \\
2 & -1.556320e+02 & 2.99e-03 & 1 & 1 \\
3 & -1.556320e+02 & 2.83e-03 & 1 & 1 \\
\BotRule
\end{tabular}
\hfill
\begin{tabular}{c|c|c|c|c}
\TopRule
\multicolumn{5}{c}{$L=64\,,\ \mu=0.5\,,\ C=-$} \\
\BotRule
$i$ & $\mathtt{E}_{i}^{-}$ & $\delta \mathtt{E}_{i}^{-}$ & $j$ & $n$ \\
\MidRule
1 & -1.558848e+02 & 7.43e-04 & 1 & 0 \\
2 & -1.556303e+02 & 2.47e-03 & 1 & 1 \\
3 & -1.556303e+02 & 3.14e-03 & 1 & 1 \\
4 & -1.555806e+02 & 4.44e-03 & 2 & 0 \\
\BotRule
\end{tabular}
\caption{Low energy spectrum of the quantum clock model at various values of $\mu = h^{15/28} L$ in the two charge-conjugation sectors at $L=48$.}
\end{table*}

\begin{table*}[h]
\centering
\renewcommand{\arraystretch}{1.3}
\setlength{\tabcolsep}{4pt}

\begin{tabular}{c|c|c|c|c}
\TopRule
\multicolumn{5}{c}{$L=64\,,\ \mu=0.75\,,\ C=+$} \\
\BotRule
$i$ & $\mathtt{E}_{i}^{+}$ & $\delta \mathtt{E}_{i}^{+}$ & $j$ & $n$ \\
\MidRule
0 & -1.559210e+02 & 1.52e-03 & 0 & 0 \\
1 & -1.558891e+02 & 1.46e-03 & 1 & 0 \\
2 & -1.556330e+02 & 3.15e-03 & 1 & 1 \\
3 & -1.556330e+02 & 2.51e-03 & 1 & 1 \\
\BotRule
\end{tabular}
\hfill
\begin{tabular}{c|c|c|c|c}
\TopRule
\multicolumn{5}{c}{$L=64\,,\ \mu=0.75\,,\ C=-$} \\
\BotRule
$i$ & $\mathtt{E}_{i}^{-}$ & $\delta \mathtt{E}_{i}^{-}$ & $j$ & $n$ \\
\MidRule
1 & -1.558831e+02 & 7.48e-04 & 1 & 0 \\
2 & -1.556294e+02 & 3.15e-03 & 1 & 1 \\
3 & -1.556294e+02 & 2.46e-03 & 1 & 1 \\
4 & -1.555807e+02 & 4.42e-03 & 2 & 0 \\
\BotRule
\end{tabular}
\hfill
\begin{tabular}{c|c|c|c|c}
\TopRule
\multicolumn{5}{c}{$L=64\,,\ \mu=1.0\,,\ C=+$} \\
\BotRule
$i$ & $\mathtt{E}_{i}^{+}$ & $\delta \mathtt{E}_{i}^{+}$ & $j$ & $n$ \\
\MidRule
0 & -1.559222e+02 & 1.46e-03 & 0 & 0 \\
1 & -1.558903e+02 & 1.46e-03 & 1 & 0 \\
2 & -1.556343e+02 & 2.94e-03 & 1 & 1 \\
3 & -1.556343e+02 & 2.78e-03 & 1 & 1 \\
\BotRule
\end{tabular}
\\[1em]
\begin{tabular}{c|c|c|c|c}
\TopRule
\multicolumn{5}{c}{$L=64\,,\ \mu=1.0\,,\ C=-$} \\
\BotRule
$i$ & $\mathtt{E}_{i}^{-}$ & $\delta \mathtt{E}_{i}^{-}$ & $j$ & $n$ \\
\MidRule
1 & -1.558807e+02 & 7.57e-04 & 1 & 0 \\
2 & -1.556282e+02 & 2.57e-03 & 1 & 1 \\
3 & -1.556282e+02 & 3.21e-03 & 1 & 1 \\
4 & -1.555808e+02 & 4.44e-03 & 2 & 0 \\
\BotRule
\end{tabular}
\hfill
\begin{tabular}{c|c|c|c|c}
\TopRule
\multicolumn{5}{c}{$L=64\,,\ \mu=2.0\,,\ C=+$} \\
\BotRule
$i$ & $\mathtt{E}_{i}^{+}$ & $\delta \mathtt{E}_{i}^{+}$ & $j$ & $n$ \\
\MidRule
0 & -1.559419e+02 & 8.83e-04 & 0 & 0 \\
1 & -1.558869e+02 & 1.43e-03 & 1 & 0 \\
2 & -1.556433e+02 & 2.78e-03 & 1 & 1 \\
3 & -1.556433e+02 & 2.58e-03 & 1 & 1 \\
\BotRule
\end{tabular}
\hfill
\begin{tabular}{c|c|c|c|c}
\TopRule
\multicolumn{5}{c}{$L=64\,,\ \mu=2.0\,,\ C=-$} \\
\BotRule
$i$ & $\mathtt{E}_{i}^{-}$ & $\delta \mathtt{E}_{i}^{-}$ & $j$ & $n$ \\
\MidRule
1 & -1.558658e+02 & 8.11e-04 & 1 & 0 \\
2 & -1.556210e+02 & 2.96e-03 & 1 & 1 \\
3 & -1.556210e+02 & 3.17e-03 & 1 & 1 \\
4 & -1.555821e+02 & 4.43e-03 & 2 & 0 \\
\BotRule
\end{tabular}

\end{table*}

\begin{table*}[h]
\centering
\renewcommand{\arraystretch}{1.3}
\setlength{\tabcolsep}{4pt}

\begin{tabular}{c|c|c|c|c}
\TopRule
\multicolumn{5}{c}{$L=64\,,\ \mu=3.0\,,\ C=+$} \\
\BotRule
$i$ & $\mathtt{E}_{i}^{+}$ & $\delta \mathtt{E}_{i}^{+}$ & $j$ & $n$ \\
\MidRule
0 & -1.559856e+02 & 4.75e-04 & 0 & 0 \\
1 & -1.558714e+02 & 1.50e-03 & 1 & 0 \\
2 & -1.556593e+02 & 2.08e-03 & 1 & 1 \\
3 & -1.556593e+02 & 2.66e-03 & 1 & 1 \\
\BotRule
\end{tabular}
\hfill
\begin{tabular}{c|c|c|c|c}
\TopRule
\multicolumn{5}{c}{$L=64\,,\ \mu=3.0\,,\ C=-$} \\
\BotRule
$i$ & $\mathtt{E}_{i}^{-}$ & $\delta \mathtt{E}_{i}^{-}$ & $j$ & $n$ \\
\MidRule
1 & -1.558430e+02 & 9.55e-04 & 1 & 0 \\
2 & -1.556122e+02 & 3.56e-03 & 1 & 1 \\
3 & -1.556122e+02 & 2.90e-03 & 1 & 1 \\
4 & -1.555856e+02 & 4.40e-03 & 2 & 0 \\
\BotRule
\end{tabular}
\hfill
\begin{tabular}{c|c|c|c|c}
\TopRule
\multicolumn{5}{c}{$L=64\,,\ \mu=4.0\,,\ C=+$} \\
\BotRule
$i$ & $\mathtt{E}_{i}^{+}$ & $\delta \mathtt{E}_{i}^{+}$ & $j$ & $n$ \\
\MidRule
0 & -1.560488e+02 & 2.71e-04 & 0 & 0 \\
1 & -1.558521e+02 & 1.72e-03 & 1 & 0 \\
2 & -1.556844e+02 & 2.04e-03 & 1 & 1 \\
3 & -1.556844e+02 & 2.18e-03 & 1 & 1 \\
\BotRule
\end{tabular}
\\[1em]
\begin{tabular}{c|c|c|c|c}
\TopRule
\multicolumn{5}{c}{$L=64\,,\ \mu=4.0\,,\ C=-$} \\
\BotRule
$i$ & $\mathtt{E}_{i}^{-}$ & $\delta \mathtt{E}_{i}^{-}$ & $j$ & $n$ \\
\MidRule
1 & -1.558134e+02 & 1.06e-03 & 1 & 0 \\
2 & -1.556057e+02 & 3.60e-03 & 1 & 1 \\
3 & -1.556055e+02 & 1.40e-02 & 1 & 1 \\
4 & -1.555933e+02 & 7.88e-03 & 2 & 0 \\
\BotRule
\end{tabular}
\hfill
\begin{tabular}{c|c|c|c|c}
\TopRule
\multicolumn{5}{c}{$L=64\,,\ \mu=5.0\,,\ C=+$} \\
\BotRule
$i$ & $\mathtt{E}_{i}^{+}$ & $\delta \mathtt{E}_{i}^{+}$ & $j$ & $n$ \\
\MidRule
0 & -1.561304e+02 & 1.71e-04 & 0 & 0 \\
1 & -1.558409e+02 & 2.06e-03 & 1 & 0 \\
\BotRule
\end{tabular}
\hfill
\begin{tabular}{c|c|c|c|c}
\TopRule
\multicolumn{5}{c}{$L=64\,,\ \mu=5.0\,,\ C=-$} \\
\BotRule
$i$ & $\mathtt{E}_{i}^{-}$ & $\delta \mathtt{E}_{i}^{-}$ & $j$ & $n$ \\
\MidRule
1 & -1.557788e+02 & 1.44e-03 & 1 & 0 \\
2 & -1.556072e+02 & 6.62e-03 & 1 & 1 \\
\BotRule
\end{tabular}

\end{table*}

\begin{table*}[h]
\centering
\renewcommand{\arraystretch}{1.3}
\setlength{\tabcolsep}{4pt}

\begin{tabular}{c|c|c|c|c}
\TopRule
\multicolumn{5}{c}{$L=64\,,\ \mu=6.0\,,\ C=+$} \\
\BotRule
$i$ & $\mathtt{E}_{i}^{+}$ & $\delta \mathtt{E}_{i}^{+}$ & $j$ & $n$ \\
\MidRule
0 & -1.562300e+02 & 1.28e-04 & 0 & 0 \\
1 & -1.558574e+02 & 1.72e-03 & 1 & 0 \\
2 & -1.557701e+02 & 1.47e-03 & 1 & 1 \\
3 & -1.557701e+02 & 1.42e-03 & 1 & 1 \\
\BotRule
\end{tabular}
\hfill
\begin{tabular}{c|c|c|c|c}
\TopRule
\multicolumn{5}{c}{$L=64\,,\ \mu=6.0\,,\ C=-$} \\
\BotRule
$i$ & $\mathtt{E}_{i}^{-}$ & $\delta \mathtt{E}_{i}^{-}$ & $j$ & $n$ \\
\MidRule
1 & -1.557441e+02 & 1.86e-03 & 1 & 0 \\
2 & -1.556231e+02 & 3.97e-03 & 1 & 1 \\
3 & -1.556229e+02 & 3.51e-03 & 1 & 1 \\
4 & -1.556229e+02 & 3.59e-03 & 2 & 0 \\
\BotRule
\end{tabular}
\hfill
\begin{tabular}{c|c|c|c|c}
\TopRule
\multicolumn{5}{c}{$L=64\,,\ \mu=7.0\,,\ C=+$} \\
\BotRule
$i$ & $\mathtt{E}_{i}^{+}$ & $\delta \mathtt{E}_{i}^{+}$ & $j$ & $n$ \\
\MidRule
0 & -1.563477e+02 & 1.22e-04 & 0 & 0 \\
1 & -1.559047e+02 & 1.32e-03 & 1 & 0 \\
\BotRule
\end{tabular}
\\[1em]
\begin{tabular}{c|c|c|c|c}
\TopRule
\multicolumn{5}{c}{$L=64\,,\ \mu=7.0\,,\ C=-$} \\
\BotRule
$i$ & $\mathtt{E}_{i}^{-}$ & $\delta \mathtt{E}_{i}^{-}$ & $j$ & $n$ \\
\MidRule
1 & -1.557266e+02 & 2.52e-03 & 1 & 0 \\
2 & -1.556557e+02 & 3.04e-03 & 1 & 1 \\
\BotRule
\end{tabular}
\hfill
\begin{tabular}{c|c|c|c|c}
\TopRule
\multicolumn{5}{c}{$L=64\,,\ \mu=8.0\,,\ C=+$} \\
\BotRule
$i$ & $\mathtt{E}_{i}^{+}$ & $\delta \mathtt{E}_{i}^{+}$ & $j$ & $n$ \\
\MidRule
0 & -1.564834e+02 & 1.34e-04 & 0 & 0 \\
1 & -1.559748e+02 & 9.94e-04 & 1 & 0 \\
2 & -1.559138e+02 & 9.34e-04 & 1 & 1 \\
3 & -1.559138e+02 & 8.39e-04 & 1 & 1 \\
\BotRule
\end{tabular}
\hfill
\begin{tabular}{c|c|c|c|c}
\TopRule
\multicolumn{5}{c}{$L=64\,,\ \mu=8.0\,,\ C=-$} \\
\BotRule
$i$ & $\mathtt{E}_{i}^{-}$ & $\delta \mathtt{E}_{i}^{-}$ & $j$ & $n$ \\
\MidRule
1 & -1.557532e+02 & 2.29e-03 & 1 & 0 \\
2 & -1.557055e+02 & 2.63e-03 & 1 & 1 \\
3 & -1.557055e+02 & 2.07e-03 & 1 & 1 \\
4 & -1.555889e+02 & 3.58e-03 & 1 & 2 \\
\BotRule
\end{tabular}
\caption{Low energy spectrum of the quantum clock model at various values of $\mu = h^{15/28} L$ in the two charge-conjugation sectors at $L=64$.}
\end{table*}

\begin{table*}[h]
\centering
\renewcommand{\arraystretch}{1.3}
\setlength{\tabcolsep}{4pt}

\begin{tabular}{c|c|c|c|c}
\TopRule
\multicolumn{5}{c}{$L=64\,,\ \mu=10.0\,,\ C=+$} \\
\BotRule
$i$ & $\mathtt{E}_{i}^{+}$ & $\delta \mathtt{E}_{i}^{+}$ & $j$ & $n$ \\
\MidRule
0 & -1.568088e+02 & 2.39e-04 & 0 & 0 \\
1 & -1.561719e+02 & 5.89e-04 & 1 & 0 \\
2 & -1.561228e+02 & 5.87e-04 & 1 & 1 \\
3 & -1.561228e+02 & 4.57e-04 & 1 & 1 \\
\BotRule
\end{tabular}
\hfill
\begin{tabular}{c|c|c|c|c}
\TopRule
\multicolumn{5}{c}{$L=64\,,\ \mu=10.0\,,\ C=-$} \\
\BotRule
$i$ & $\mathtt{E}_{i}^{-}$ & $\delta \mathtt{E}_{i}^{-}$ & $j$ & $n$ \\
\MidRule
1 & -1.558925e+02 & 2.18e-03 & 1 & 0 \\
2 & -1.558575e+02 & 1.85e-03 & 1 & 1 \\
3 & -1.558575e+02 & 1.68e-03 & 1 & 1 \\
4 & -1.557604e+02 & 2.18e-03 & 1 & 2 \\
\BotRule
\end{tabular}
\hfill
\begin{tabular}{c|c|c|c|c}
\TopRule
\multicolumn{5}{c}{$L=64\,,\ \mu=12.0\,,\ C=+$} \\
\BotRule
$i$ & $\mathtt{E}_{i}^{+}$ & $\delta \mathtt{E}_{i}^{+}$ & $j$ & $n$ \\
\MidRule
0 & -1.572059e+02 & 1.25e-04 & 0 & 0 \\
1 & -1.564413e+02 & 3.67e-04 & 1 & 0 \\
2 & -1.564000e+02 & 3.69e-04 & 1 & 1 \\
3 & -1.564000e+02 & 3.01e-04 & 1 & 1 \\
\BotRule
\end{tabular}
\\[1em]
\begin{tabular}{c|c|c|c|c}
\TopRule
\multicolumn{5}{c}{$L=64\,,\ \mu=12.0\,,\ C=-$} \\
\BotRule
$i$ & $\mathtt{E}_{i}^{-}$ & $\delta \mathtt{E}_{i}^{-}$ & $j$ & $n$ \\
\MidRule
1 & -1.561090e+02 & 1.06e-03 & 1 & 0 \\
2 & -1.560799e+02 & 9.92e-04 & 1 & 1 \\
3 & -1.560799e+02 & 9.29e-04 & 1 & 1 \\
4 & -1.559969e+02 & 1.29e-03 & 1 & 2 \\
\BotRule
\end{tabular}
\hfill
\begin{tabular}{c|c|c|c|c}
\TopRule
\multicolumn{5}{c}{$L=64\,,\ \mu=14.0\,,\ C=+$} \\
\BotRule
$i$ & $\mathtt{E}_{i}^{+}$ & $\delta \mathtt{E}_{i}^{+}$ & $j$ & $n$ \\
\MidRule
0 & -1.576747e+02 & 1.24e-04 & 0 & 0 \\
1 & -1.567820e+02 & 5.15e-03 & 1 & 0 \\
2 & -1.567467e+02 & 4.87e-03 & 1 & 1 \\
3 & -1.567468e+02 & 7.58e-03 & 1 & 1 \\
4 & -1.566478e+02 & 1.41e-03 & 1 & 2 \\
5 & -1.566478e+02 & 9.51e-03 & 1 & 2 \\
\BotRule
\end{tabular}
\hfill
\begin{tabular}{c|c|c|c|c}
\TopRule
\multicolumn{5}{c}{$L=64\,,\ \mu=14.0\,,\ C=-$} \\
\BotRule
$i$ & $\mathtt{E}_{i}^{-}$ & $\delta \mathtt{E}_{i}^{-}$ & $j$ & $n$ \\
\MidRule
1 & -1.563982e+02 & 7.24e-04 & 1 & 0 \\
2 & -1.563732e+02 & 6.01e-04 & 1 & 1 \\
3 & -1.563732e+02 & 6.14e-04 & 1 & 1 \\
4 & -1.563009e+02 & 8.28e-04 & 1 & 2 \\
5 & -1.563009e+02 & 8.15e-04 & 1 & 2 \\
6 & -1.561885e+02 & 1.14e-03 & 1 & 3 \\
\BotRule
\end{tabular}
\caption{Low energy spectrum of the quantum clock model at various values of $\mu = h^{15/28} L$ in the two charge-conjugation sectors at $L=64$.}
\end{table*}

\clearpage

\begin{table*}[h]
\centering
\renewcommand{\arraystretch}{1.3}
\setlength{\tabcolsep}{4pt}

\begin{tabular}{c|c|c|c|c}
\TopRule
\multicolumn{5}{c}{$L=80\,,\ \mu=0.0\,,\ C=+$} \\
\BotRule
$i$ & $\mathtt{E}_{i}^{+}$ & $\delta \mathtt{E}_{i}^{+}$ & $j$ & $n$ \\
\MidRule
0 & -1.948929e+02 & 2.99e-03 & 0 & 0 \\
1 & -1.948656e+02 & 2.50e-03 & 1 & 0 \\
2 & -1.946615e+02 & 4.71e-03 & 1 & 1 \\
3 & -1.947282e+02 & 6.46e-03 & 2 & 0 \\
\BotRule
\end{tabular}
\hfill
\begin{tabular}{c|c|c|c|c}
\TopRule
\multicolumn{5}{c}{$L=80\,,\ \mu=0.0\,,\ C=-$} \\
\BotRule
$i$ & $\mathtt{E}_{i}^{-}$ & $\delta \mathtt{E}_{i}^{-}$ & $j$ & $n$ \\
\MidRule
1 & -1.948656e+02 & 1.41e-03 & 1 & 0 \\
2 & -1.946615e+02 & 5.52e-03 & 1 & 1 \\
3 & -1.946615e+02 & 4.46e-03 & 1 & 1 \\
4 & -1.946209e+02 & 7.85e-03 & 2 & 0 \\
\BotRule
\end{tabular}
\hfill
\begin{tabular}{c|c|c|c|c}
\TopRule
\multicolumn{5}{c}{$L=80\,,\ \mu=0.5\,,\ C=+$} \\
\BotRule
$i$ & $\mathtt{E}_{i}^{+}$ & $\delta \mathtt{E}_{i}^{+}$ & $j$ & $n$ \\
\MidRule
0 & -1.948930e+02 & 2.98e-03 & 0 & 0 \\
1 & -1.948668e+02 & 2.58e-03 & 1 & 0 \\
2 & -1.946621e+02 & 5.50e-03 & 1 & 1 \\
3 & -1.947282e+02 & 6.47e-03 & 2 & 0 \\
\BotRule
\end{tabular}
\\[1em]
\begin{tabular}{c|c|c|c|c}
\TopRule
\multicolumn{5}{c}{$L=80\,,\ \mu=0.5\,,\ C=-$} \\
\BotRule
$i$ & $\mathtt{E}_{i}^{-}$ & $\delta \mathtt{E}_{i}^{-}$ & $j$ & $n$ \\
\MidRule
1 & -1.948644e+02 & 1.42e-03 & 1 & 0 \\
2 & -1.946608e+02 & 5.36e-03 & 1 & 1 \\
3 & -1.946608e+02 & 4.81e-03 & 1 & 1 \\
4 & -1.946210e+02 & 7.85e-03 & 2 & 0 \\
\BotRule
\end{tabular}
\hfill
\begin{tabular}{c|c|c|c|c}
\TopRule
\multicolumn{5}{c}{$L=80\,,\ \mu=0.75\,,\ C=+$} \\
\BotRule
$i$ & $\mathtt{E}_{i}^{+}$ & $\delta \mathtt{E}_{i}^{+}$ & $j$ & $n$ \\
\MidRule
0 & -1.948934e+02 & 2.88e-03 & 0 & 0 \\
1 & -1.948679e+02 & 2.73e-03 & 1 & 0 \\
\BotRule
\end{tabular}
\hfill
\begin{tabular}{c|c|c|c|c}
\TopRule
\multicolumn{5}{c}{$L=80\,,\ \mu=0.75\,,\ C=-$} \\
\BotRule
$i$ & $\mathtt{E}_{i}^{-}$ & $\delta \mathtt{E}_{i}^{-}$ & $j$ & $n$ \\
\MidRule
1 & -1.948630e+02 & 1.80e-03 & 1 & 0 \\
2 & -1.946601e+02 & 5.56e-03 & 1 & 1 \\
\BotRule
\end{tabular}

\end{table*}

\begin{table*}[h]
\centering
\renewcommand{\arraystretch}{1.3}
\setlength{\tabcolsep}{4pt}

\begin{tabular}{c|c|c|c|c}
\TopRule
\multicolumn{5}{c}{$L=80\,,\ \mu=1.0\,,\ C=+$} \\
\BotRule
$i$ & $\mathtt{E}_{i}^{+}$ & $\delta \mathtt{E}_{i}^{+}$ & $j$ & $n$ \\
\MidRule
0 & -1.948943e+02 & 2.78e-03 & 0 & 0 \\
1 & -1.948688e+02 & 2.76e-03 & 1 & 0 \\
2 & -1.946640e+02 & 5.49e-03 & 1 & 1 \\
3 & -1.946640e+02 & 4.53e-03 & 1 & 1 \\
\BotRule
\end{tabular}
\hfill
\begin{tabular}{c|c|c|c|c}
\TopRule
\multicolumn{5}{c}{$L=80\,,\ \mu=1.0\,,\ C=-$} \\
\BotRule
$i$ & $\mathtt{E}_{i}^{-}$ & $\delta \mathtt{E}_{i}^{-}$ & $j$ & $n$ \\
\MidRule
1 & -1.948611e+02 & 1.44e-03 & 1 & 0 \\
2 & -1.946591e+02 & 5.53e-03 & 1 & 1 \\
3 & -1.946591e+02 & 4.77e-03 & 1 & 1 \\
4 & -1.946211e+02 & 7.85e-03 & 2 & 0 \\
\BotRule
\end{tabular}
\hfill
\begin{tabular}{c|c|c|c|c}
\TopRule
\multicolumn{5}{c}{$L=80\,,\ \mu=2.0\,,\ C=+$} \\
\BotRule
$i$ & $\mathtt{E}_{i}^{+}$ & $\delta \mathtt{E}_{i}^{+}$ & $j$ & $n$ \\
\MidRule
0 & -1.949101e+02 & 1.69e-03 & 0 & 0 \\
1 & -1.948661e+02 & 2.70e-03 & 1 & 0 \\
2 & -1.946712e+02 & 5.18e-03 & 1 & 1 \\
3 & -1.946712e+02 & 4.34e-03 & 1 & 1 \\
\BotRule
\end{tabular}
\\[1em]
\begin{tabular}{c|c|c|c|c}
\TopRule
\multicolumn{5}{c}{$L=80\,,\ \mu=2.0\,,\ C=-$} \\
\BotRule
$i$ & $\mathtt{E}_{i}^{-}$ & $\delta \mathtt{E}_{i}^{-}$ & $j$ & $n$ \\
\MidRule
1 & -1.948492e+02 & 1.53e-03 & 1 & 0 \\
2 & -1.946534e+02 & 4.74e-03 & 1 & 1 \\
3 & -1.946534e+02 & 5.94e-03 & 1 & 1 \\
4 & -1.946220e+02 & 7.84e-03 & 2 & 0 \\
\BotRule
\end{tabular}
\hfill
\begin{tabular}{c|c|c|c|c}
\TopRule
\multicolumn{5}{c}{$L=80\,,\ \mu=3.0\,,\ C=+$} \\
\BotRule
$i$ & $\mathtt{E}_{i}^{+}$ & $\delta \mathtt{E}_{i}^{+}$ & $j$ & $n$ \\
\MidRule
0 & -1.949451e+02 & 9.21e-04 & 0 & 0 \\
1 & -1.948537e+02 & 2.88e-03 & 1 & 0 \\
2 & -1.946840e+02 & 4.63e-03 & 1 & 1 \\
3 & -1.946840e+02 & 4.13e-03 & 1 & 1 \\
\BotRule
\end{tabular}
\hfill
\begin{tabular}{c|c|c|c|c}
\TopRule
\multicolumn{5}{c}{$L=80\,,\ \mu=3.0\,,\ C=-$} \\
\BotRule
$i$ & $\mathtt{E}_{i}^{-}$ & $\delta \mathtt{E}_{i}^{-}$ & $j$ & $n$ \\
\MidRule
1 & -1.948309e+02 & 1.70e-03 & 1 & 0 \\
2 & -1.946463e+02 & 6.36e-03 & 1 & 1 \\
3 & -1.946463e+02 & 5.11e-03 & 1 & 1 \\
4 & -1.946248e+02 & 7.82e-03 & 2 & 0 \\
\BotRule
\end{tabular}
\caption{Low energy spectrum of the quantum clock model at various values of $\mu = h^{15/28} L$ in the two charge-conjugation sectors at $L=80$.}
\end{table*}

\begin{table*}[h]
\centering
\renewcommand{\arraystretch}{1.3}
\setlength{\tabcolsep}{4pt}

\begin{tabular}{c|c|c|c|c}
\TopRule
\multicolumn{5}{c}{$L=80\,,\ \mu=4.0\,,\ C=+$} \\
\BotRule
$i$ & $\mathtt{E}_{i}^{+}$ & $\delta \mathtt{E}_{i}^{+}$ & $j$ & $n$ \\
\MidRule
0 & -1.949957e+02 & 5.27e-04 & 0 & 0 \\
1 & -1.948383e+02 & 3.26e-03 & 1 & 0 \\
2 & -1.947042e+02 & 4.23e-03 & 1 & 1 \\
3 & -1.947042e+02 & 3.28e-03 & 1 & 1 \\
\BotRule
\end{tabular}
\hfill
\begin{tabular}{c|c|c|c|c}
\TopRule
\multicolumn{5}{c}{$L=80\,,\ \mu=4.0\,,\ C=-$} \\
\BotRule
$i$ & $\mathtt{E}_{i}^{-}$ & $\delta \mathtt{E}_{i}^{-}$ & $j$ & $n$ \\
\MidRule
1 & -1.948072e+02 & 1.95e-03 & 1 & 0 \\
2 & -1.946410e+02 & 6.46e-03 & 1 & 1 \\
3 & -1.946410e+02 & 5.89e-03 & 1 & 1 \\
4 & -1.946306e+02 & 7.75e-03 & 2 & 0 \\
\BotRule
\end{tabular}
\hfill
\begin{tabular}{c|c|c|c|c}
\TopRule
\multicolumn{5}{c}{$L=80\,,\ \mu=5.0\,,\ C=+$} \\
\BotRule
$i$ & $\mathtt{E}_{i}^{+}$ & $\delta \mathtt{E}_{i}^{+}$ & $j$ & $n$ \\
\MidRule
0 & -1.950609e+02 & 3.17e-04 & 0 & 0 \\
1 & -1.948294e+02 & 3.72e-03 & 1 & 0 \\
\BotRule
\end{tabular}
\\[1em]
\begin{tabular}{c|c|c|c|c}
\TopRule
\multicolumn{5}{c}{$L=80\,,\ \mu=5.0\,,\ C=-$} \\
\BotRule
$i$ & $\mathtt{E}_{i}^{-}$ & $\delta \mathtt{E}_{i}^{-}$ & $j$ & $n$ \\
\MidRule
1 & -1.947795e+02 & 2.62e-03 & 1 & 0 \\
2 & -1.946420e+02 & 7.13e-03 & 1 & 1 \\
\BotRule
\end{tabular}
\hfill
\begin{tabular}{c|c|c|c|c}
\TopRule
\multicolumn{5}{c}{$L=80\,,\ \mu=6.0\,,\ C=+$} \\
\BotRule
$i$ & $\mathtt{E}_{i}^{+}$ & $\delta \mathtt{E}_{i}^{+}$ & $j$ & $n$ \\
\MidRule
0 & -1.951407e+02 & 2.08e-04 & 0 & 0 \\
1 & -1.948427e+02 & 3.23e-03 & 1 & 0 \\
2 & -1.947729e+02 & 2.43e-03 & 1 & 1 \\
3 & -1.947729e+02 & 3.06e-03 & 1 & 1 \\
\BotRule
\end{tabular}
\hfill
\begin{tabular}{c|c|c|c|c}
\TopRule
\multicolumn{5}{c}{$L=80\,,\ \mu=6.0\,,\ C=-$} \\
\BotRule
$i$ & $\mathtt{E}_{i}^{-}$ & $\delta \mathtt{E}_{i}^{-}$ & $j$ & $n$ \\
\MidRule
1 & -1.947516e+02 & 3.31e-03 & 1 & 0 \\
2 & -1.946544e+02 & 7.17e-03 & 1 & 1 \\
3 & -1.946544e+02 & 6.60e-03 & 1 & 1 \\
4 & -1.946544e+02 & 5.41e-03 & 2 & 0 \\
\BotRule
\end{tabular}%
\caption{Low energy spectrum of the quantum clock model at various values of $\mu = h^{15/28} L$ in the two charge-conjugation sectors at $L=80$.}
\end{table*}

\clearpage

\begin{table*}[h]
\centering
\renewcommand{\arraystretch}{1.3}
\setlength{\tabcolsep}{4pt}
%
\caption{Low energy spectrum of the quantum clock model at various values of $\mu = h^{15/28} L$ in the two charge-conjugation sectors at $L=96$.}
\end{table*}

\clearpage
\subsection{String tension data tables}

\begin{table*}[h!]
\centering
\renewcommand{\arraystretch}{1.4}
\setlength{\tabcolsep}{4pt}

%
\caption{Static quark-antiquark potential as a function of $\txtw$ at various $L$ and $\mu$.}
\end{table*}

\begin{table*}[h]
\centering
\renewcommand{\arraystretch}{1.4}
\setlength{\tabcolsep}{4pt}
%
\caption{Static quark-antiquark potential as a function of $\txtw$ at various $L$ and $\mu$.}
\end{table*}

\clearpage

\subsection{Plots that determine \texorpdfstring{$R(\mu)$}{R(\textmu)} from the DMRG data}

\begin{figure}[h!]
    \centering
    \subfigure[]{\includegraphics[width=0.45\textwidth]{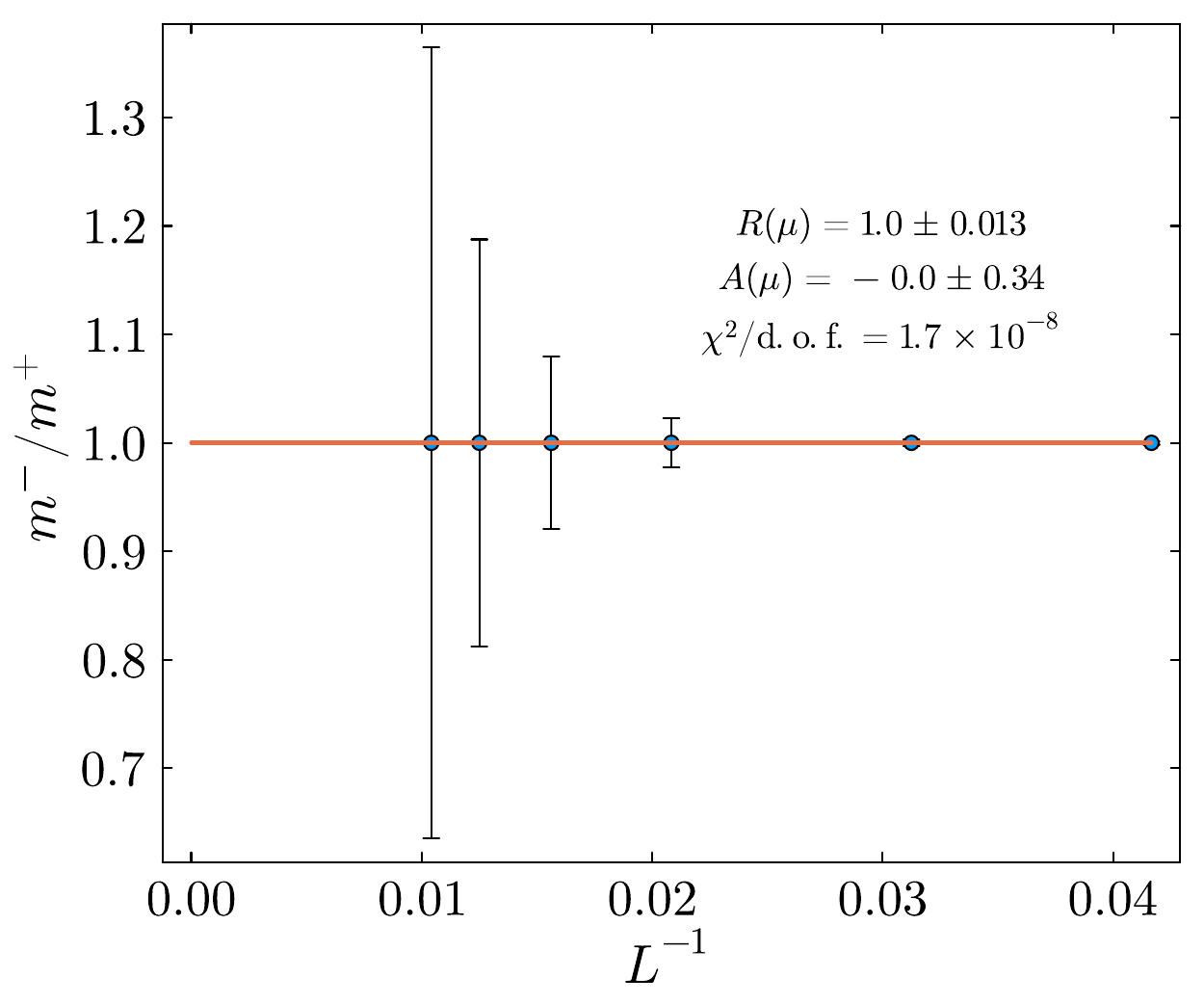}}
    \subfigure[]{\includegraphics[width=0.45\textwidth]{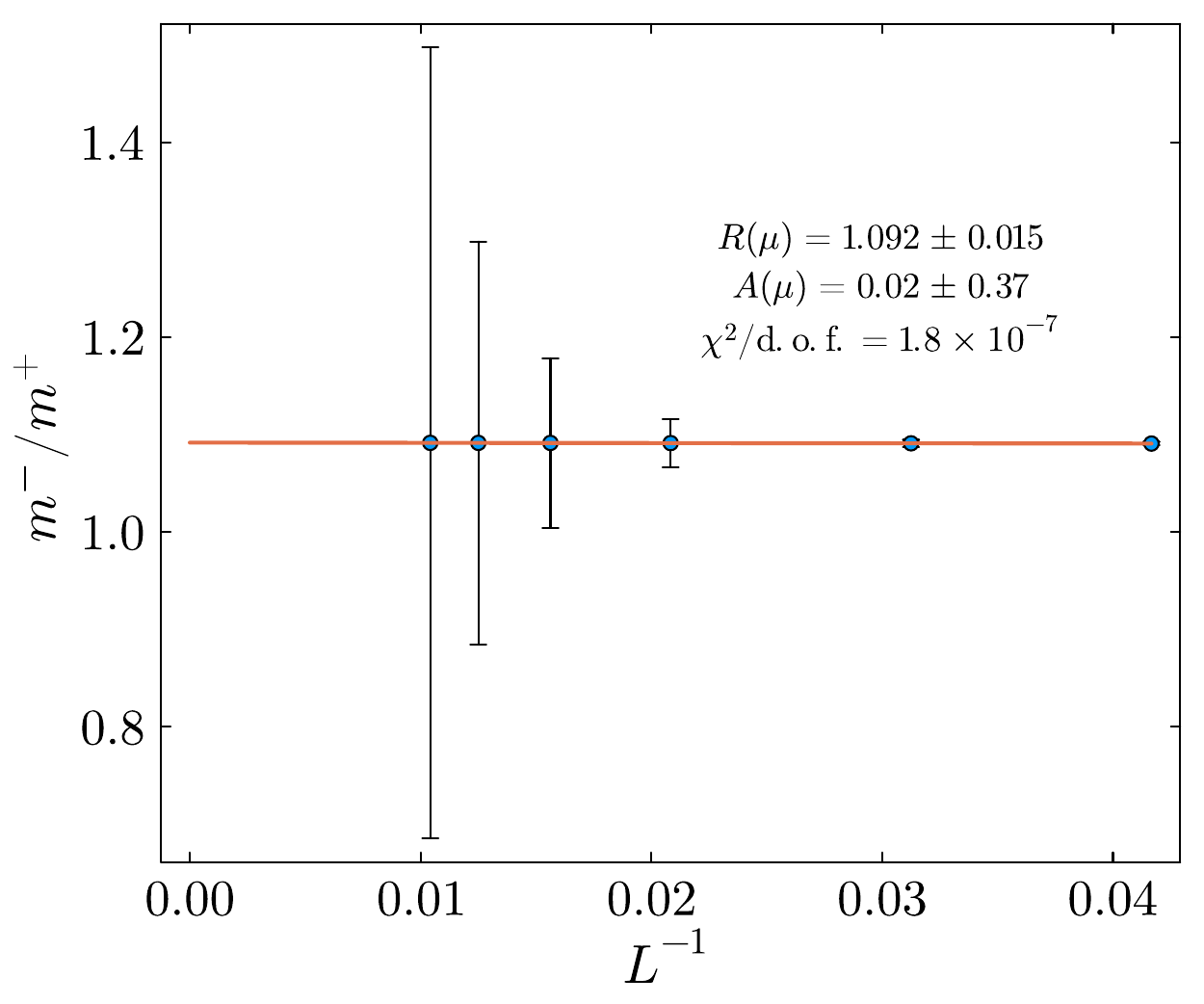}}
    \subfigure[]{\includegraphics[width=0.45\textwidth]{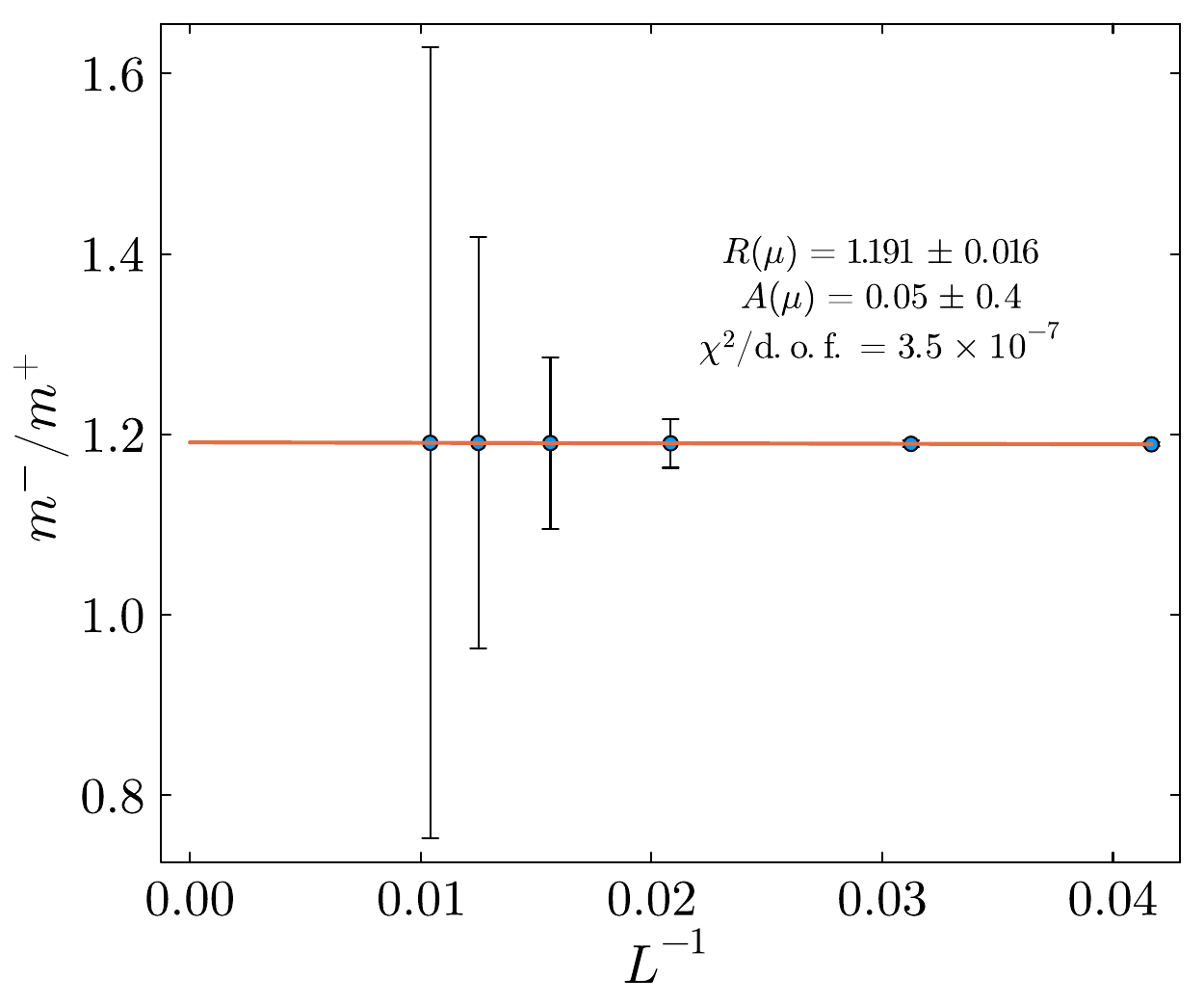}}
    \subfigure[]{\includegraphics[width=0.45\textwidth]{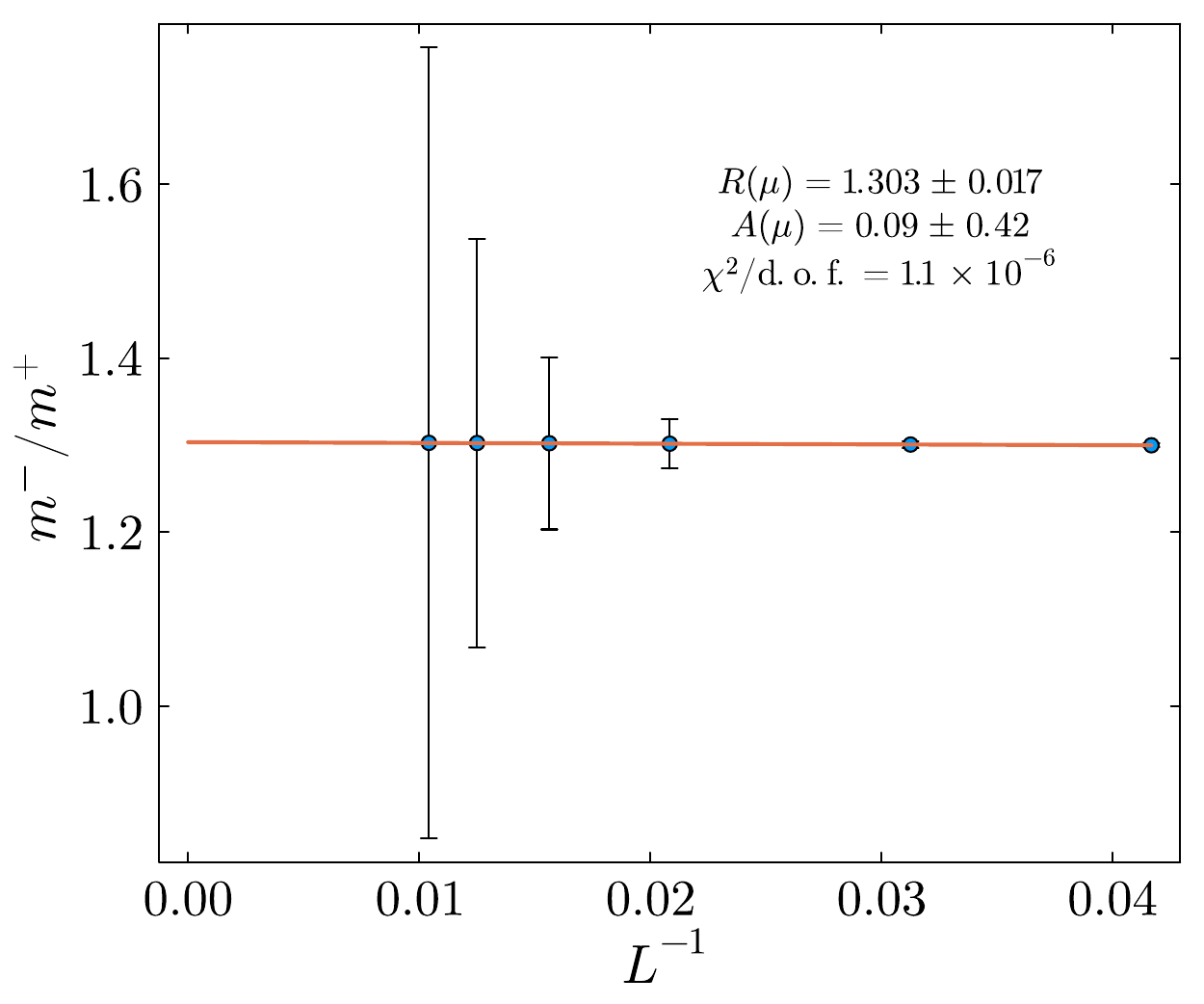}}
    \caption{Mass ratio scaling at (a) $\mu=0.0$, (b) $\mu=0.5$, (c) $\mu=0.75$, (d) $\mu=1.0$. The masses of the lightest particles in the charge-conjugation-even and -odd sectors are given by the expression $m^{\pm}= \mathsf{E}^{\pm}_{1}-\mathsf{E}^+_0$, using the values given in \cref{subsec:lowspectab}.}
    \label{fig:massratioscaling_1}
\end{figure}

\begin{figure}[h!]
    \centering
    \subfigure[]{\includegraphics[width=0.45\textwidth]{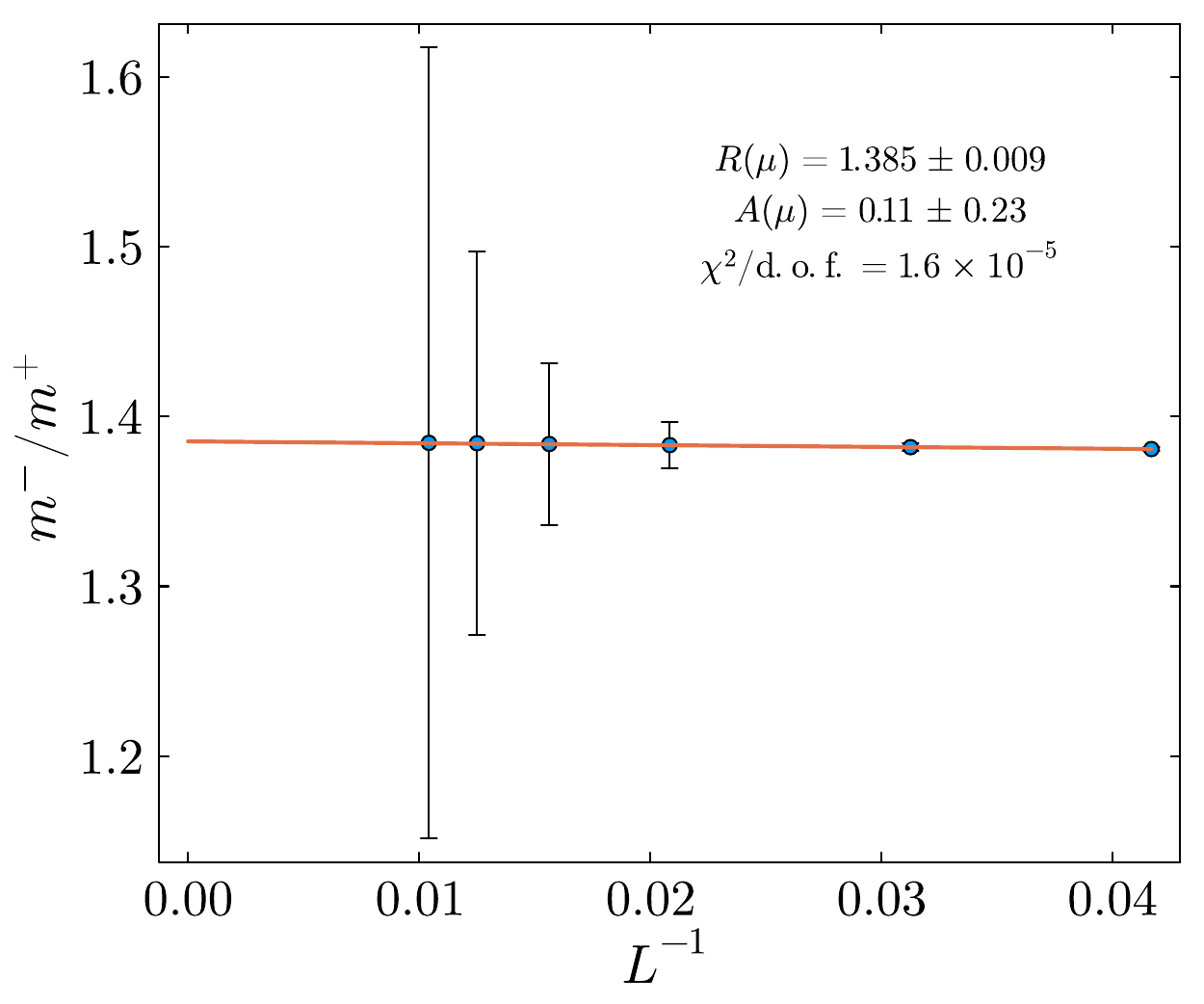}}
    \subfigure[]{\includegraphics[width=0.45\textwidth]{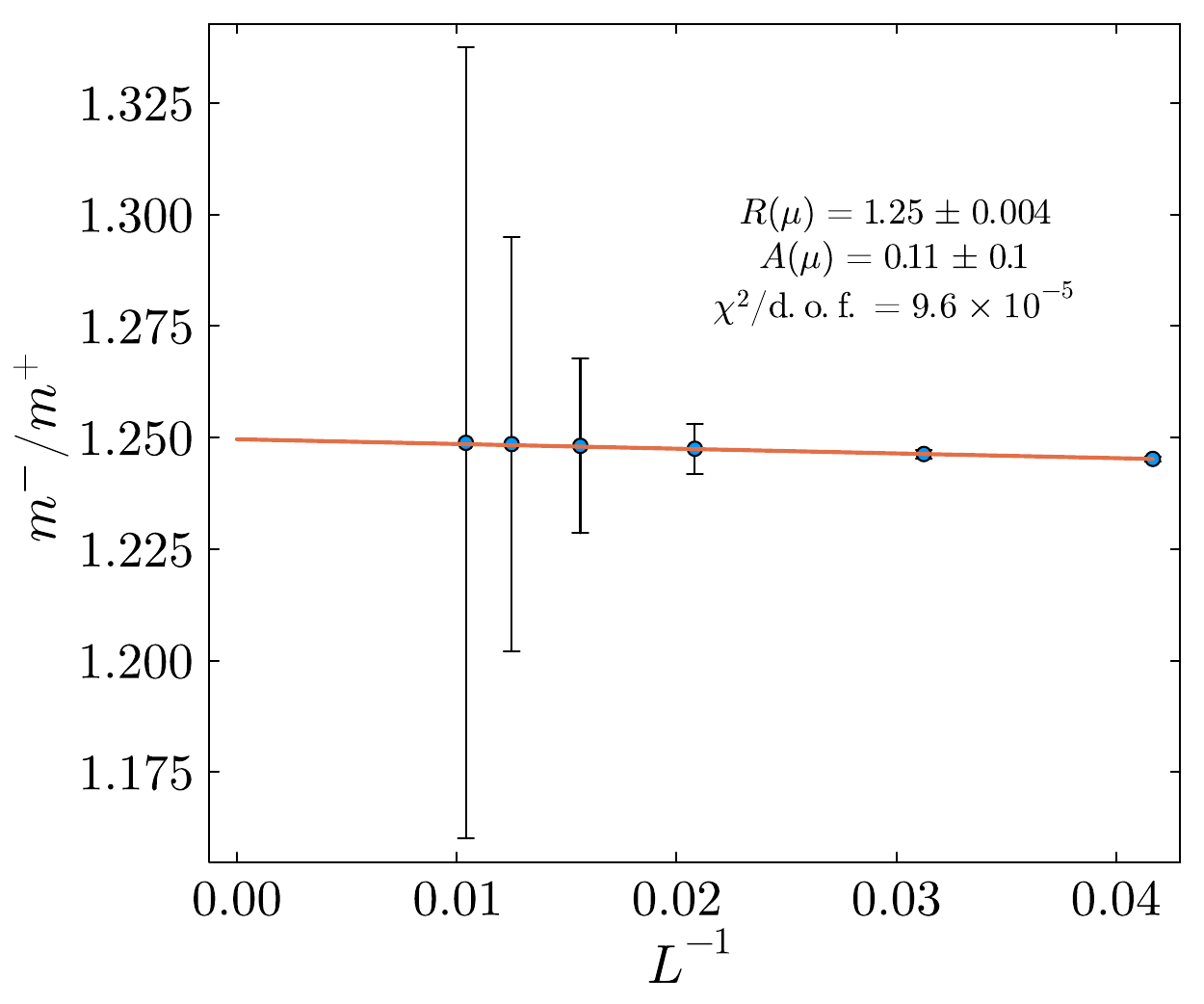}}
    \subfigure[]{\includegraphics[width=0.45\textwidth]{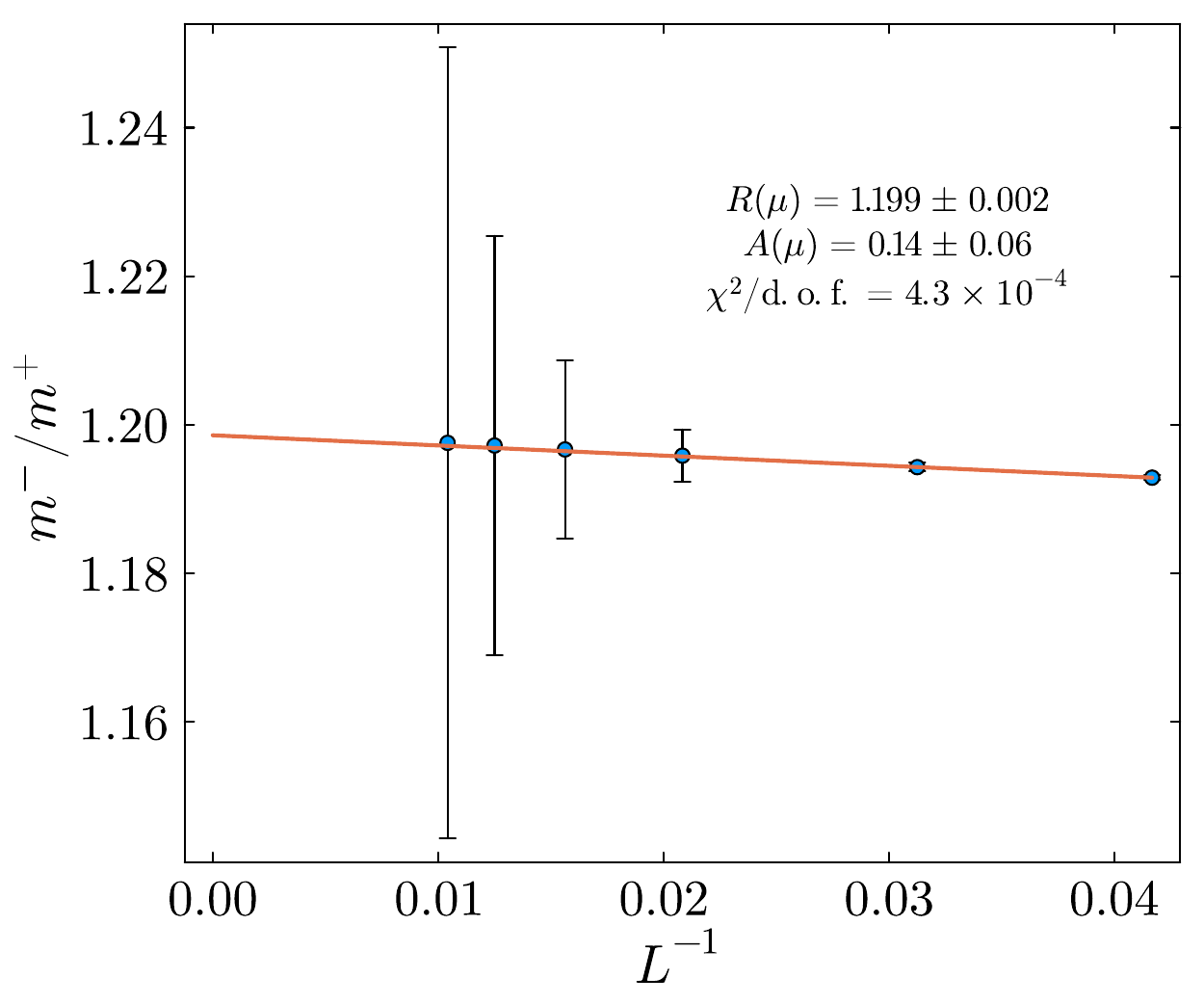}}
    \subfigure[]{\includegraphics[width=0.45\textwidth]{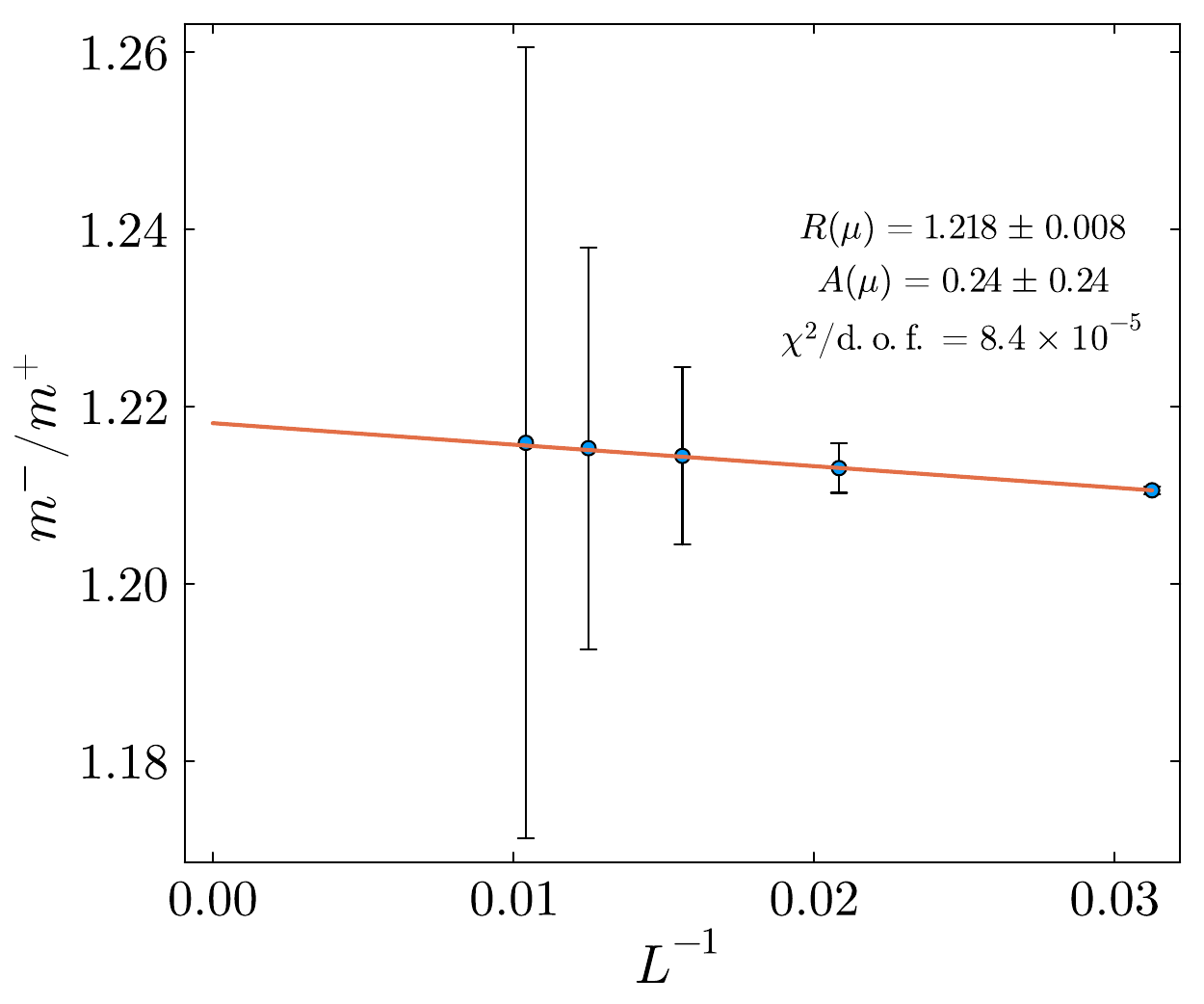}}
    \caption{Mass ratio scaling at (a) $\mu=2.0$, (b) $\mu=3.0$, (c) $\mu=4.0$, (d) $\mu=5.0$. The masses of the lightest particles in the charge-conjugation-even and -odd sectors are given by the expression $m^{\pm}= \mathsf{E}^{\pm}_{1}-\mathsf{E}^+_0$, using the values given in \cref{subsec:lowspectab}.}
    \label{fig:massratioscaling_2}
\end{figure}

\begin{figure}[h!]
    \centering
    \subfigure[]{\includegraphics[width=0.44\textwidth]{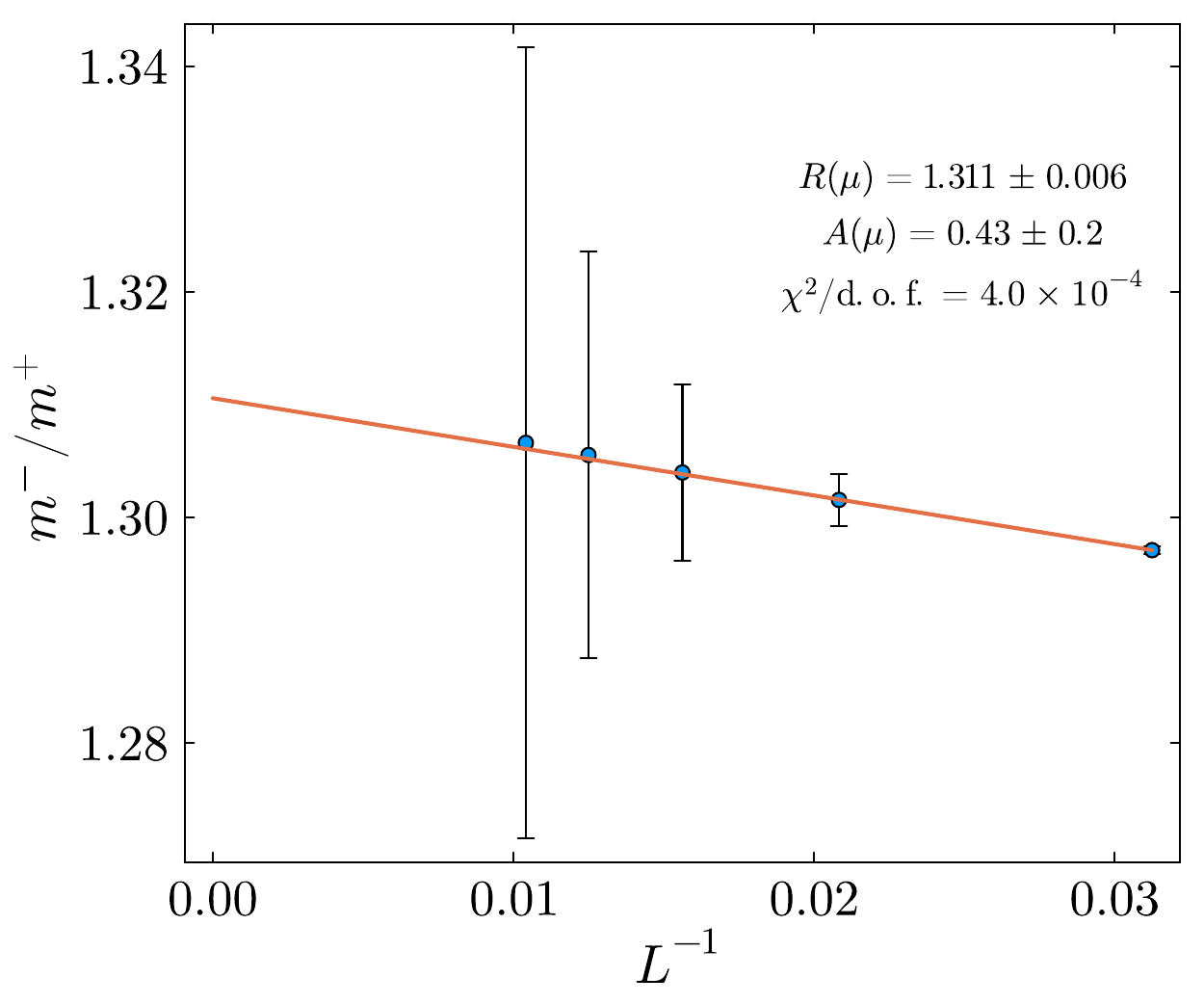}}
    \subfigure[]{\includegraphics[width=0.44\textwidth]{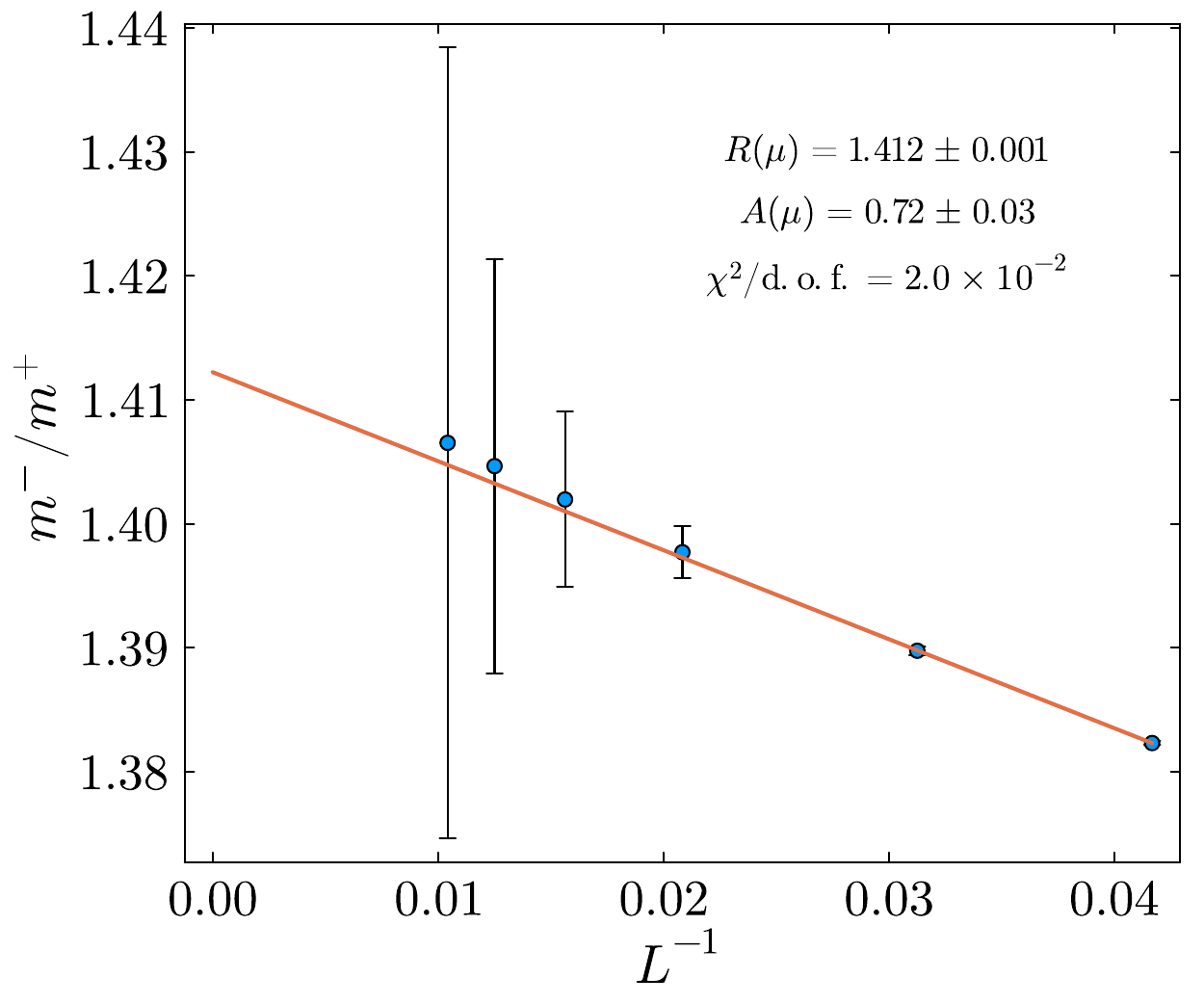}}
    \subfigure[]{\includegraphics[width=0.44\textwidth]{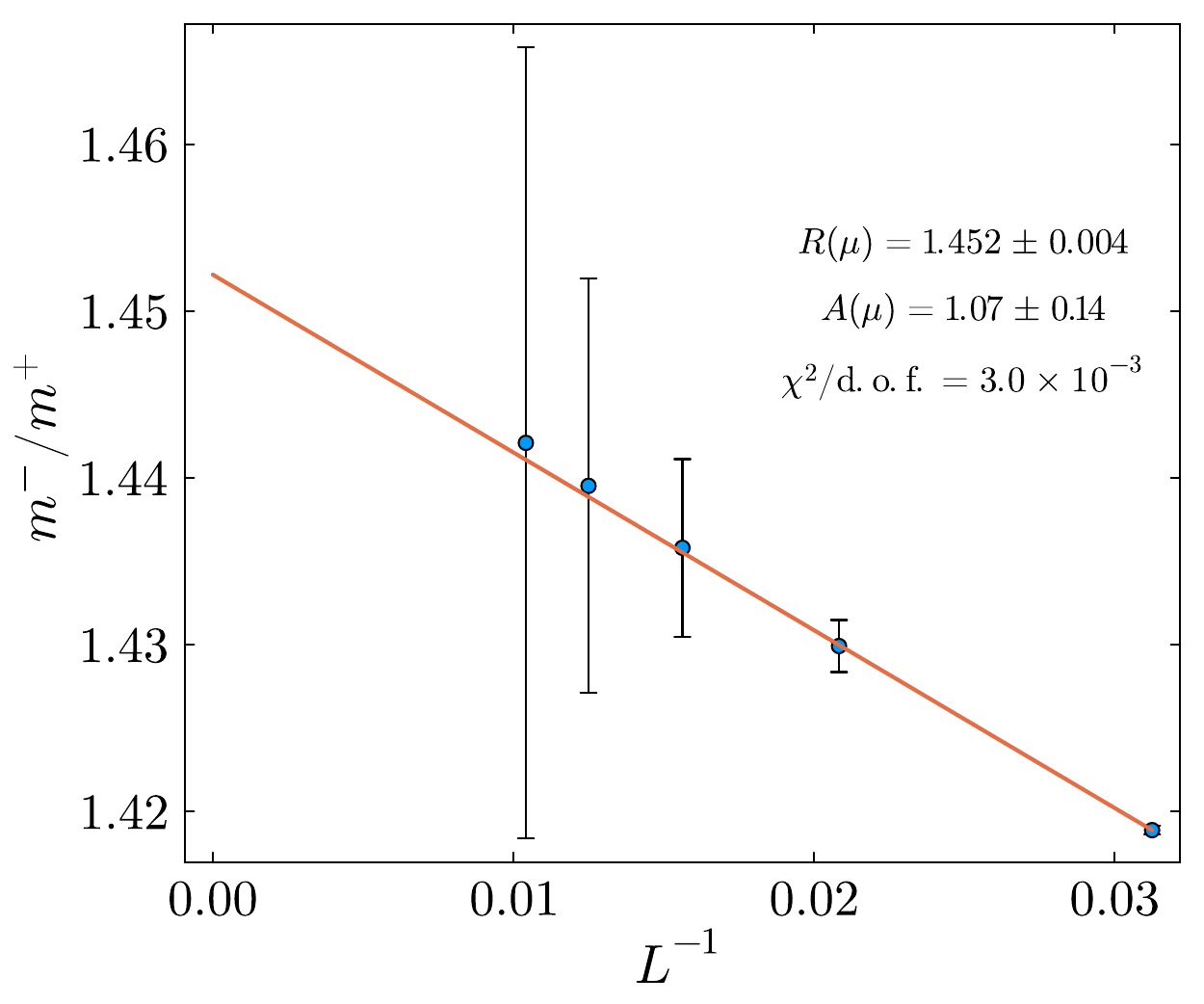}}
    \subfigure[]{\includegraphics[width=0.44\textwidth]{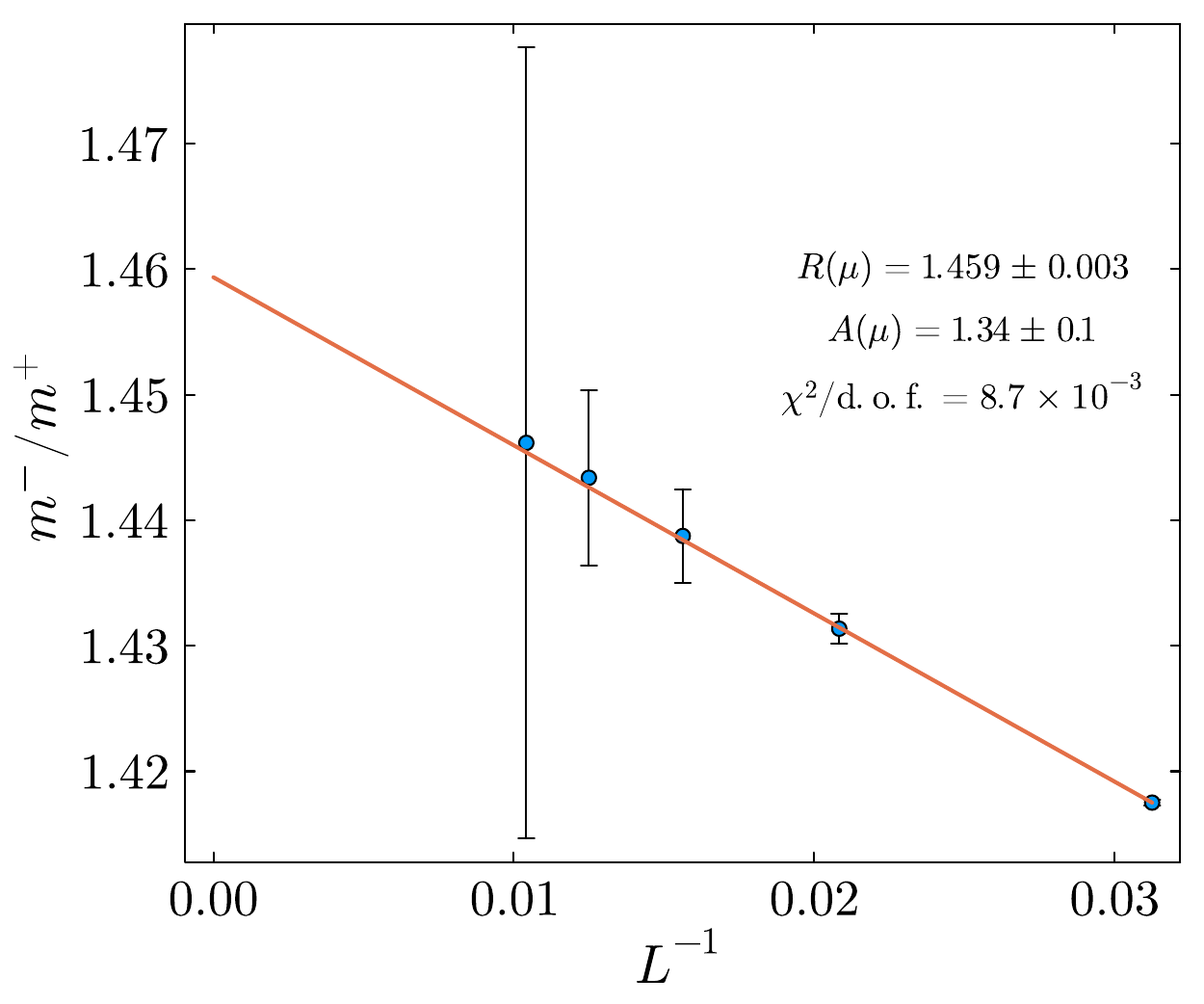}}
    \subfigure[]{\includegraphics[width=0.44\textwidth]{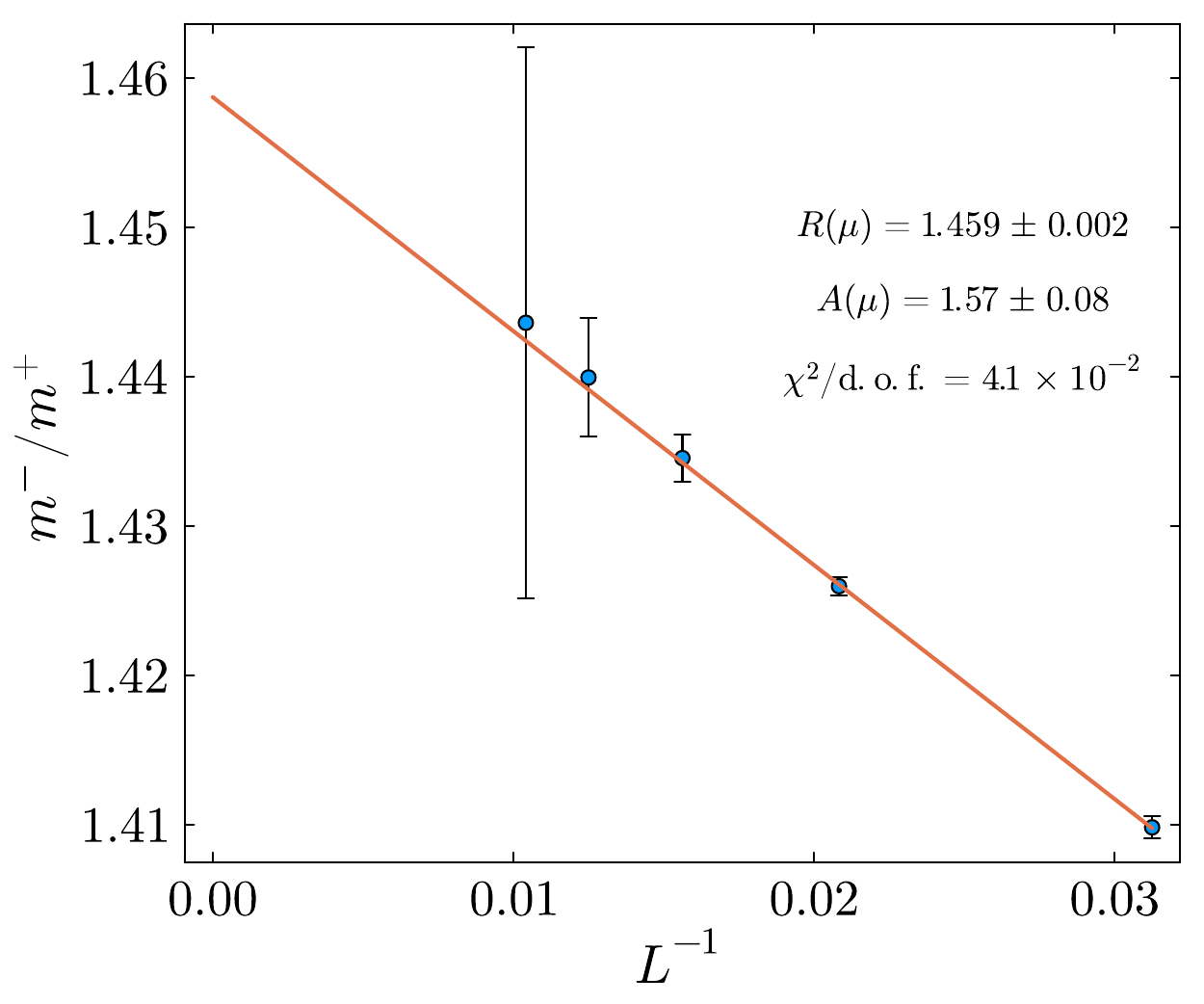}}
    \subfigure[]{\includegraphics[width=0.44\textwidth]{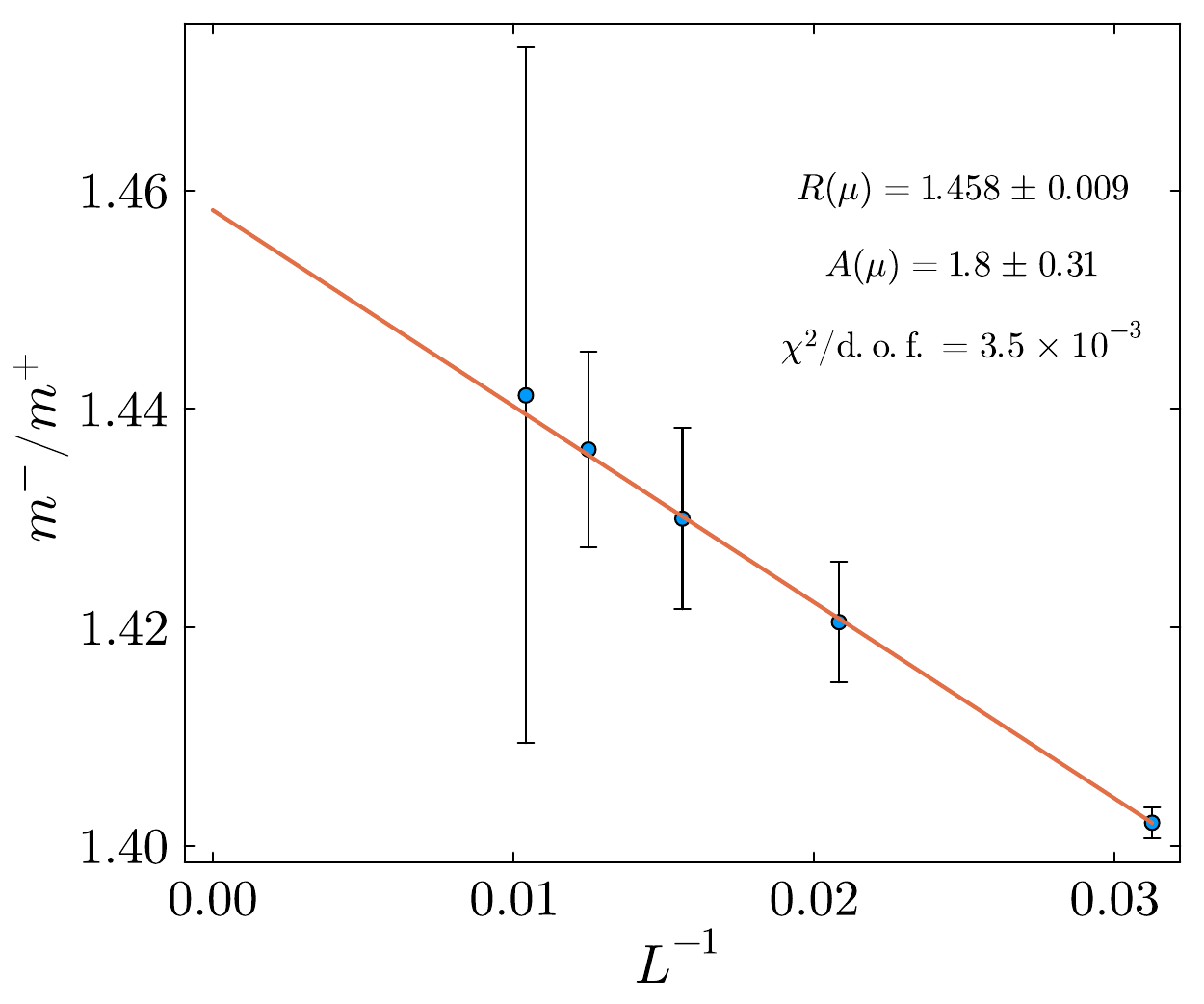}}
    \caption{Mass ratio scaling at (a) $\mu=6.0$, (b) $\mu=7.0$, (c) $\mu=8.0$, (d) $\mu=10.0$, (e) $
    \mu=12.0$, (f) $\mu=14.0$. The masses of the lightest particles in the charge-conjugation-even and -odd sectors are given by the expression $m^{\pm}= \mathsf{E}^{\pm}_{1}-\mathsf{E}^+_0$, using the values given in \cref{subsec:lowspectab}.}
    \label{fig:massratioscaling_3}
\end{figure}

\clearpage

\subsection{Plots that determine \texorpdfstring{$\bar{\sigma}$}{sigma-bar} from the DMRG data}\label{subsec:tension}

\begin{figure}[h!]
    \centering
    \subfigure[]{\includegraphics[width=0.45\textwidth]{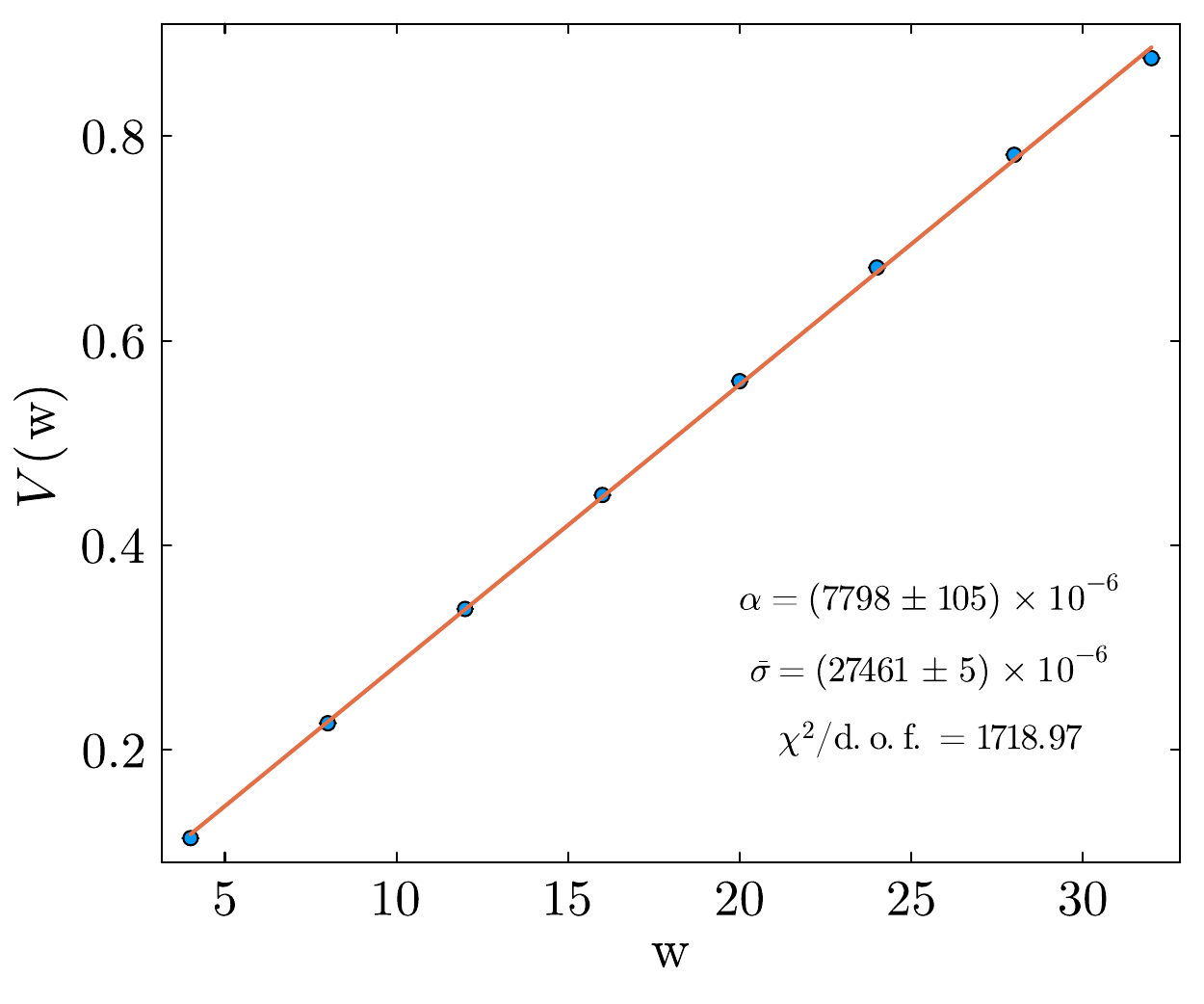}}
    \subfigure[]{\includegraphics[width=0.45\textwidth]{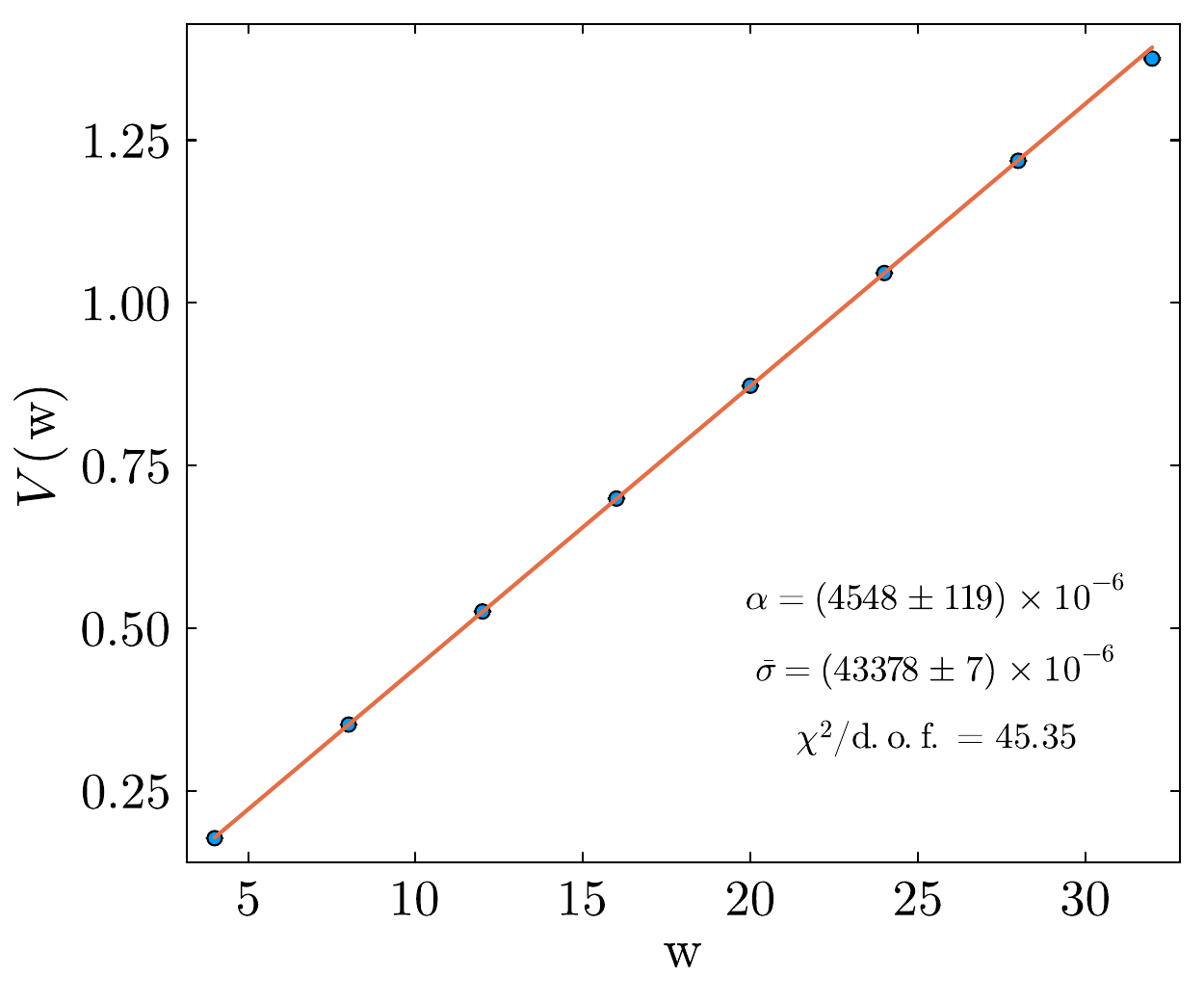}}
    \subfigure[]{\includegraphics[width=0.45\textwidth]{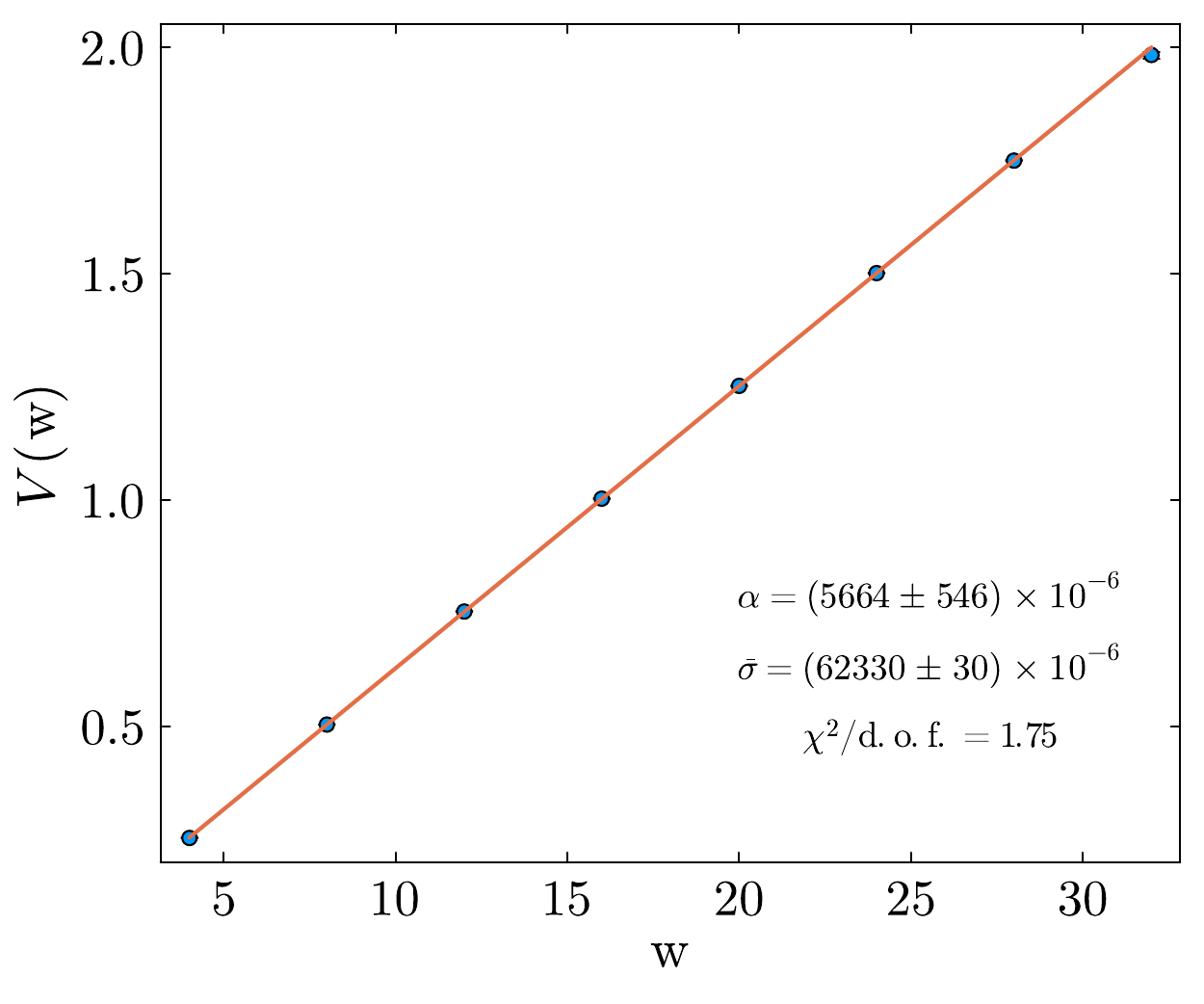}}
    \subfigure[]{\includegraphics[width=0.45\textwidth]{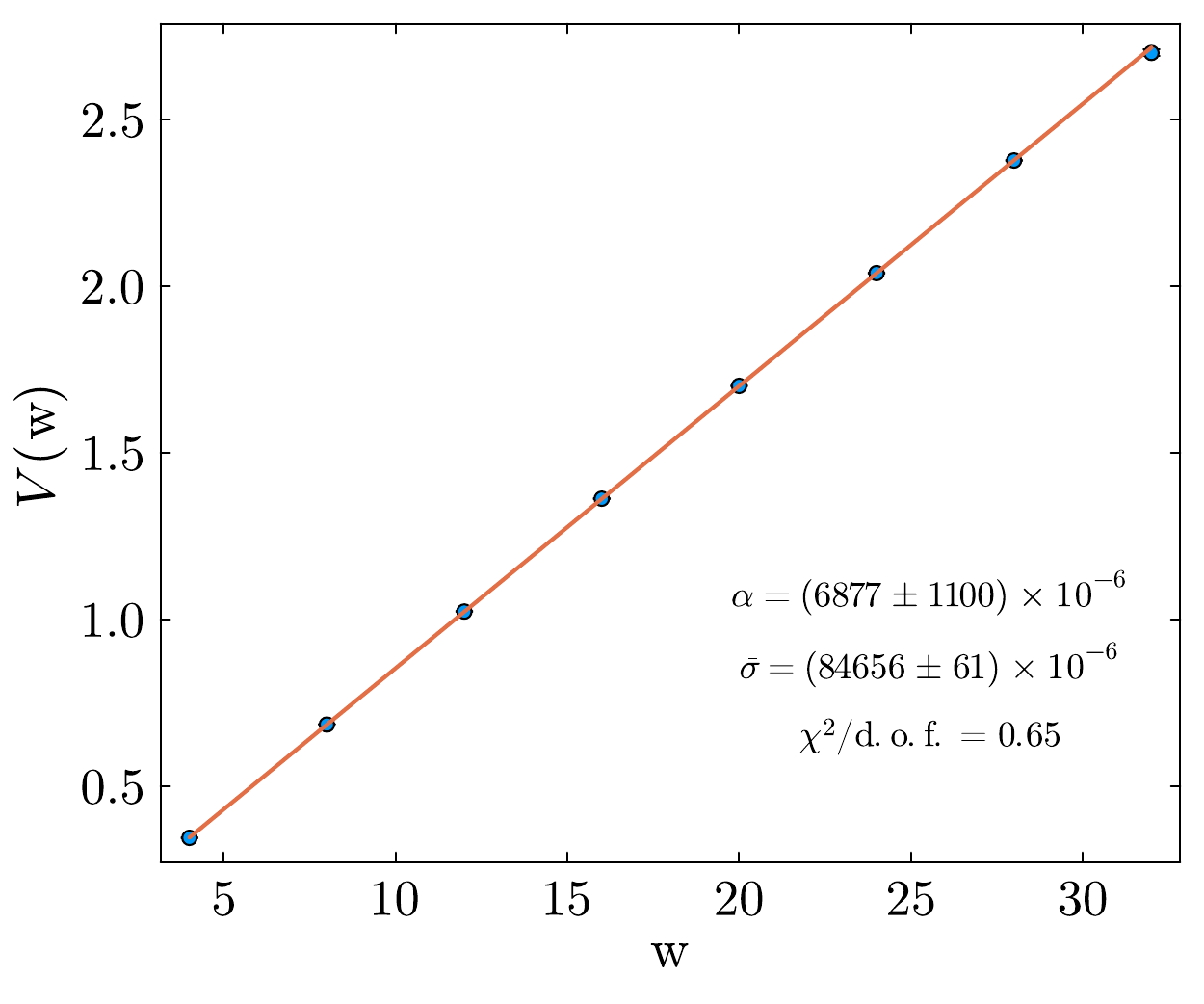}}
    \caption{Static quark potential $V(\mathsf{w})$ versus separation $\mathsf{w}$ for $L=32$ at (a) $\mu=8.0$, (b) $\mu=10.0$, (c) $\mu=12.0$, (d) $\mu=14.0$.}
    \label{fig:stringtension_L32}
\end{figure}

\begin{figure}
    \centering
    \subfigure[]{\includegraphics[width=0.45\textwidth]{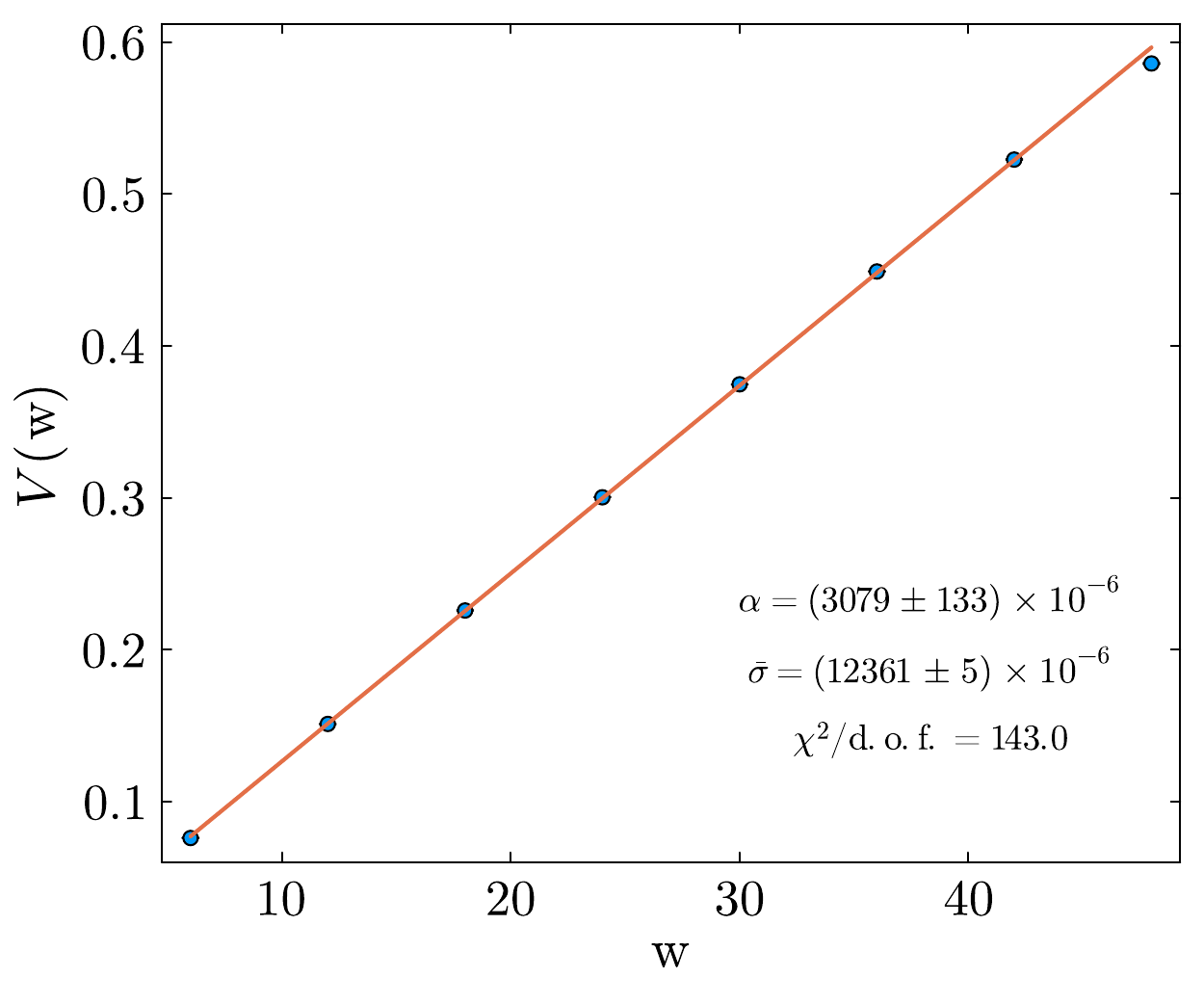}}
    \subfigure[]{\includegraphics[width=0.45\textwidth]{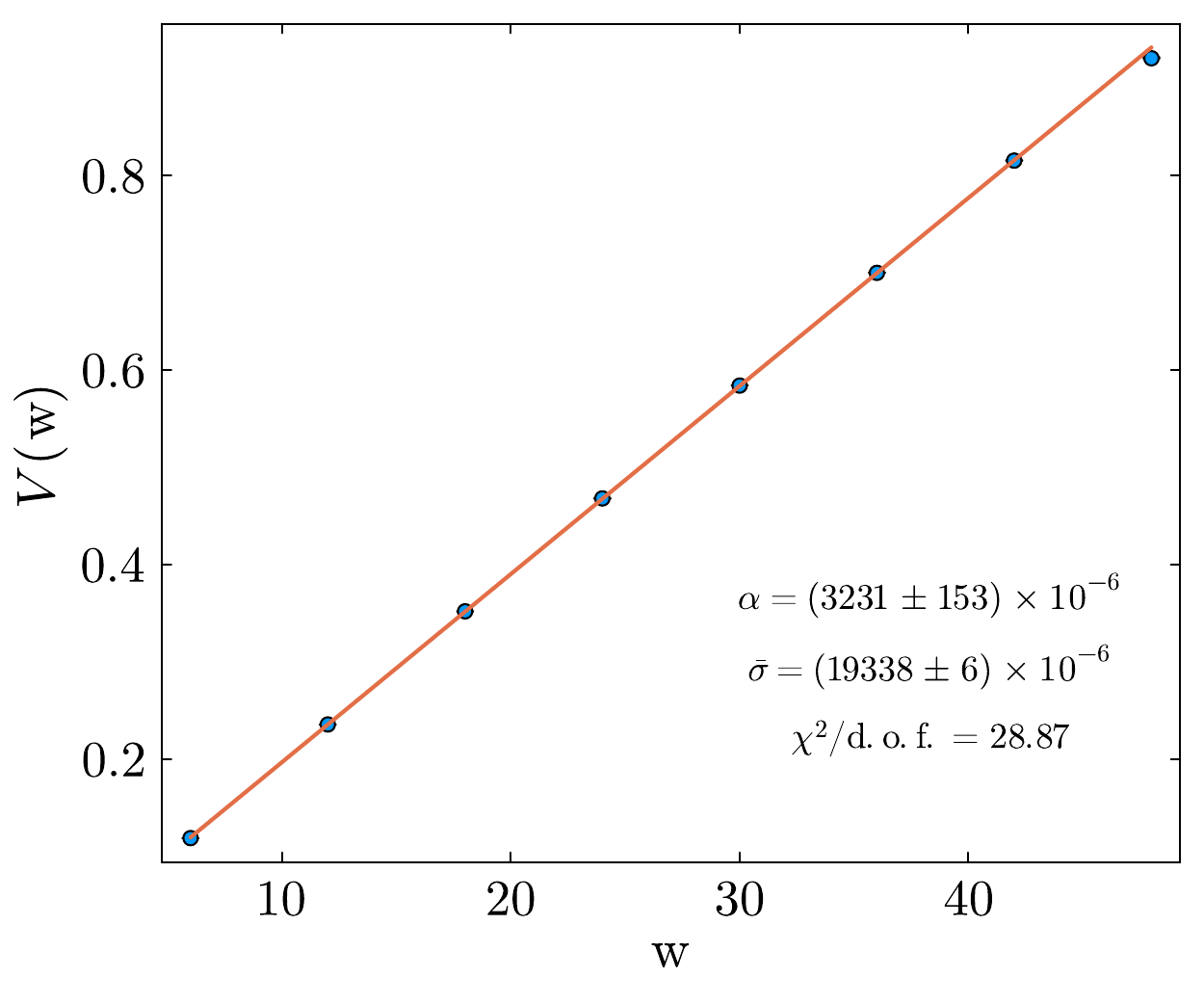}}
    \subfigure[]{\includegraphics[width=0.45\textwidth]{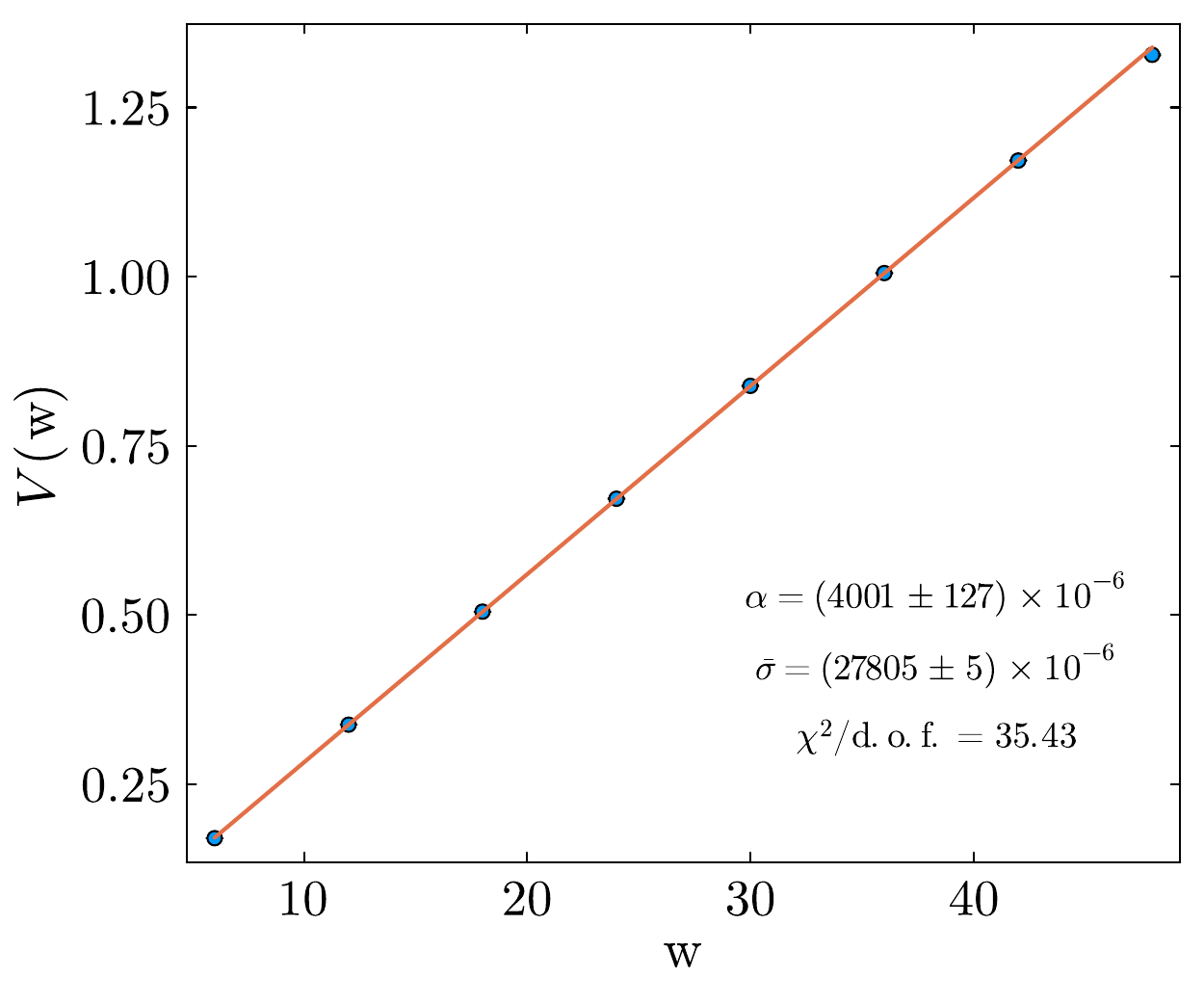}}
    \subfigure[]{\includegraphics[width=0.45\textwidth]{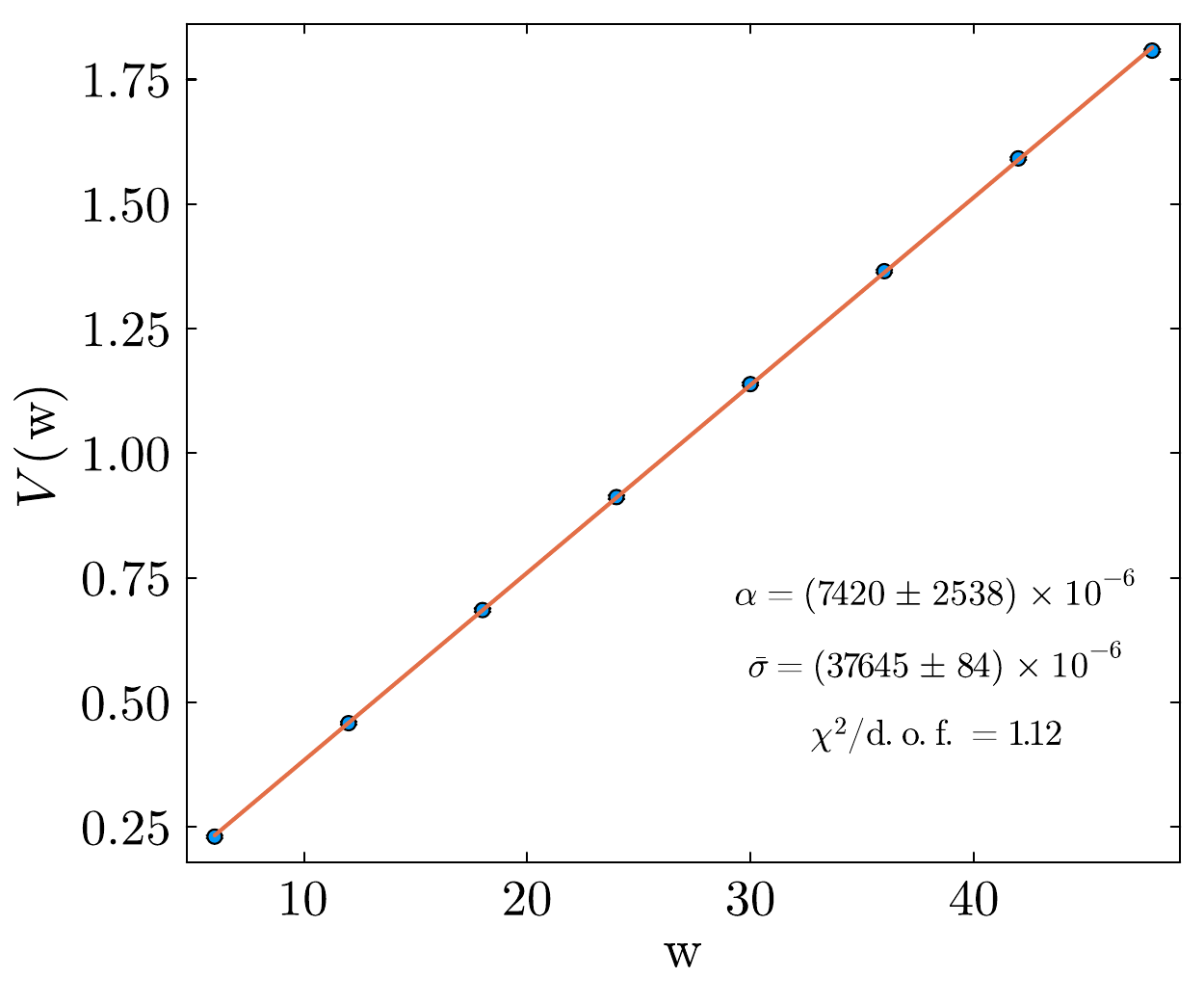}}
    \caption{Static quark potential $V(\mathsf{w})$ versus separation $\mathsf{w}$ for $L=48$ at (a) $\mu=8.0$, (b) $\mu=10.0$, (c) $\mu=12.0$, (d) $\mu=14.0$.}
    \label{fig:stringtension_L48}
\end{figure}

\begin{figure}
    \centering
    \subfigure[]{\includegraphics[width=0.45\textwidth]{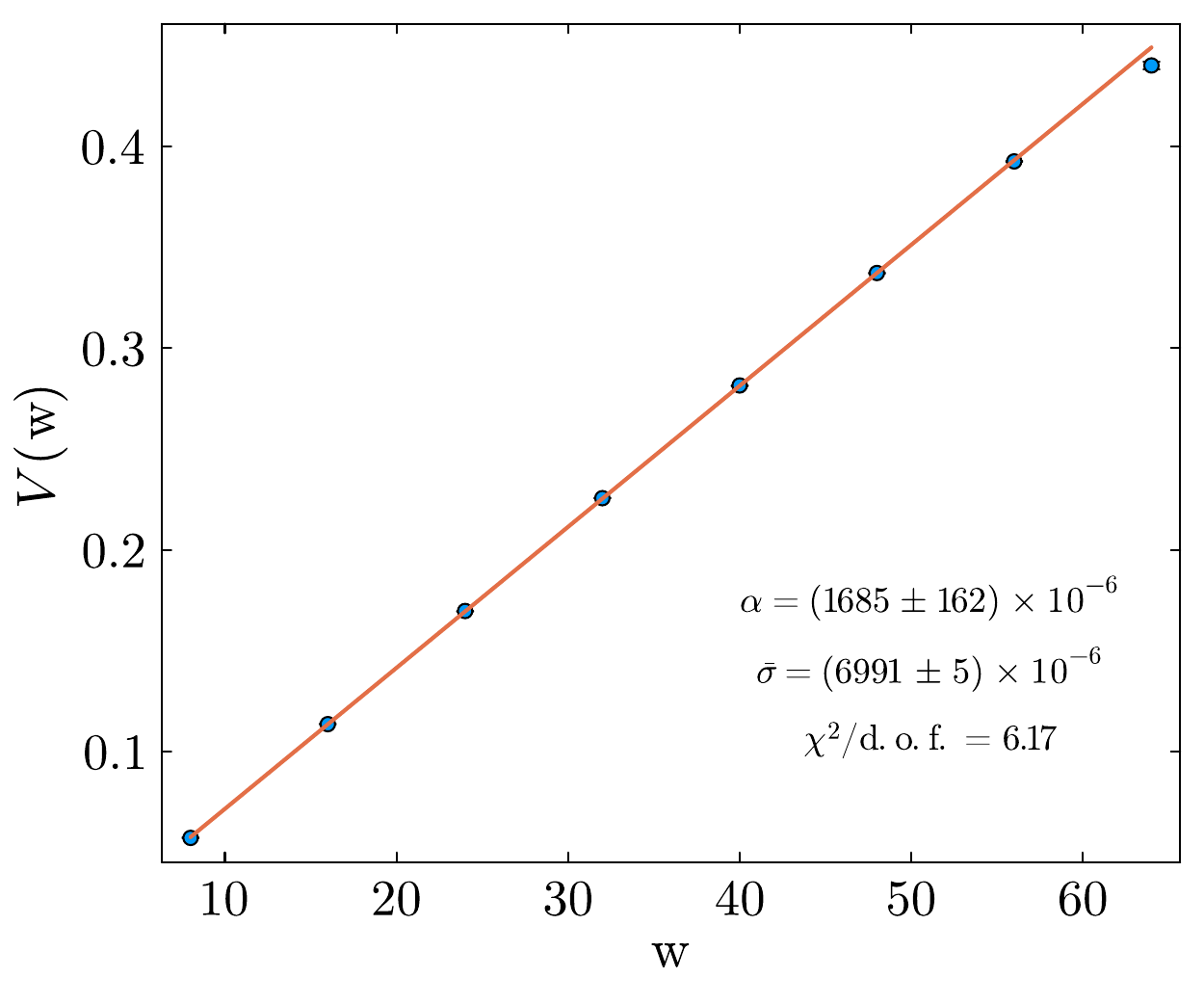}}
    \subfigure[]{\includegraphics[width=0.45\textwidth]{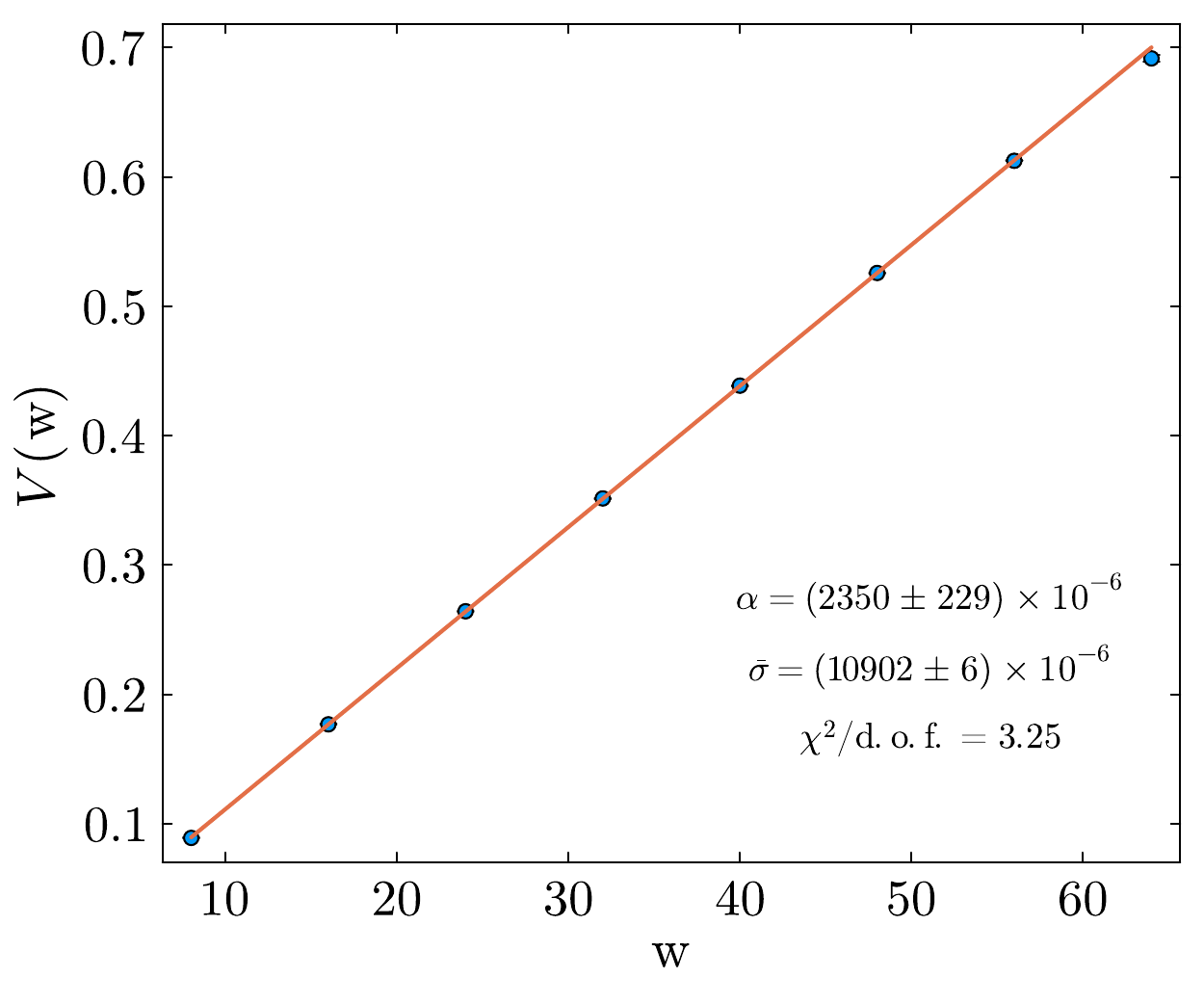}}
    \subfigure[]{\includegraphics[width=0.45\textwidth]{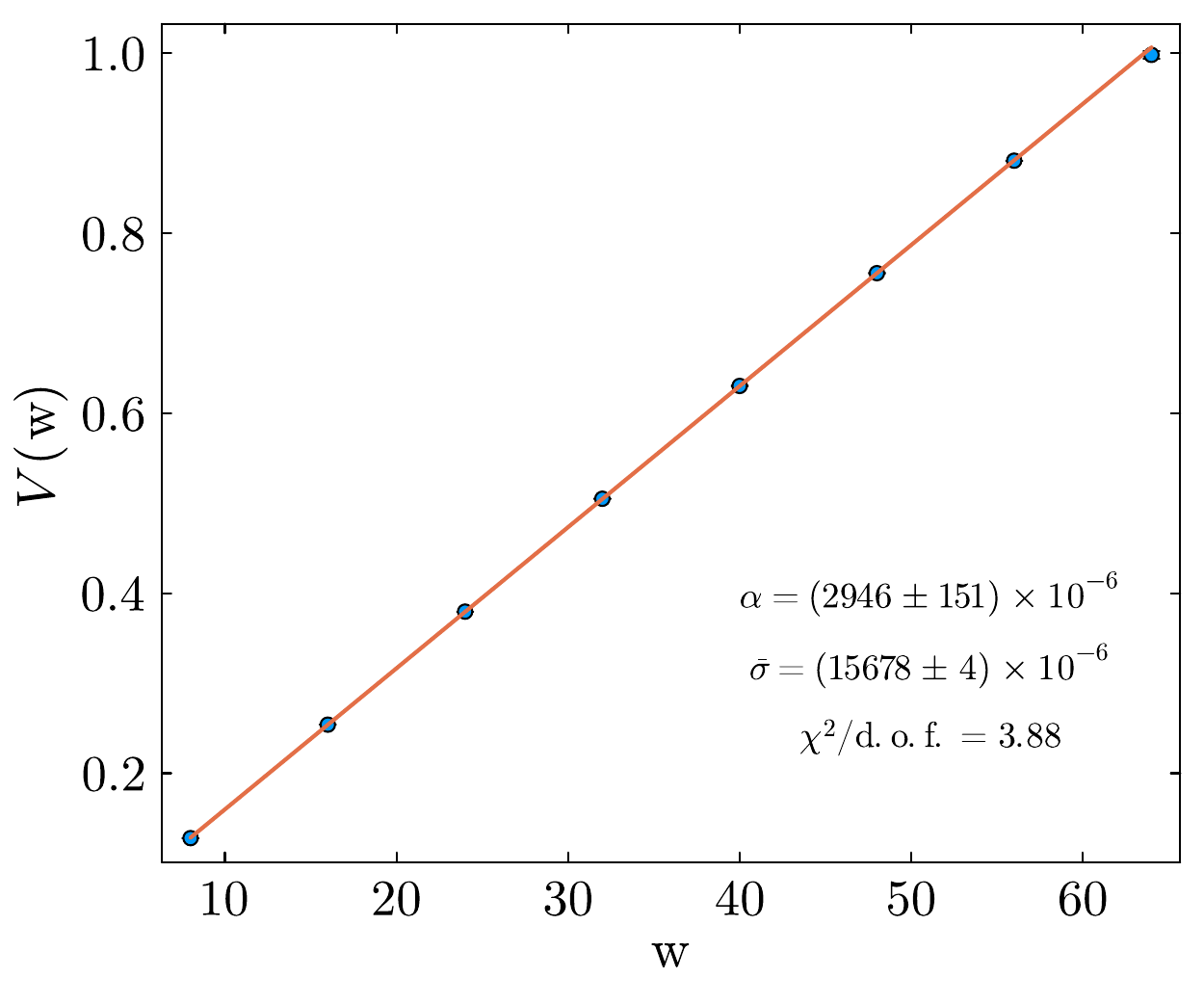}}
    \subfigure[]{\includegraphics[width=0.45\textwidth]{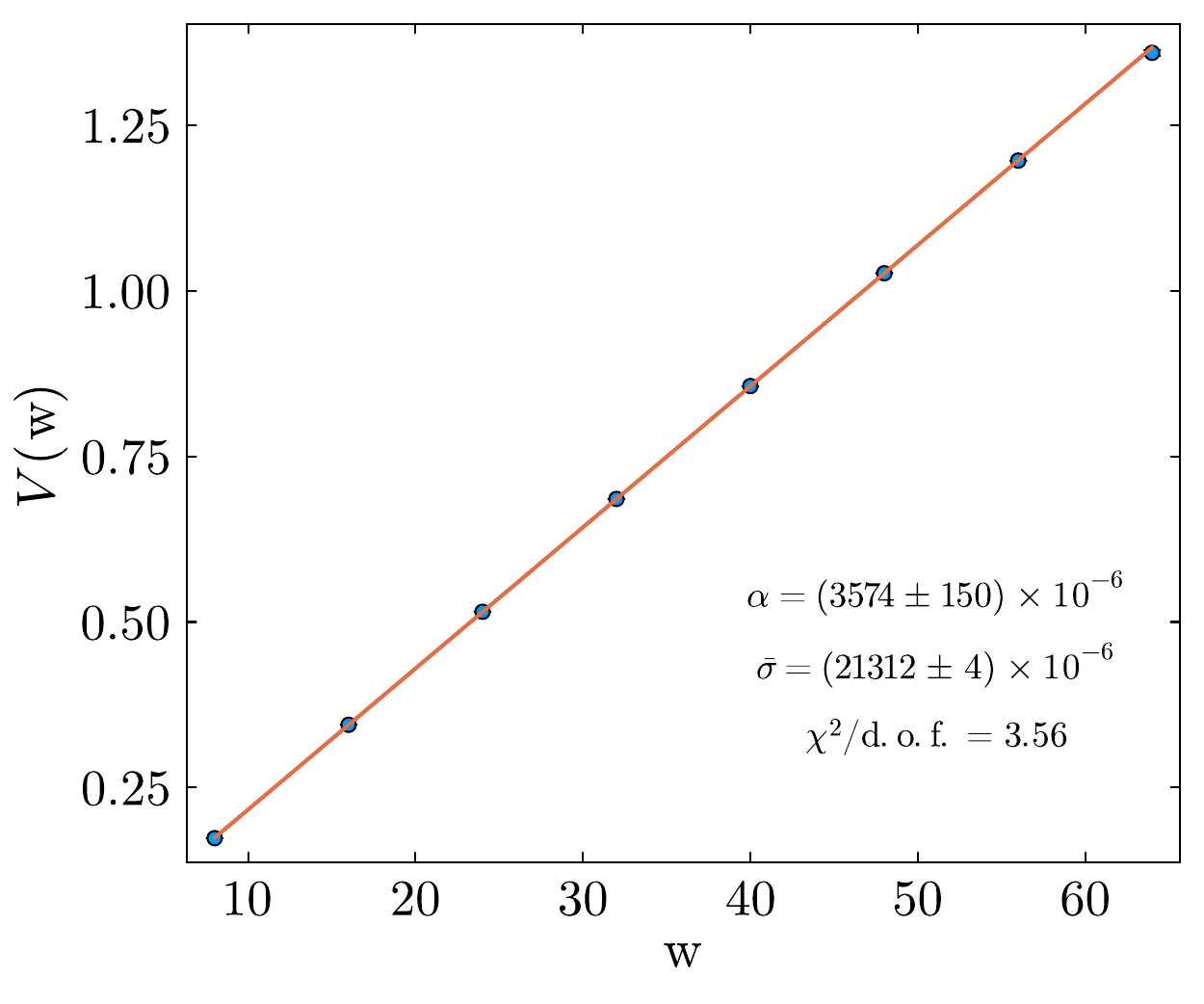}}
    \caption{Static quark potential $V(\mathsf{w})$ versus separation $\mathsf{w}$ for $L=64$ at (a) $\mu=8.0$, (b) $\mu=10.0$, (c) $\mu=12.0$, (d) $\mu=14.0$.}
    \label{fig:stringtension_L64}
\end{figure}

\begin{figure}
    \centering
    \subfigure[]{\includegraphics[width=0.45\textwidth]{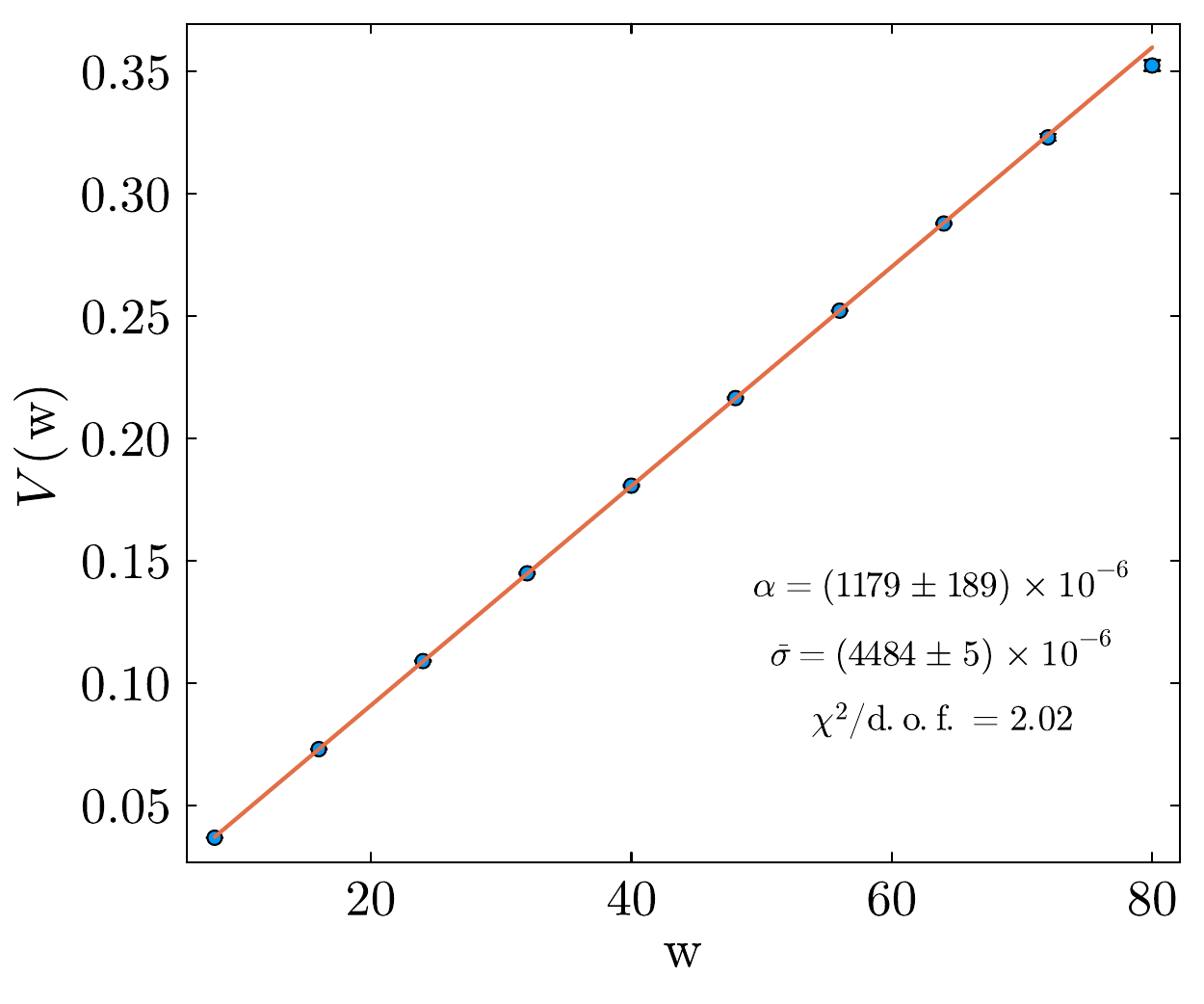}}
    \subfigure[]{\includegraphics[width=0.45\textwidth]{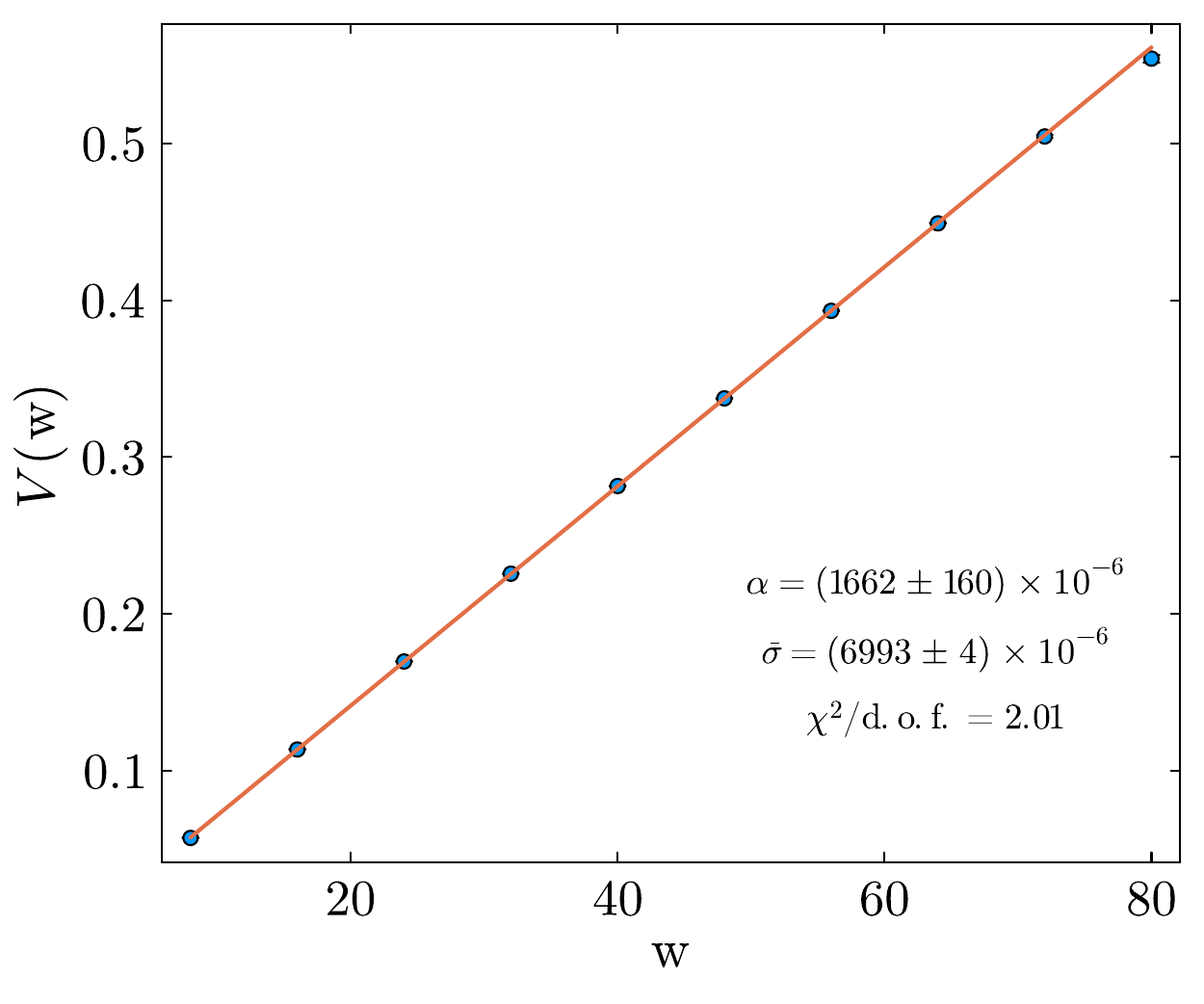}}
    \subfigure[]{\includegraphics[width=0.45\textwidth]{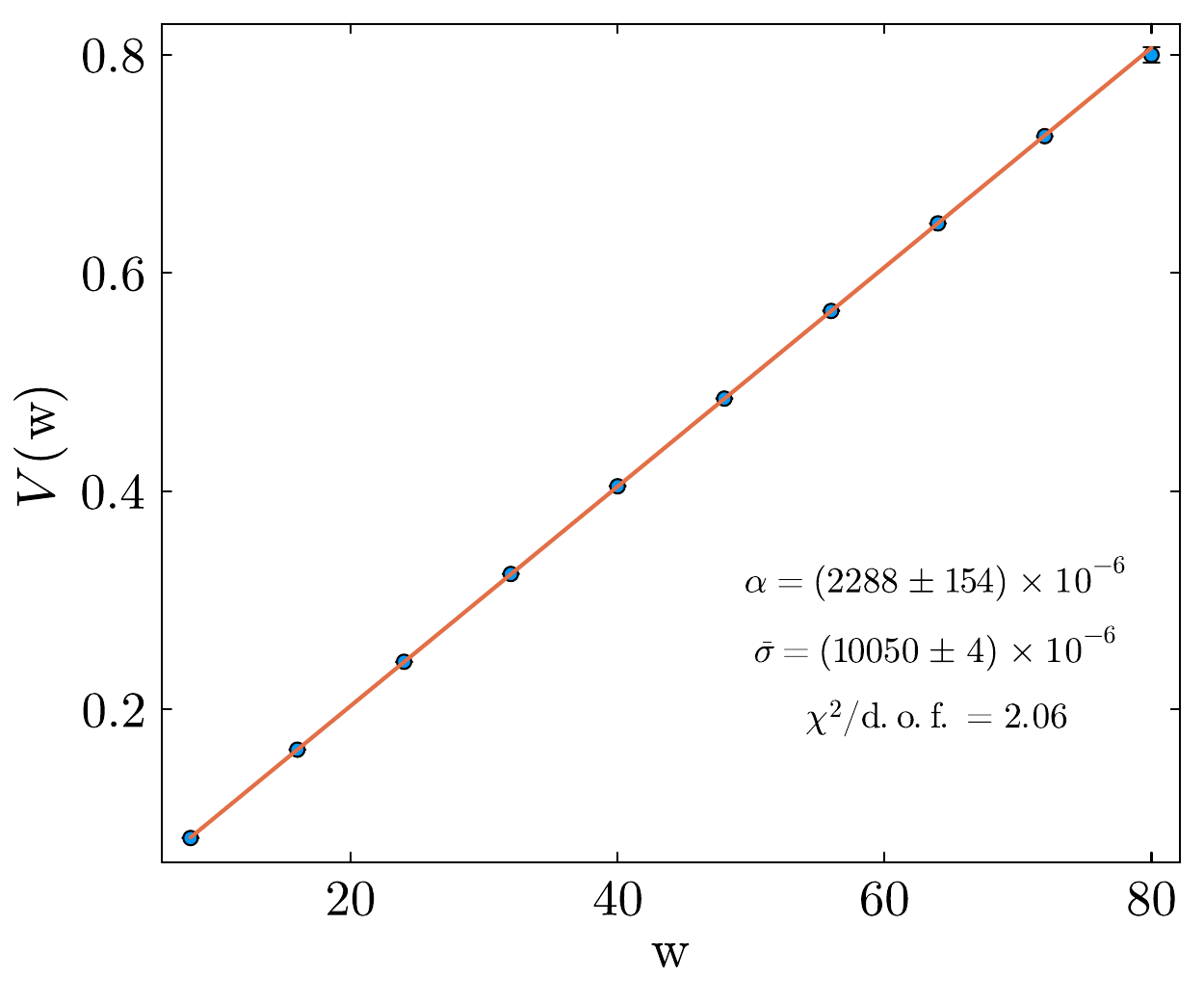}}
    \subfigure[]{\includegraphics[width=0.45\textwidth]{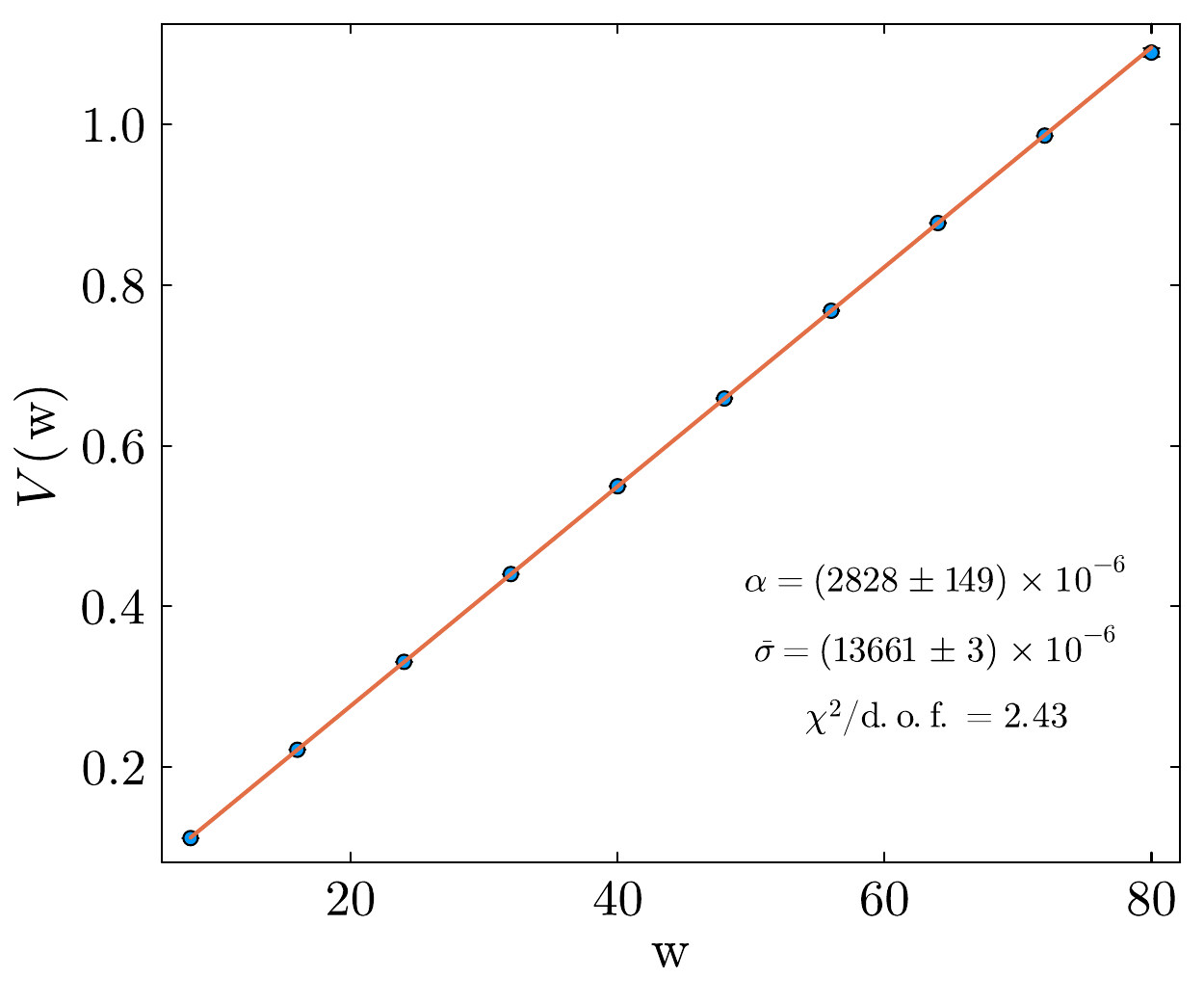}}
    \caption{Static quark potential $V(\mathsf{w})$ versus separation $\mathsf{w}$ for $L=80$ at (a) $\mu=8.0$, (b) $\mu=10.0$, (c) $\mu=12.0$, (d) $\mu=14.0$.}
    \label{fig:stringtension_L80}
\end{figure}

\begin{figure}
    \centering
    \subfigure[]{\includegraphics[width=0.45\textwidth]{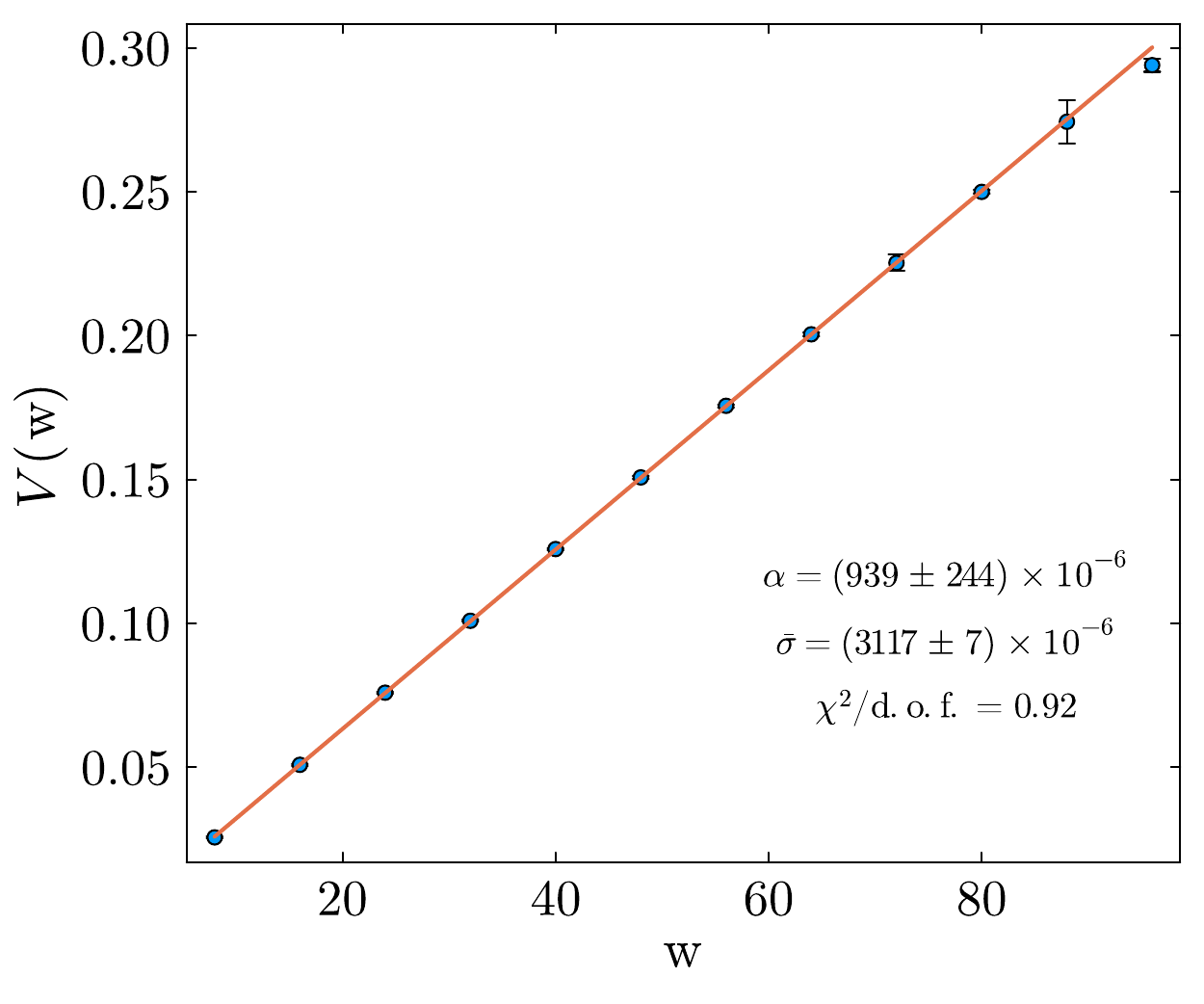}}
    \subfigure[]{\includegraphics[width=0.45\textwidth]{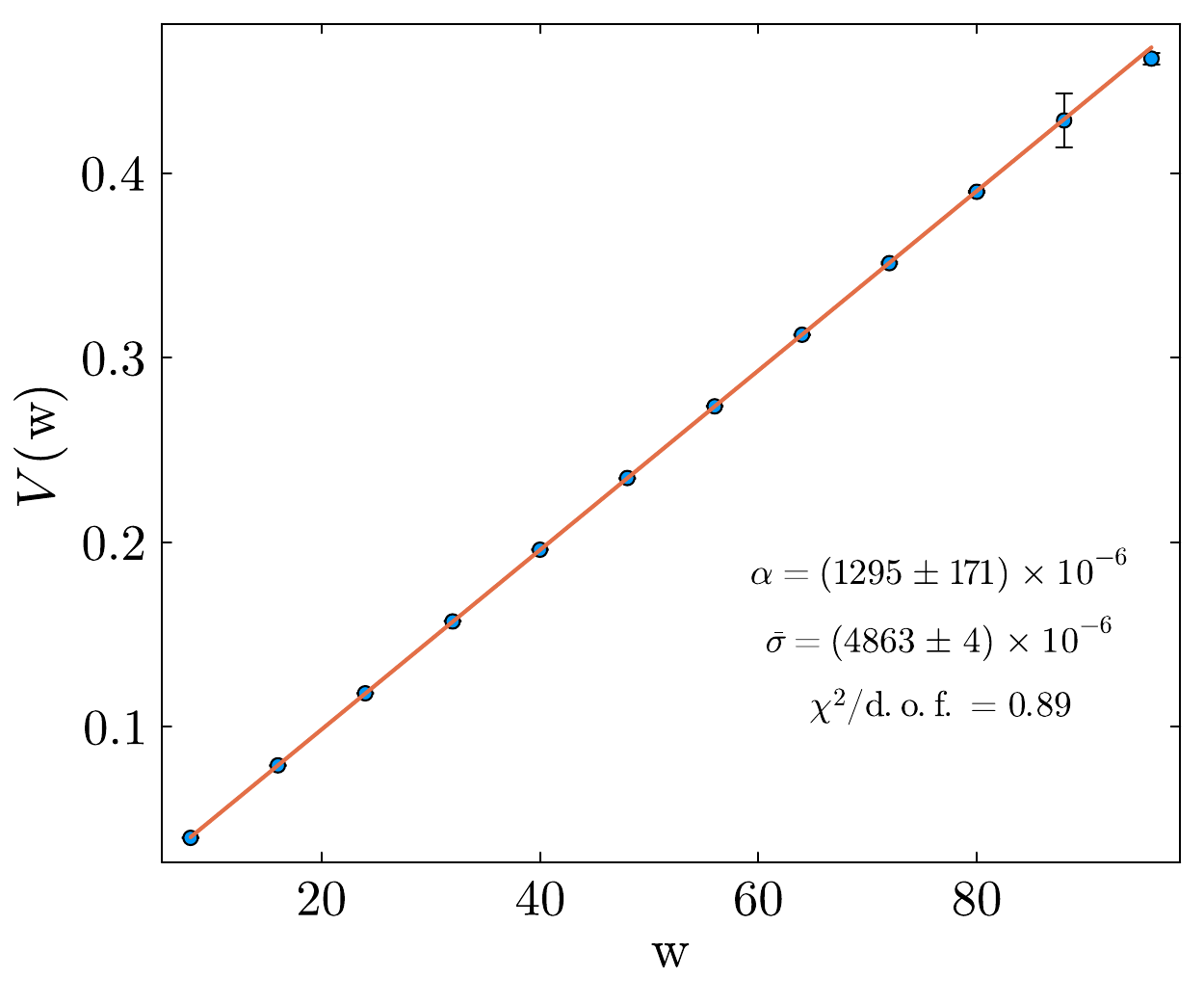}}
    \subfigure[]{\includegraphics[width=0.45\textwidth]{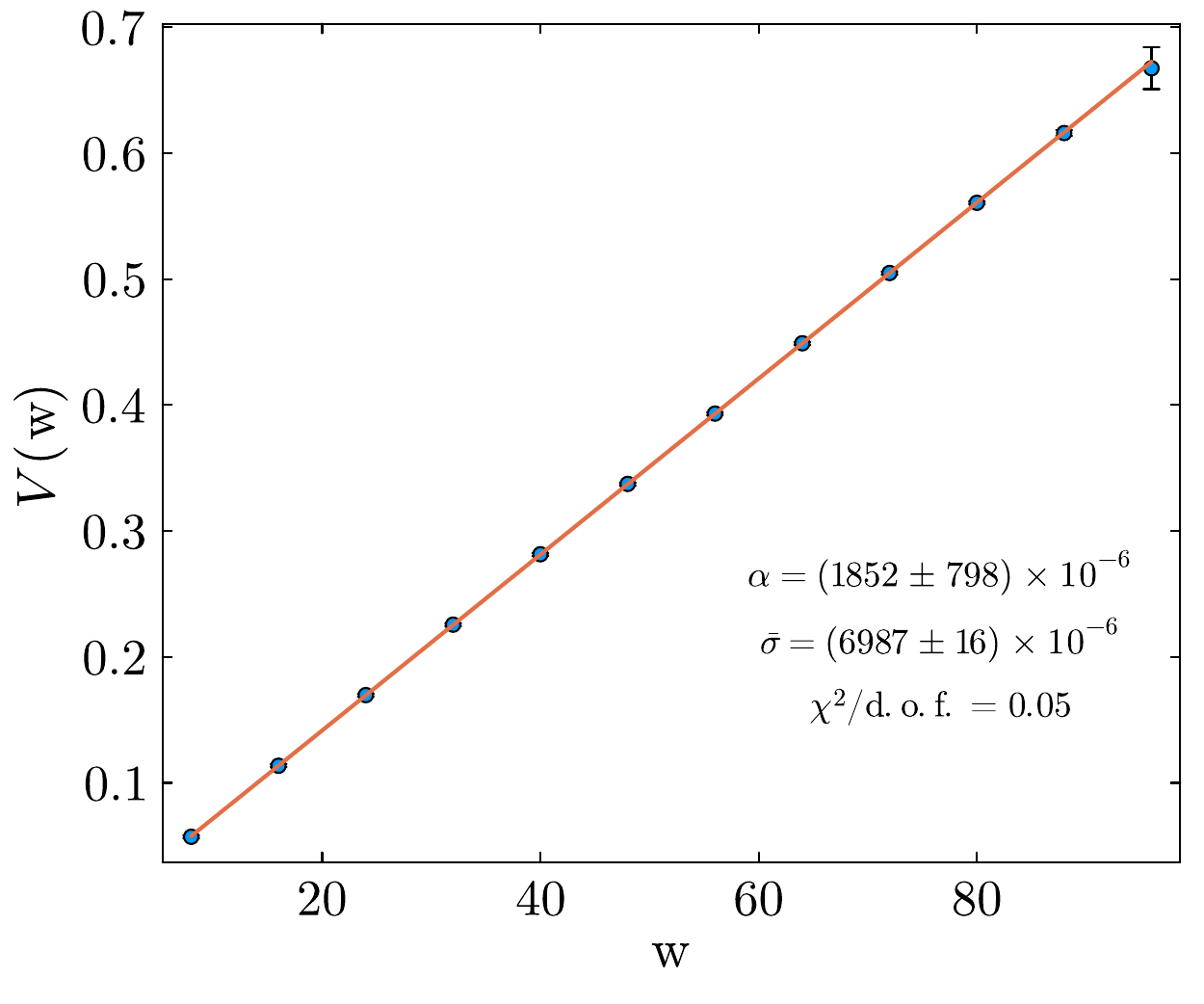}}
    \subfigure[]{\includegraphics[width=0.45\textwidth]{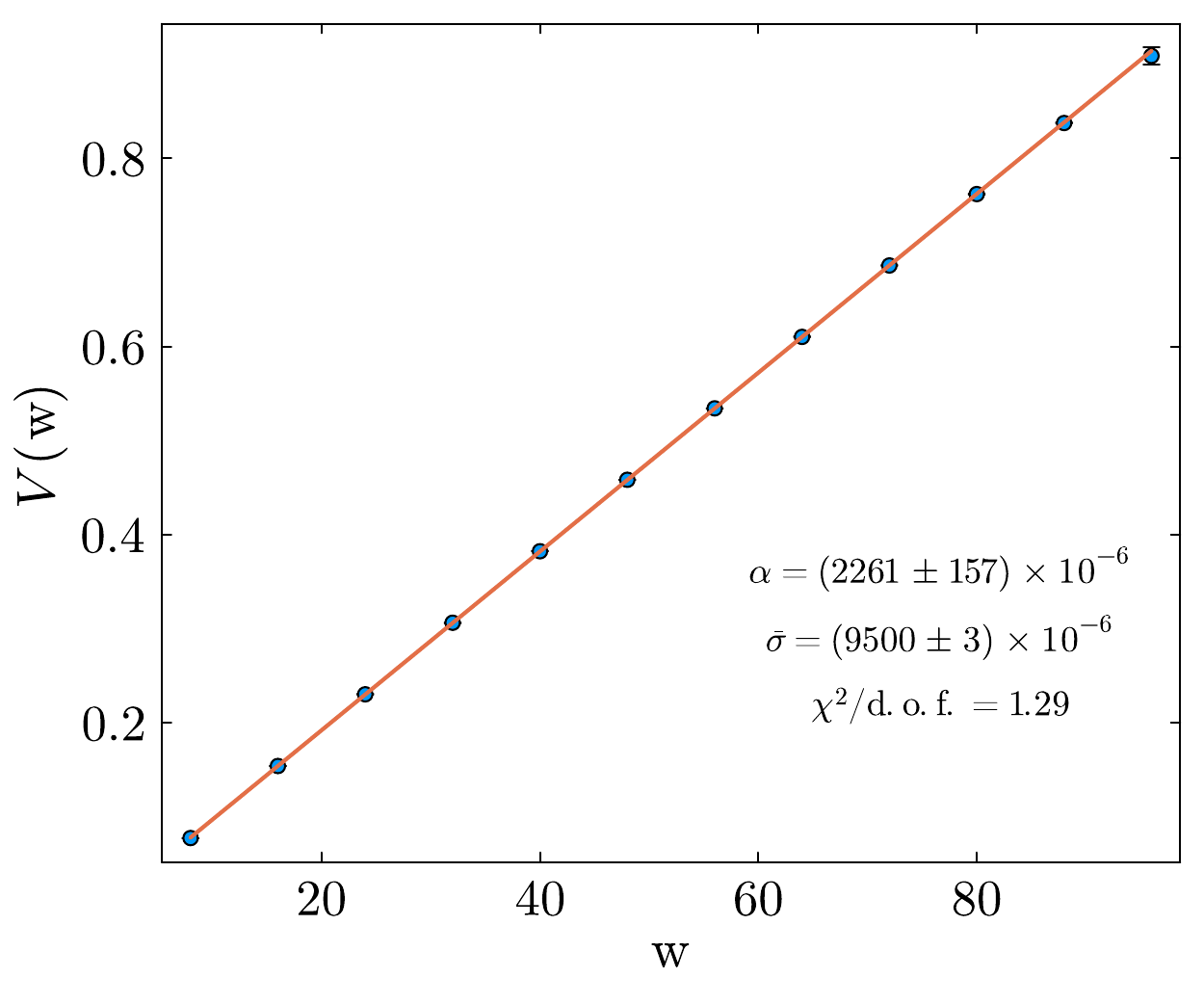}}
    \caption{Static quark potential $V(\mathsf{w})$ versus separation $\mathsf{w}$ for $L=96$ at (a) $\mu=8.0$, (b) $\mu=10.0$, (c) $\mu=12.0$, (d) $\mu=14.0$.}
    \label{fig:stringtension_L96}
\end{figure}

\clearpage

\subsection{Plots that determine \texorpdfstring{$\sqrt{\sigma}/m^+$}{sqrt(sigma/m+)}}

\begin{figure}[h!]
    \centering
    \subfigure[]{\includegraphics[width=0.45\textwidth]{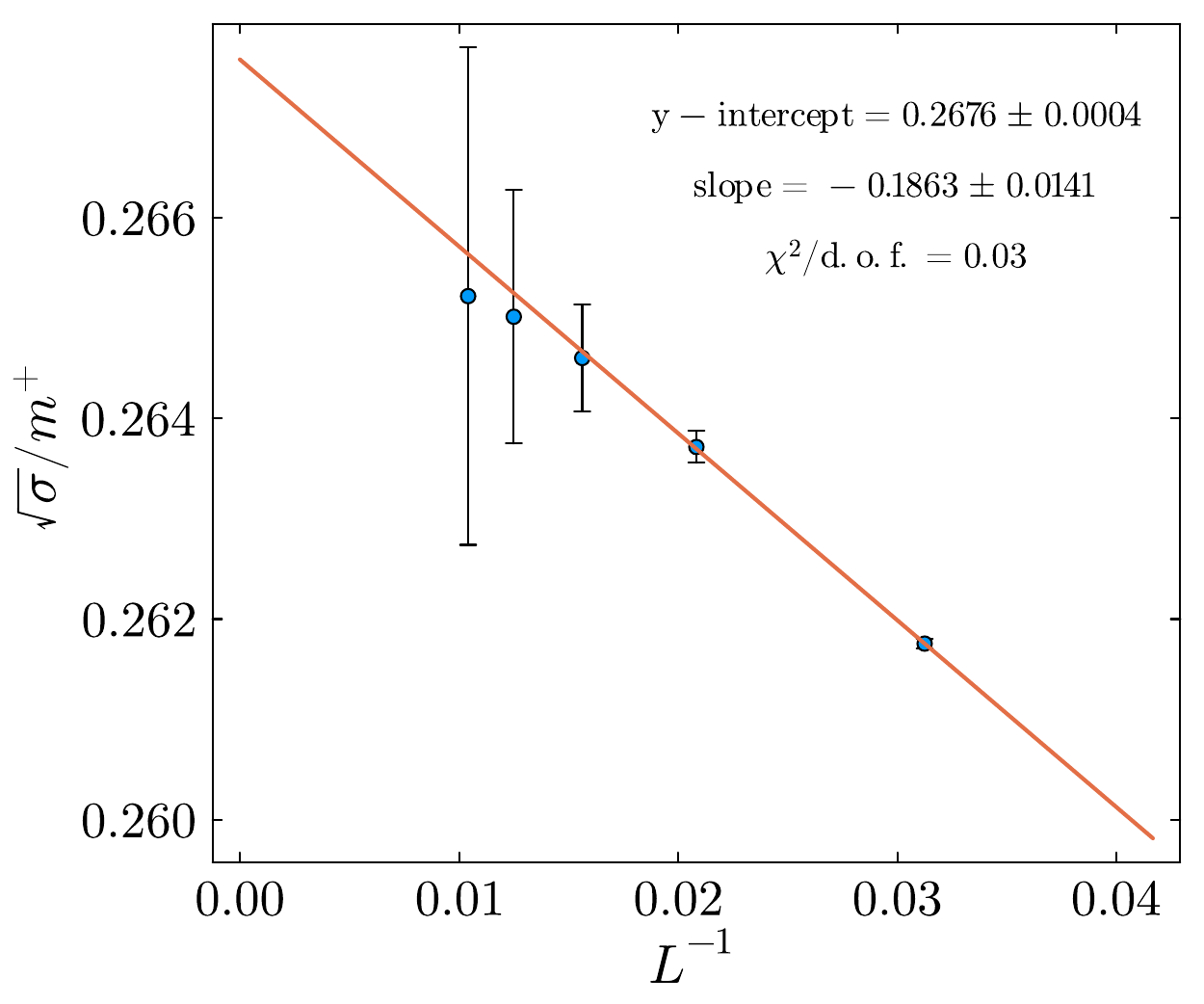}}
    \subfigure[]{\includegraphics[width=0.45\textwidth]{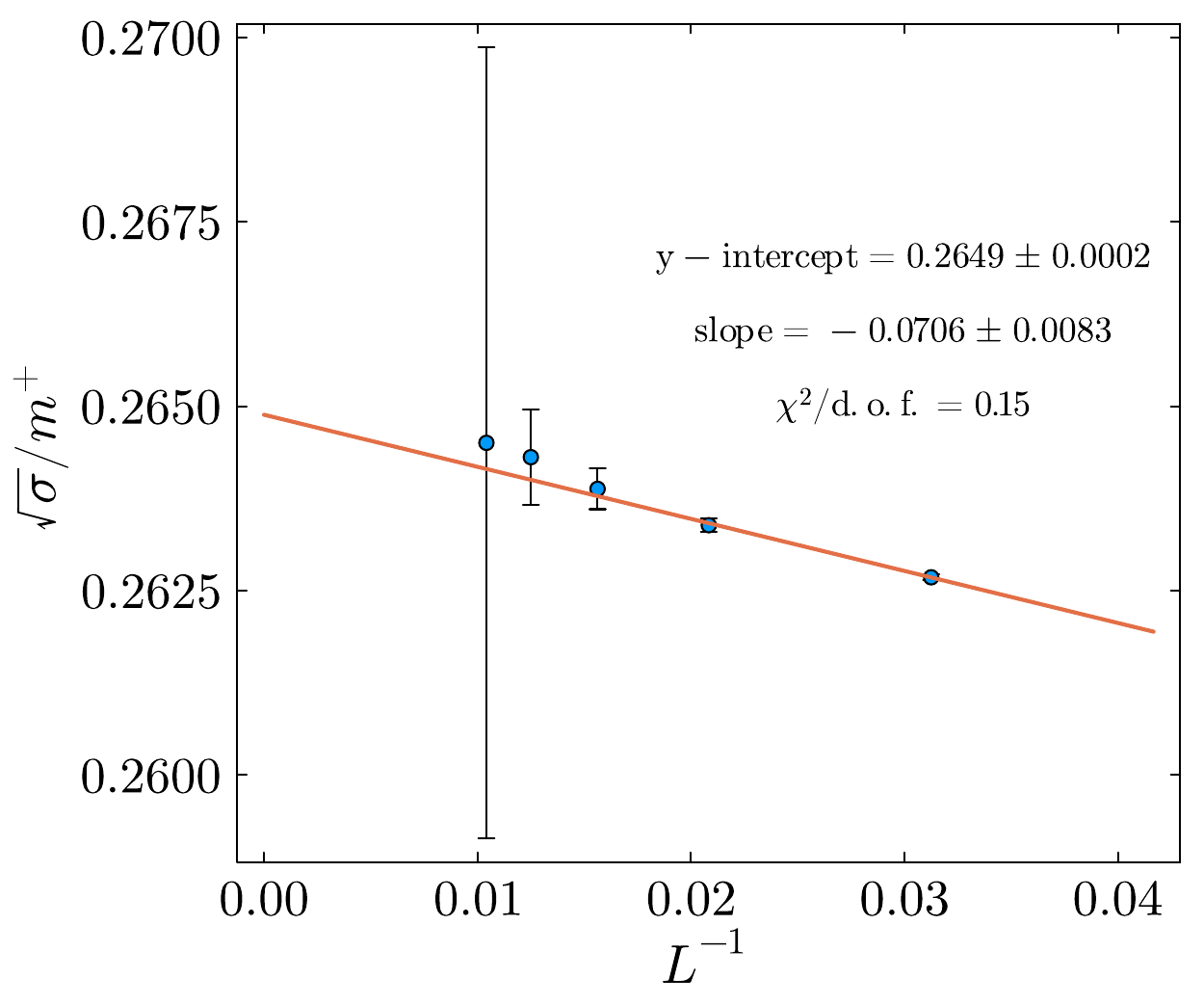}}
    \subfigure[]{\includegraphics[width=0.45\textwidth]{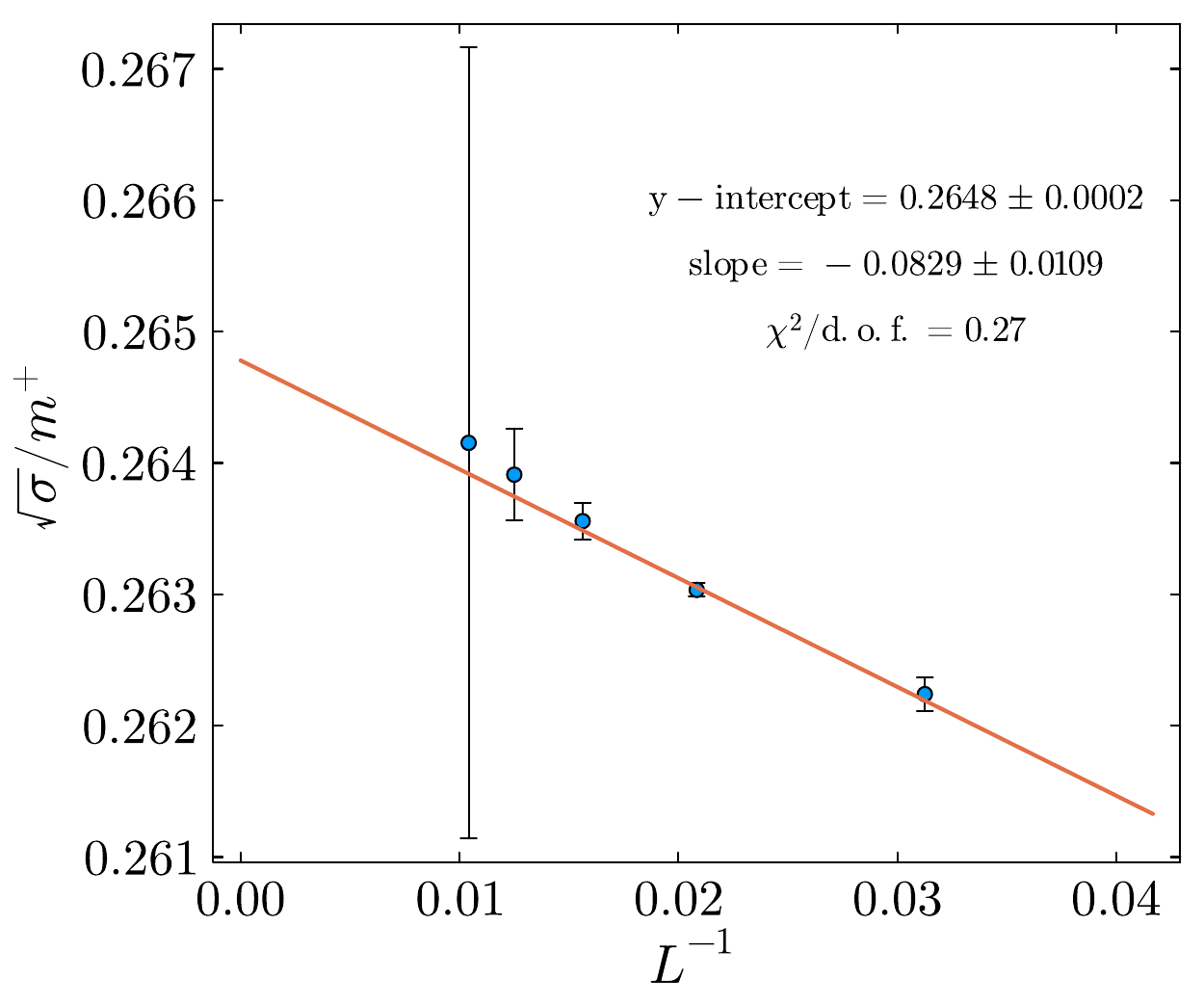}}
    \subfigure[]{\includegraphics[width=0.45\textwidth]{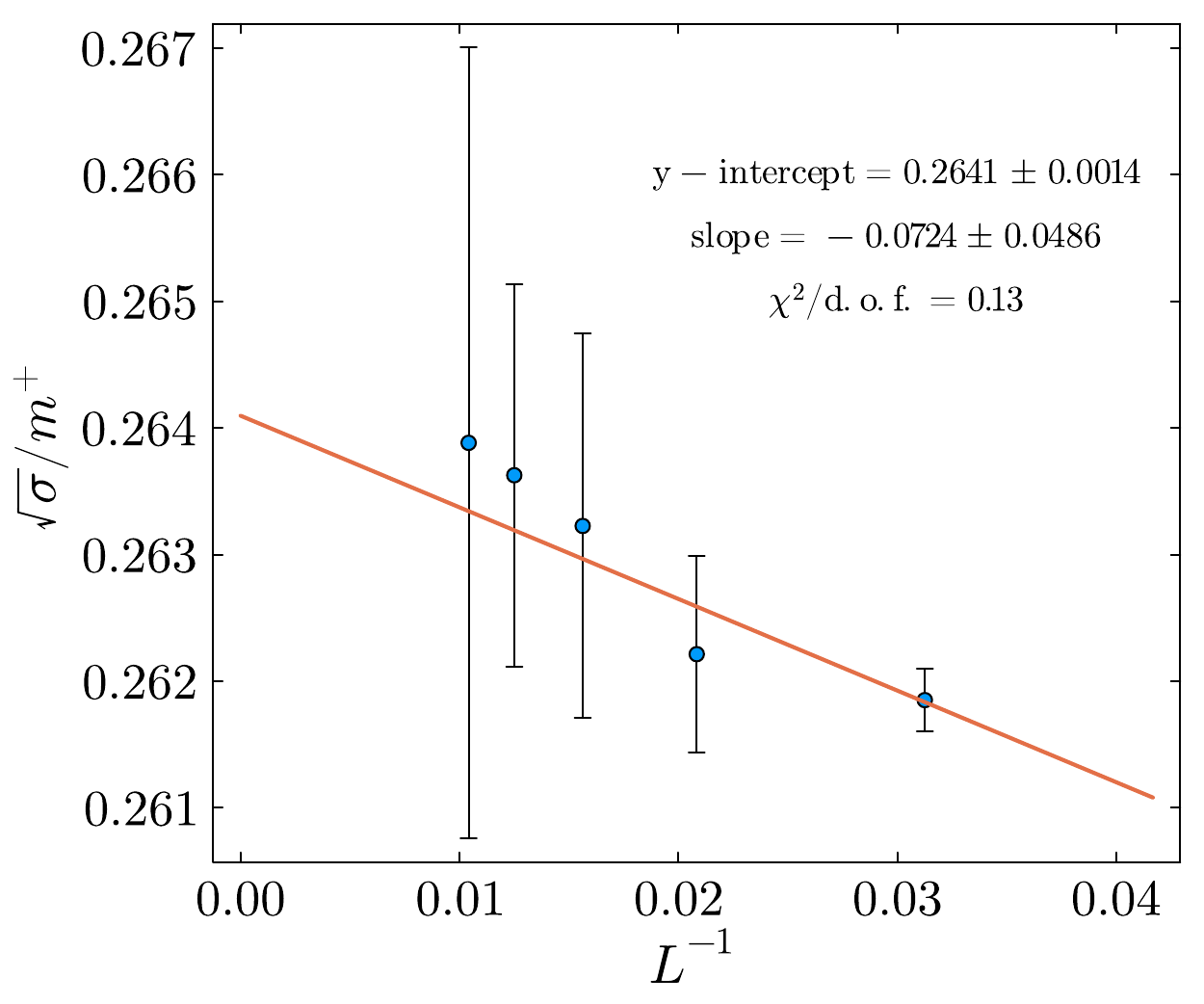}}
    \caption{String tension scaling at (a) $\mu=8.0$, (b) $\mu=10.0$, (c) $\mu=12.0$, (d) $\mu=14.0$. Each data point is computed using $\zeta = 2.5907(2)$, values of $\bar{\sigma}$ obtained in \cref{subsec:tension}, the definition $\sigma= \bar{\sigma}\zeta$, and the values of $m^+ = \mathsf{E}^+_1-\mathsf{E}^+_0$ from \cref{subsec:lowspectab}.} 
    \label{fig:stringtensionscaling}
\end{figure}


\begin{thebibliography}{31}%
\makeatletter
\providecommand \@ifxundefined [1]{%
 \@ifx{#1\undefined}
}%
\providecommand \@ifnum [1]{%
 \ifnum #1\expandafter \@firstoftwo
 \else \expandafter \@secondoftwo
 \fi
}%
\providecommand \@ifx [1]{%
 \ifx #1\expandafter \@firstoftwo
 \else \expandafter \@secondoftwo
 \fi
}%
\providecommand \natexlab [1]{#1}%
\providecommand \enquote  [1]{``#1''}%
\providecommand \bibnamefont  [1]{#1}%
\providecommand \bibfnamefont [1]{#1}%
\providecommand \citenamefont [1]{#1}%
\providecommand \href@noop [0]{\@secondoftwo}%
\providecommand \href [0]{\begingroup \@sanitize@url \@href}%
\providecommand \@href[1]{\@@startlink{#1}\@@href}%
\providecommand \@@href[1]{\endgroup#1\@@endlink}%
\providecommand \@sanitize@url [0]{\catcode `\\12\catcode `\$12\catcode `\&12\catcode `\#12\catcode `\^12\catcode `\_12\catcode `\%12\relax}%
\providecommand \@@startlink[1]{}%
\providecommand \@@endlink[0]{}%
\providecommand \url  [0]{\begingroup\@sanitize@url \@url }%
\providecommand \@url [1]{\endgroup\@href {#1}{\urlprefix }}%
\providecommand \urlprefix  [0]{URL }%
\providecommand \Eprint [0]{\href }%
\providecommand \doibase [0]{https://doi.org/}%
\providecommand \selectlanguage [0]{\@gobble}%
\providecommand \bibinfo  [0]{\@secondoftwo}%
\providecommand \bibfield  [0]{\@secondoftwo}%
\providecommand \translation [1]{[#1]}%
\providecommand \BibitemOpen [0]{}%
\providecommand \bibitemStop [0]{}%
\providecommand \bibitemNoStop [0]{.\EOS\space}%
\providecommand \EOS [0]{\spacefactor3000\relax}%
\providecommand \BibitemShut  [1]{\csname bibitem#1\endcsname}%
\let\auto@bib@innerbib\@empty
\bibitem [{\citenamefont {Athenodorou}\ and\ \citenamefont {Teper}(2020)}]{Athenodorou:2020ani}%
  \BibitemOpen
  \bibfield  {author} {\bibinfo {author} {\bibfnamefont {A.}~\bibnamefont {Athenodorou}}\ and\ \bibinfo {author} {\bibfnamefont {M.}~\bibnamefont {Teper}},\ }\bibfield  {title} {\bibinfo {title} {{The glueball spectrum of SU(3) gauge theory in 3 + 1 dimensions}},\ }\href {https://doi.org/10.1007/JHEP11(2020)172} {\bibfield  {journal} {\bibinfo  {journal} {J. High Energ. Phys.}\ }\textbf {\bibinfo {volume} {2020}}\bibfield  {number} {\bibinfo  {number} { (11)},\ \bibinfo {pages} {172}},\ }\Eprint {https://arxiv.org/abs/2007.06422} {arXiv:2007.06422 [hep-lat]} \BibitemShut {NoStop}%
\bibitem [{\citenamefont {Chandrasekharan}\ \emph {et~al.}(2025)\citenamefont {Chandrasekharan}, \citenamefont {Siew},\ and\ \citenamefont {Bhattacharya}}]{Chandrasekharan:2025smw}%
  \BibitemOpen
  \bibfield  {author} {\bibinfo {author} {\bibfnamefont {S.}~\bibnamefont {Chandrasekharan}}, \bibinfo {author} {\bibfnamefont {R.~X.}\ \bibnamefont {Siew}},\ and\ \bibinfo {author} {\bibfnamefont {T.}~\bibnamefont {Bhattacharya}},\ }\bibfield  {title} {\bibinfo {title} {Monomer-dimer tensor-network basis for qubit-regularized lattice gauge theories},\ }\href {https://doi.org/10.1103/gns9-l3gk} {\bibfield  {journal} {\bibinfo  {journal} {Phys. Rev. D}\ }\textbf {\bibinfo {volume} {111}},\ \bibinfo {pages} {114502} (\bibinfo {year} {2025})},\ \Eprint {https://arxiv.org/abs/2502.14175} {arXiv:2502.14175 [hep-lat]} \BibitemShut {NoStop}%
\bibitem [{\citenamefont {Chandrasekharan}(2025)}]{Chandrasekharan:2025Cb}%
  \BibitemOpen
  \bibfield  {author} {\bibinfo {author} {\bibfnamefont {S.}~\bibnamefont {Chandrasekharan}},\ }\bibfield  {title} {\bibinfo {title} {{Qubit regularization of quantum field theories}},\ }\href {https://doi.org/10.22323/1.466.0001} {\bibfield  {journal} {\bibinfo  {journal} {PoS}\ }\textbf {\bibinfo {volume} {LATTICE2024}},\ \bibinfo {pages} {001} (\bibinfo {year} {2025})},\ \Eprint {https://arxiv.org/abs/2502.05716} {arXiv:2502.05716 [hep-lat]} \BibitemShut {NoStop}%
\bibitem [{\citenamefont {Bauer}\ \emph {et~al.}(2023)\citenamefont {Bauer}, \citenamefont {Davoudi}, \citenamefont {Klco},\ and\ \citenamefont {Savage}}]{Bauer:2023qgm}%
  \BibitemOpen
  \bibfield  {author} {\bibinfo {author} {\bibfnamefont {C.~W.}\ \bibnamefont {Bauer}}, \bibinfo {author} {\bibfnamefont {Z.}~\bibnamefont {Davoudi}}, \bibinfo {author} {\bibfnamefont {N.}~\bibnamefont {Klco}},\ and\ \bibinfo {author} {\bibfnamefont {M.~J.}\ \bibnamefont {Savage}},\ }\bibfield  {title} {\bibinfo {title} {{Quantum simulation of fundamental particles and forces}},\ }\href {https://doi.org/10.1038/s42254-023-00599-8} {\bibfield  {journal} {\bibinfo  {journal} {Nature Rev. Phys.}\ }\textbf {\bibinfo {volume} {5}},\ \bibinfo {pages} {420} (\bibinfo {year} {2023})},\ \Eprint {https://arxiv.org/abs/2404.06298} {arXiv:2404.06298 [hep-ph]} \BibitemShut {NoStop}%
\bibitem [{\citenamefont {Kapustin}\ and\ \citenamefont {Witten}(2007)}]{Kapustin:2006pk}%
  \BibitemOpen
  \bibfield  {author} {\bibinfo {author} {\bibfnamefont {A.}~\bibnamefont {Kapustin}}\ and\ \bibinfo {author} {\bibfnamefont {E.}~\bibnamefont {Witten}},\ }\bibfield  {title} {\bibinfo {title} {{Electric-magnetic duality and the geometric Langlands program}},\ }\href {https://doi.org/10.4310/CNTP.2007.v1.n1.a1} {\bibfield  {journal} {\bibinfo  {journal} {Commun. Num. Theor. Phys.}\ }\textbf {\bibinfo {volume} {1}},\ \bibinfo {pages} {1} (\bibinfo {year} {2007})},\ \Eprint {https://arxiv.org/abs/hep-th/0604151} {arXiv:hep-th/0604151} \BibitemShut {NoStop}%
\bibitem [{\citenamefont {Kogut}\ and\ \citenamefont {Susskind}(1975)}]{Kogut:1974ag}%
  \BibitemOpen
  \bibfield  {author} {\bibinfo {author} {\bibfnamefont {J.~B.}\ \bibnamefont {Kogut}}\ and\ \bibinfo {author} {\bibfnamefont {L.}~\bibnamefont {Susskind}},\ }\bibfield  {title} {\bibinfo {title} {{H}amiltonian formulation of {W}ilson's lattice gauge theories},\ }\href {https://doi.org/10.1103/PhysRevD.11.395} {\bibfield  {journal} {\bibinfo  {journal} {Phys.\ Rev.\ D}\ }\textbf {\bibinfo {volume} {11}},\ \bibinfo {pages} {395} (\bibinfo {year} {1975})}\BibitemShut {NoStop}%
\bibitem [{\citenamefont {Bhattacharya}\ \emph {et~al.}(2021)\citenamefont {Bhattacharya}, \citenamefont {Buser}, \citenamefont {Chandrasekharan}, \citenamefont {Gupta},\ and\ \citenamefont {Singh}}]{Bhattacharya:2020gpm}%
  \BibitemOpen
  \bibfield  {author} {\bibinfo {author} {\bibfnamefont {T.}~\bibnamefont {Bhattacharya}}, \bibinfo {author} {\bibfnamefont {A.~J.}\ \bibnamefont {Buser}}, \bibinfo {author} {\bibfnamefont {S.}~\bibnamefont {Chandrasekharan}}, \bibinfo {author} {\bibfnamefont {R.}~\bibnamefont {Gupta}},\ and\ \bibinfo {author} {\bibfnamefont {H.}~\bibnamefont {Singh}},\ }\bibfield  {title} {\bibinfo {title} {Qubit regularization of asymptotic freedom},\ }\href {https://doi.org/10.1103/PhysRevLett.126.172001} {\bibfield  {journal} {\bibinfo  {journal} {Phys. Rev. Lett.}\ }\textbf {\bibinfo {volume} {126}},\ \bibinfo {pages} {172001} (\bibinfo {year} {2021})},\ \Eprint {https://arxiv.org/abs/2012.02153} {arXiv:2012.02153 [hep-lat]} \BibitemShut {NoStop}%
\bibitem [{\citenamefont {Maiti}\ \emph {et~al.}(2024)\citenamefont {Maiti}, \citenamefont {Banerjee}, \citenamefont {Chandrasekharan},\ and\ \citenamefont {Marinkovic}}]{Maiti:2023kpn}%
  \BibitemOpen
  \bibfield  {author} {\bibinfo {author} {\bibfnamefont {S.}~\bibnamefont {Maiti}}, \bibinfo {author} {\bibfnamefont {D.}~\bibnamefont {Banerjee}}, \bibinfo {author} {\bibfnamefont {S.}~\bibnamefont {Chandrasekharan}},\ and\ \bibinfo {author} {\bibfnamefont {M.~K.}\ \bibnamefont {Marinkovic}},\ }\bibfield  {title} {\bibinfo {title} {Asymptotic freedom at the {B}erezinskii-{K}osterlitz-{T}houless transition without fine-tuning using a qubit regularization},\ }\href {https://doi.org/10.1103/PhysRevLett.132.041601} {\bibfield  {journal} {\bibinfo  {journal} {Phys. Rev. Lett.}\ }\textbf {\bibinfo {volume} {132}},\ \bibinfo {pages} {041601} (\bibinfo {year} {2024})},\ \Eprint {https://arxiv.org/abs/2307.06117} {arXiv:2307.06117 [hep-lat]} \BibitemShut {NoStop}%
\bibitem [{\citenamefont {Siew}\ \emph {et~al.}(2026)\citenamefont {Siew}, \citenamefont {Chandrasekharan},\ and\ \citenamefont {Bhattacharya}}]{2cy4-cp7f}%
  \BibitemOpen
  \bibfield  {author} {\bibinfo {author} {\bibfnamefont {R.~X.}\ \bibnamefont {Siew}}, \bibinfo {author} {\bibfnamefont {S.}~\bibnamefont {Chandrasekharan}},\ and\ \bibinfo {author} {\bibfnamefont {T.}~\bibnamefont {Bhattacharya}},\ }\bibfield  {title} {\bibinfo {title} {{Asymptotic freedom and massive glueballs in a qubit-regularized SU(2) gauge theory}},\ }\href {https://doi.org/10.1103/2cy4-cp7f} {\bibfield  {journal} {\bibinfo  {journal} {Phys. Rev. D}\ }\textbf {\bibinfo {volume} {113}},\ \bibinfo {pages} {114522} (\bibinfo {year} {2026})}\BibitemShut {NoStop}%
\bibitem [{\citenamefont {Shahmoon}\ \emph {et~al.}(2014)\citenamefont {Shahmoon}, \citenamefont {Mazets},\ and\ \citenamefont {Kurizki}}]{wgphoton}%
  \BibitemOpen
  \bibfield  {author} {\bibinfo {author} {\bibfnamefont {E.}~\bibnamefont {Shahmoon}}, \bibinfo {author} {\bibfnamefont {I.}~\bibnamefont {Mazets}},\ and\ \bibinfo {author} {\bibfnamefont {G.}~\bibnamefont {Kurizki}},\ }\bibfield  {title} {\bibinfo {title} {Giant vacuum forces via transmission lines},\ }\href {https://doi.org/10.1073/pnas.1401346111} {\bibfield  {journal} {\bibinfo  {journal} {Proc.\ Natl.\ Acad. Sci.\ U S A.}\ }\textbf {\bibinfo {volume} {111}},\ \bibinfo {pages} {10485} (\bibinfo {year} {2014})}\BibitemShut {NoStop}%
\bibitem [{\citenamefont {Del~Debbio}\ \emph {et~al.}(2012)\citenamefont {Del~Debbio}, \citenamefont {Hart},\ and\ \citenamefont {Rinaldi}}]{DelDebbio:2012mr}%
  \BibitemOpen
  \bibfield  {author} {\bibinfo {author} {\bibfnamefont {L.}~\bibnamefont {Del~Debbio}}, \bibinfo {author} {\bibfnamefont {A.}~\bibnamefont {Hart}},\ and\ \bibinfo {author} {\bibfnamefont {E.}~\bibnamefont {Rinaldi}},\ }\bibfield  {title} {\bibinfo {title} {Light scalars in strongly-coupled extra-dimensional theories},\ }\href {https://doi.org/10.1007/JHEP07(2012)178} {\bibfield  {journal} {\bibinfo  {journal} {JHEP}\ }\textbf {\bibinfo {volume} {07}},\ \bibinfo {pages} {178}},\ \Eprint {https://arxiv.org/abs/1203.2116} {arXiv:1203.2116 [hep-lat]} \BibitemShut {NoStop}%
\bibitem [{\citenamefont {Antonuccio}\ and\ \citenamefont {Dalley}(1996)}]{Antonuccio:1995rb}%
  \BibitemOpen
  \bibfield  {author} {\bibinfo {author} {\bibfnamefont {F.}~\bibnamefont {Antonuccio}}\ and\ \bibinfo {author} {\bibfnamefont {S.}~\bibnamefont {Dalley}},\ }\bibfield  {title} {\bibinfo {title} {{Glueballs from (1+1)-dimensional gauge theories with transverse degrees of freedom}},\ }\href {https://doi.org/10.1016/0550-3213(95)00662-1} {\bibfield  {journal} {\bibinfo  {journal} {Nucl. Phys. B}\ }\textbf {\bibinfo {volume} {461}},\ \bibinfo {pages} {275} (\bibinfo {year} {1996})},\ \Eprint {https://arxiv.org/abs/hep-ph/9506456} {arXiv:hep-ph/9506456} \BibitemShut {NoStop}%
\bibitem [{\citenamefont {Nyhegn}\ \emph {et~al.}(2021)\citenamefont {Nyhegn}, \citenamefont {Chung},\ and\ \citenamefont {Burrello}}]{Nyhegn:2020cxu}%
  \BibitemOpen
  \bibfield  {author} {\bibinfo {author} {\bibfnamefont {J.}~\bibnamefont {Nyhegn}}, \bibinfo {author} {\bibfnamefont {C.-M.}\ \bibnamefont {Chung}},\ and\ \bibinfo {author} {\bibfnamefont {M.}~\bibnamefont {Burrello}},\ }\bibfield  {title} {\bibinfo {title} {{$\mathbb{Z}_N$ lattice gauge theory in a ladder geometry}},\ }\href {https://doi.org/10.1103/PhysRevResearch.3.013133} {\bibfield  {journal} {\bibinfo  {journal} {Phys. Rev. Res.}\ }\textbf {\bibinfo {volume} {3}},\ \bibinfo {pages} {013133} (\bibinfo {year} {2021})},\ \Eprint {https://arxiv.org/abs/2011.06534} {arXiv:2011.06534 [quant-ph]} \BibitemShut {NoStop}%
\bibitem [{\citenamefont {Pradhan}\ \emph {et~al.}(2024)\citenamefont {Pradhan}, \citenamefont {Maroncelli},\ and\ \citenamefont {Ercolessi}}]{Pradhan:2022lzo}%
  \BibitemOpen
  \bibfield  {author} {\bibinfo {author} {\bibfnamefont {S.}~\bibnamefont {Pradhan}}, \bibinfo {author} {\bibfnamefont {A.}~\bibnamefont {Maroncelli}},\ and\ \bibinfo {author} {\bibfnamefont {E.}~\bibnamefont {Ercolessi}},\ }\bibfield  {title} {\bibinfo {title} {Discrete {A}belian lattice gauge theories on a ladder and their dualities with quantum clock models},\ }\href {https://doi.org/10.1103/PhysRevB.109.064410} {\bibfield  {journal} {\bibinfo  {journal} {Phys. Rev. B}\ }\textbf {\bibinfo {volume} {109}},\ \bibinfo {pages} {064410} (\bibinfo {year} {2024})},\ \Eprint {https://arxiv.org/abs/2208.04182} {arXiv:2208.04182 [hep-lat]} \BibitemShut {NoStop}%
\bibitem [{\citenamefont {Yao}(2023)}]{Yao:2023pht}%
  \BibitemOpen
  \bibfield  {author} {\bibinfo {author} {\bibfnamefont {X.}~\bibnamefont {Yao}},\ }\bibfield  {title} {\bibinfo {title} {{SU(2)} gauge theory in {$2+1$} dimensions on a plaquette chain obeys the eigenstate thermalization hypothesis},\ }\href {https://doi.org/10.1103/PhysRevD.108.L031504} {\bibfield  {journal} {\bibinfo  {journal} {Phys. Rev. D}\ }\textbf {\bibinfo {volume} {108}},\ \bibinfo {pages} {L031504} (\bibinfo {year} {2023})},\ \Eprint {https://arxiv.org/abs/2303.14264} {arXiv:2303.14264 [hep-lat]} \BibitemShut {NoStop}%
\bibitem [{\citenamefont {Ebner}\ \emph {et~al.}(2024)\citenamefont {Ebner}, \citenamefont {Sch\"afer}, \citenamefont {Seidl}, \citenamefont {M\"uller},\ and\ \citenamefont {Yao}}]{PhysRevD.110.014505}%
  \BibitemOpen
  \bibfield  {author} {\bibinfo {author} {\bibfnamefont {L.}~\bibnamefont {Ebner}}, \bibinfo {author} {\bibfnamefont {A.}~\bibnamefont {Sch\"afer}}, \bibinfo {author} {\bibfnamefont {C.}~\bibnamefont {Seidl}}, \bibinfo {author} {\bibfnamefont {B.}~\bibnamefont {M\"uller}},\ and\ \bibinfo {author} {\bibfnamefont {X.}~\bibnamefont {Yao}},\ }\bibfield  {title} {\bibinfo {title} {Entanglement entropy of {($2+1$)}-dimensional {SU(2)} lattice gauge theory on plaquette chains},\ }\href {https://doi.org/10.1103/PhysRevD.110.014505} {\bibfield  {journal} {\bibinfo  {journal} {Phys. Rev. D}\ }\textbf {\bibinfo {volume} {110}},\ \bibinfo {pages} {014505} (\bibinfo {year} {2024})}\BibitemShut {NoStop}%
\bibitem [{\citenamefont {Popov}\ \emph {et~al.}(2025)\citenamefont {Popov}, \citenamefont {Ballini}, \citenamefont {Bottarelli}, \citenamefont {Burrello}, \citenamefont {Silvi}, \citenamefont {Wauters},\ and\ \citenamefont {Hauke}}]{Popov:2025byi}%
  \BibitemOpen
  \bibfield  {author} {\bibinfo {author} {\bibfnamefont {P.~P.}\ \bibnamefont {Popov}}, \bibinfo {author} {\bibfnamefont {E.}~\bibnamefont {Ballini}}, \bibinfo {author} {\bibfnamefont {A.}~\bibnamefont {Bottarelli}}, \bibinfo {author} {\bibfnamefont {M.}~\bibnamefont {Burrello}}, \bibinfo {author} {\bibfnamefont {P.}~\bibnamefont {Silvi}}, \bibinfo {author} {\bibfnamefont {M.~M.}\ \bibnamefont {Wauters}},\ and\ \bibinfo {author} {\bibfnamefont {P.}~\bibnamefont {Hauke}},\ }\bibfield  {title} {\bibinfo {title} {{When the center matters: color screening and gluelumps in dihedral lattice gauge theories}},\ }\href@noop {} {\bibfield  {journal} {\bibinfo  {journal} {Preprint}\ } (\bibinfo {year} {2025})},\ \Eprint {https://arxiv.org/abs/2512.00152} {arXiv:2512.00152 [hep-lat]} \BibitemShut {NoStop}%
\bibitem [{\citenamefont {Ciavarella}\ \emph {et~al.}(2021)\citenamefont {Ciavarella}, \citenamefont {Klco},\ and\ \citenamefont {Savage}}]{PhysRevD.103.094501}%
  \BibitemOpen
  \bibfield  {author} {\bibinfo {author} {\bibfnamefont {A.}~\bibnamefont {Ciavarella}}, \bibinfo {author} {\bibfnamefont {N.}~\bibnamefont {Klco}},\ and\ \bibinfo {author} {\bibfnamefont {M.~J.}\ \bibnamefont {Savage}},\ }\bibfield  {title} {\bibinfo {title} {{Trailhead for quantum simulation of SU(3) Yang-Mills lattice gauge theory in the local multiplet basis}},\ }\href {https://doi.org/10.1103/PhysRevD.103.094501} {\bibfield  {journal} {\bibinfo  {journal} {Phys. Rev. D}\ }\textbf {\bibinfo {volume} {103}},\ \bibinfo {pages} {094501} (\bibinfo {year} {2021})}\BibitemShut {NoStop}%
\bibitem [{\citenamefont {Kadam}\ \emph {et~al.}(2025)\citenamefont {Kadam}, \citenamefont {Naskar}, \citenamefont {Raychowdhury},\ and\ \citenamefont {Stryker}}]{PhysRevD.111.074516}%
  \BibitemOpen
  \bibfield  {author} {\bibinfo {author} {\bibfnamefont {S.~V.}\ \bibnamefont {Kadam}}, \bibinfo {author} {\bibfnamefont {A.}~\bibnamefont {Naskar}}, \bibinfo {author} {\bibfnamefont {I.}~\bibnamefont {Raychowdhury}},\ and\ \bibinfo {author} {\bibfnamefont {J.~R.}\ \bibnamefont {Stryker}},\ }\bibfield  {title} {\bibinfo {title} {{Loop-string-hadron approach to SU(3) lattice Yang-Mills theory: Hilbert space of a trivalent vertex}},\ }\href {https://doi.org/10.1103/PhysRevD.111.074516} {\bibfield  {journal} {\bibinfo  {journal} {Phys. Rev. D}\ }\textbf {\bibinfo {volume} {111}},\ \bibinfo {pages} {074516} (\bibinfo {year} {2025})}\BibitemShut {NoStop}%
\bibitem [{\citenamefont {Chen}\ \emph {et~al.}(2026)\citenamefont {Chen}, \citenamefont {M{\"u}ller},\ and\ \citenamefont {Yao}}]{Chen:2026hnh}%
  \BibitemOpen
  \bibfield  {author} {\bibinfo {author} {\bibfnamefont {V.}~\bibnamefont {Chen}}, \bibinfo {author} {\bibfnamefont {B.}~\bibnamefont {M{\"u}ller}},\ and\ \bibinfo {author} {\bibfnamefont {X.}~\bibnamefont {Yao}},\ }\bibfield  {title} {\bibinfo {title} {{Minimally truncated SU(3) lattice gauge theory and string tension}},\ }\href@noop {} {\bibfield  {journal} {\bibinfo  {journal} {Preprint}\ } (\bibinfo {year} {2026})},\ \Eprint {https://arxiv.org/abs/2601.10065} {arXiv:2601.10065 [hep-lat]} \BibitemShut {NoStop}%
\bibitem [{\citenamefont {Elitzur}\ \emph {et~al.}(1979)\citenamefont {Elitzur}, \citenamefont {Pearson},\ and\ \citenamefont {Shigemitsu}}]{PhysRevD.19.3698}%
  \BibitemOpen
  \bibfield  {author} {\bibinfo {author} {\bibfnamefont {S.}~\bibnamefont {Elitzur}}, \bibinfo {author} {\bibfnamefont {R.~B.}\ \bibnamefont {Pearson}},\ and\ \bibinfo {author} {\bibfnamefont {J.}~\bibnamefont {Shigemitsu}},\ }\bibfield  {title} {\bibinfo {title} {{Phase structure of discrete Abelian spin and gauge systems}},\ }\href {https://doi.org/10.1103/PhysRevD.19.3698} {\bibfield  {journal} {\bibinfo  {journal} {Phys. Rev. D}\ }\textbf {\bibinfo {volume} {19}},\ \bibinfo {pages} {3698} (\bibinfo {year} {1979})}\BibitemShut {NoStop}%
\bibitem [{\citenamefont {Fateev}\ and\ \citenamefont {Zamolodchikov}(1985)}]{Fateev:1985mm}%
  \BibitemOpen
  \bibfield  {author} {\bibinfo {author} {\bibfnamefont {V.~A.}\ \bibnamefont {Fateev}}\ and\ \bibinfo {author} {\bibfnamefont {A.~B.}\ \bibnamefont {Zamolodchikov}},\ }\bibfield  {title} {\bibinfo {title} {{Parafermionic currents in the two-dimensional conformal quantum field theory and self-dual critical points in $Z_N$ symmetric statistical systems}},\ }\href@noop {} {\bibfield  {journal} {\bibinfo  {journal} {Sov. Phys. JETP}\ }\textbf {\bibinfo {volume} {62}},\ \bibinfo {pages} {215} (\bibinfo {year} {1985})}\BibitemShut {NoStop}%
\bibitem [{\citenamefont {Lepori}\ \emph {et~al.}(2009)\citenamefont {Lepori}, \citenamefont {Toth},\ and\ \citenamefont {Delfino}}]{Lepori:2009ip}%
  \BibitemOpen
  \bibfield  {author} {\bibinfo {author} {\bibfnamefont {L.}~\bibnamefont {Lepori}}, \bibinfo {author} {\bibfnamefont {G.~Z.}\ \bibnamefont {Toth}},\ and\ \bibinfo {author} {\bibfnamefont {G.}~\bibnamefont {Delfino}},\ }\bibfield  {title} {\bibinfo {title} {Particle spectrum of the 3-state {P}otts field theory: A numerical study},\ }\href {https://doi.org/10.1088/1742-5468/2009/11/P11007} {\bibfield  {journal} {\bibinfo  {journal} {J. Stat. Mech.}\ }\textbf {\bibinfo {volume} {0911}},\ \bibinfo {pages} {P11007} (\bibinfo {year} {2009})},\ \Eprint {https://arxiv.org/abs/0909.2192} {arXiv:0909.2192 [hep-th]} \BibitemShut {NoStop}%
\bibitem [{\citenamefont {Cardy}(1996)}]{cardy1996scaling}%
  \BibitemOpen
  \bibfield  {author} {\bibinfo {author} {\bibfnamefont {J.}~\bibnamefont {Cardy}},\ }\href@noop {} {\emph {\bibinfo {title} {Scaling and renormalization in statistical physics}}},\ Vol.~\bibinfo {volume} {5}\ (\bibinfo  {publisher} {Cambridge University Press},\ \bibinfo {year} {1996})\BibitemShut {NoStop}%
\bibitem [{\citenamefont {Mong}\ \emph {et~al.}(2014)\citenamefont {Mong}, \citenamefont {Clarke}, \citenamefont {Alicea}, \citenamefont {Lindner},\ and\ \citenamefont {Fendley}}]{Mong_2014}%
  \BibitemOpen
  \bibfield  {author} {\bibinfo {author} {\bibfnamefont {R.~S.~K.}\ \bibnamefont {Mong}}, \bibinfo {author} {\bibfnamefont {D.~J.}\ \bibnamefont {Clarke}}, \bibinfo {author} {\bibfnamefont {J.}~\bibnamefont {Alicea}}, \bibinfo {author} {\bibfnamefont {N.~H.}\ \bibnamefont {Lindner}},\ and\ \bibinfo {author} {\bibfnamefont {P.}~\bibnamefont {Fendley}},\ }\bibfield  {title} {\bibinfo {title} {Parafermionic conformal field theory on the lattice},\ }\href {https://doi.org/10.1088/1751-8113/47/45/452001} {\bibfield  {journal} {\bibinfo  {journal} {J. Phys. A: Math. Theor.}\ }\textbf {\bibinfo {volume} {47}},\ \bibinfo {pages} {452001} (\bibinfo {year} {2014})}\BibitemShut {NoStop}%
\bibitem [{\citenamefont {Cardy}(1986)}]{CARDY1986186}%
  \BibitemOpen
  \bibfield  {author} {\bibinfo {author} {\bibfnamefont {J.~L.}\ \bibnamefont {Cardy}},\ }\bibfield  {title} {\bibinfo {title} {Operator content of two-dimensional conformally invariant theories},\ }\href {https://doi.org/https://doi.org/10.1016/0550-3213(86)90552-3} {\bibfield  {journal} {\bibinfo  {journal} {Nuclear Physics B}\ }\textbf {\bibinfo {volume} {270}},\ \bibinfo {pages} {186} (\bibinfo {year} {1986})}\BibitemShut {NoStop}%
\bibitem [{\citenamefont {Fishman}\ \emph {et~al.}(2022{\natexlab{a}})\citenamefont {Fishman}, \citenamefont {White},\ and\ \citenamefont {Stoudenmire}}]{ITensor}%
  \BibitemOpen
  \bibfield  {author} {\bibinfo {author} {\bibfnamefont {M.}~\bibnamefont {Fishman}}, \bibinfo {author} {\bibfnamefont {S.~R.}\ \bibnamefont {White}},\ and\ \bibinfo {author} {\bibfnamefont {E.~M.}\ \bibnamefont {Stoudenmire}},\ }\bibfield  {title} {\bibinfo {title} {{The ITensor software library for tensor network calculations}},\ }\href {https://doi.org/10.21468/SciPostPhysCodeb.4} {\bibfield  {journal} {\bibinfo  {journal} {SciPost Phys. Codebases}\ ,\ \bibinfo {pages} {4}} (\bibinfo {year} {2022}{\natexlab{a}})}\BibitemShut {NoStop}%
\bibitem [{\citenamefont {Fishman}\ \emph {et~al.}(2022{\natexlab{b}})\citenamefont {Fishman}, \citenamefont {White},\ and\ \citenamefont {Stoudenmire}}]{ITensor-r0.3}%
  \BibitemOpen
  \bibfield  {author} {\bibinfo {author} {\bibfnamefont {M.}~\bibnamefont {Fishman}}, \bibinfo {author} {\bibfnamefont {S.~R.}\ \bibnamefont {White}},\ and\ \bibinfo {author} {\bibfnamefont {E.~M.}\ \bibnamefont {Stoudenmire}},\ }\bibfield  {title} {\bibinfo {title} {{Codebase release 0.3 for ITensor}},\ }\href {https://doi.org/10.21468/SciPostPhysCodeb.4-r0.3} {\bibfield  {journal} {\bibinfo  {journal} {SciPost Phys. Codebases}\ ,\ \bibinfo {pages} {4}} (\bibinfo {year} {2022}{\natexlab{b}})}\BibitemShut {NoStop}%
\bibitem [{sup(2026)}]{supp}%
  \BibitemOpen
  \href@noop {} {} \bibinfo {note} {See the Supplementary Material at [url] for the fitted data and plots.}\BibitemShut {Stop}%
\bibitem [{\citenamefont {Heller}(1995)}]{Heller:1995bz}%
  \BibitemOpen
  \bibfield  {author} {\bibinfo {author} {\bibfnamefont {U.~M.}\ \bibnamefont {Heller}},\ }\bibfield  {title} {\bibinfo {title} {{SU(3)} lattice gauge theory in the fundamental adjoint plane and scaling along the {W}ilson axis},\ }\href {https://doi.org/10.1016/0370-2693(95)01186-T} {\bibfield  {journal} {\bibinfo  {journal} {Phys. Lett. B}\ }\textbf {\bibinfo {volume} {362}},\ \bibinfo {pages} {123} (\bibinfo {year} {1995})},\ \Eprint {https://arxiv.org/abs/hep-lat/9508009} {arXiv:hep-lat/9508009} \BibitemShut {NoStop}%
\bibitem [{\citenamefont {Fateev}(1991)}]{Fateev91}%
  \BibitemOpen
  \bibfield  {author} {\bibinfo {author} {\bibfnamefont {V.}~\bibnamefont {Fateev}},\ }\bibfield  {title} {\bibinfo {title} {Integrable deformations in {$Z_N$}-symmetrical models of the conformal quantum field theory},\ }\href {https://doi.org/10.1142/S0217751X91001052} {\bibfield  {journal} {\bibinfo  {journal} {Int. J. Mod. Phys. A}\ }\textbf {\bibinfo {volume} {06}},\ \bibinfo {pages} {2109} (\bibinfo {year} {1991})},\ \Eprint {https://arxiv.org/abs/https://doi.org/10.1142/S0217751X91001052} {https://doi.org/10.1142/S0217751X91001052} \BibitemShut {NoStop}%
\end{thebibliography}
\end{document}